\documentclass[prd,aps,twocolumn,preprintnumbers,amsmath,amssymb,nofootinbib,superscriptaddress,notitlepage]{revtex4-1}

\usepackage{float}
\usepackage{epsfig}
\usepackage[utf8]{inputenc}
\usepackage{color}
\usepackage{dsfont} 
\usepackage{amsbsy} 
\usepackage{mathrsfs} 
\usepackage{graphicx}
\usepackage[colorlinks=true,
linkcolor=blue,
breaklinks=true,
urlcolor=blue,
citecolor=blue]{hyperref}
\usepackage{ulem}

\newcommand{\be}{\begin{equation}}
\newcommand{\ee}{\end{equation}}
\newcommand{\bea}{\begin{eqnarray}}
\newcommand{\eea}{\end{eqnarray}}
\newcommand{\ba}{\begin{array}}
\newcommand{\ea}{\end{array}}
\newcommand{\bi}{\begin{itemize}}
\newcommand{\ei}{\end{itemize}}

\renewcommand{\vec}[1]{\mbox{\boldmath $#1 \!\!$ \unboldmath}}

\newcommand{\lf}{\left}
\newcommand{\rg}{\right}

\newcommand{\ucas}{\affiliation{University of Chinese Academy of Sciences, Beijing 100049, China}}

\newcommand{\imp}{\affiliation{Institute of Modern Physics, Chinese Academy of Sciences, Lanzhou 730000, China}}

\newcommand{\csr}{\affiliation{Research Center for Hadron and CSR Physics, Lanzhou University and Institute of Modern Physics of CAS, Lanzhou 730000, China}}

\newcommand{\hebei}{\affiliation{College of Physics Science and Technology, Hebei University, Baoding 071002, China}}

\begin{document}

\title{Manipulating Bell nonlocality and entanglement in polarized electron-positron annihilation}

\author{Hong-Wei Zhang} \email{hweiz0929@163.com}
\hebei
\imp

\author{Xu Cao} \email{caoxu@impcas.ac.cn}
\imp
\ucas
\csr

\author{Tai-Fu Feng} \email{fengtf@hbu.edu.cn}
\hebei

\date{\today}

\begin{abstract}
  \rule{0ex}{3ex}
    The hyperon-antihyperon pairs produced in electron-positron annihilation as a massive two-qubit quantum system can be used to study the quantum correlations at high energies.
    This paper is theoretically dedicated to how polarization of lepton beams manipulate the Bell nonlocality and entanglement of hyperon pairs system.
    The response of CHSH parameter, concurrence, and negativity to the polarization degree of beam is numerically calculated by exploiting the joint spin density matrix of hyperon-antihyperon pairs.
    Different influences of longitudinal and transverse polarization of beams on entanglement are found and compared.
    The results provide alternative perspectives for the decay of charmonium to hyperon pairs.
\end{abstract}

\maketitle

\section{Introduction}

High energy particle physics enables a new avenue of investigating entanglement and Bell non-locality, two defining characteristics of quantum mechanics which have been convincingly tested in low-energy photonic and atomic systems \cite{Aspect:1981zz,Wiseman:2007hyt,Giustina:2015yza}.
Elementary particles, e.g. top quarks \cite{Afik:2020onf,Fabbrichesi:2021npl,ATLAS:2023fsd,CMS:2024pts} and neutrinos \cite{Formaggio:2016cuh} was exploited to witness the entanglement at high energy experiment.
Plenty of new proposals, for instance top-quark and $\tau$-lepton pair at unpolarized hadron and lepton colliders \cite{Fabbrichesi:2022ovb} and polarized lepton colliders \cite{Altakach:2026fpl,Guo:2026yhz}, massless quark pair in dihadron pair production at lepton colliders \cite{Cheng:2025cuv}, and electron-quark or quark-antiquark pairs at Electron-Ion Colliders \cite{Fucilla:2025kit,Qi:2025onf,Cheng:2025zaw,Hatta:2025obw}, was declared as alternative environment for explore quantum entanglement.

From another landscape, maximal entanglement is conjectured as a probe of many interesting physics in hadronic field.
Maximum entanglement between helicity states has been previously proposed to constrain quantum electrodynamics and weak mixing angle \cite{Cervera-Lierta:2017tdt}.
Maximal entanglement of partons seems to provide a new perspective in color confinement and evolution of quantum chromodynamics (QCD) \cite{Kharzeev:2017qzs,Tu:2019ouv,Hentschinski:2023izh,Hentschinski:2024gaa} as well as hadronization \cite{Datta:2024hpn}.
Maximal entanglement would also facilitate the study of rich spin-dependent proton-proton interactions \cite{Shen:2025aqf}.

The connection between entanglement and (fundamental and approximate) symmetries of particle interactions is enlightening.
Entanglement suppression in the strong interaction is found to be correlated with approximate spin-flavor symmetries, leading to emergent symmetries in low-energy baryon interactions \cite{Beane:2018oxh}.
Entangled elementary particle systems would constrain new physics beyond the Standard Model \cite{Fabbrichesi:2022ovb,LoChiatto:2024dmx}.
The parity-violating interactions in spin half bipartite systems is shown to disentangle the entanglement of particle pairs \cite{Du:2024sly}.
The influence of spin and lifetime entanglement on the decay of entangled particles is also investigated \cite{Tang:2025oav}.
The quantum correlated $D\bar{D}$ pairs are used to measure the strong-phase differences between their decay amplitudes \cite{BESIII:2025xed,BESIII:2025pod}.

Hyperon-antihyperon pairs produced in electron-positron annihilation can be regarded as massive qubit system \cite{Tornqvist:1980af}.
Several scenarios for the observation of quantum entanglement at electron-positron colliders are suggested \cite{Qian:2020ini,Khan:2020seu,Pei:2025yvr,Fabbrichesi:2024rec,Wu:2024mtj,Wu:2024asu,Hong:2025drg,Jaloum:2025bkx,Wu:2025dds},
leading to the recent experimental claim of the exclusion of local hidden variable theory (LHVT) after addressing several loopholes by BESIII collaboration \cite{BESIII:2025vsr}.
Although experiments at colliders, in principle, do not rule out all types of hidden variable theories as discussed by several critical appraisals
\cite{Abel:1992kz,Dreiner:1992gt,Li:2024luk,Low:2025aqq,Bechtle:2025ugc,Abel:2025skj,Ai:2025wnt}, the study of entanglement of elementary and composite particles is of its own interest from theoretical perspective.
The non-trivial correlation of helcicity states of virtual photons or charmonium produced in electron-positron annihilation results from the fact that helicity of lepton is effectively conserved in the vertex in the massless limit \cite{Cao:2024tvz}.
This is a feature of Quantum Electrodynamics (QED) related to the chiral nature of the interaction that
the chiral eigenstates are the same as the helicity eigenstates for both massless particle and antiparticle spinors.
Spin density matrix of hyperon and antihyperon carries forward those correlation of helcicity states by virtue of the non-zero relative phase between electro and magnetic form factor in timelike region.
In all previous studies, hyperon-antihyperon pairs are considered to be produced in unpolarized electron-positron collisions.
This work will study how the longitudinal and transverse beam polarization manipulate the Bell
nonlocality and entanglement of hyperon–antihyperon pairs.

This paper is organized as follows.
In Sec.~\ref{sec:preparation}, the formalism of Bell nonlocality, concurrence, and negativity are introduced  based on the spin density matrix incorporating beam polarization effects.
The influence of longitudinal and transverse beam polarization is present respectively in Sec.~\ref{sec:PL} and Sec.~\ref{sec:PT}.
Hierarchy of quantum correlations and separable state are discussed in Sec.~\ref{sec:disc}, and the conclusion is briefly given in Sec.~\ref{sec:conclusion}.

\section{Theoretical setup} \label{sec:preparation}
\subsection{Spin Density Matrix}

\begin{figure}[!t]
  \includegraphics[scale=0.42]{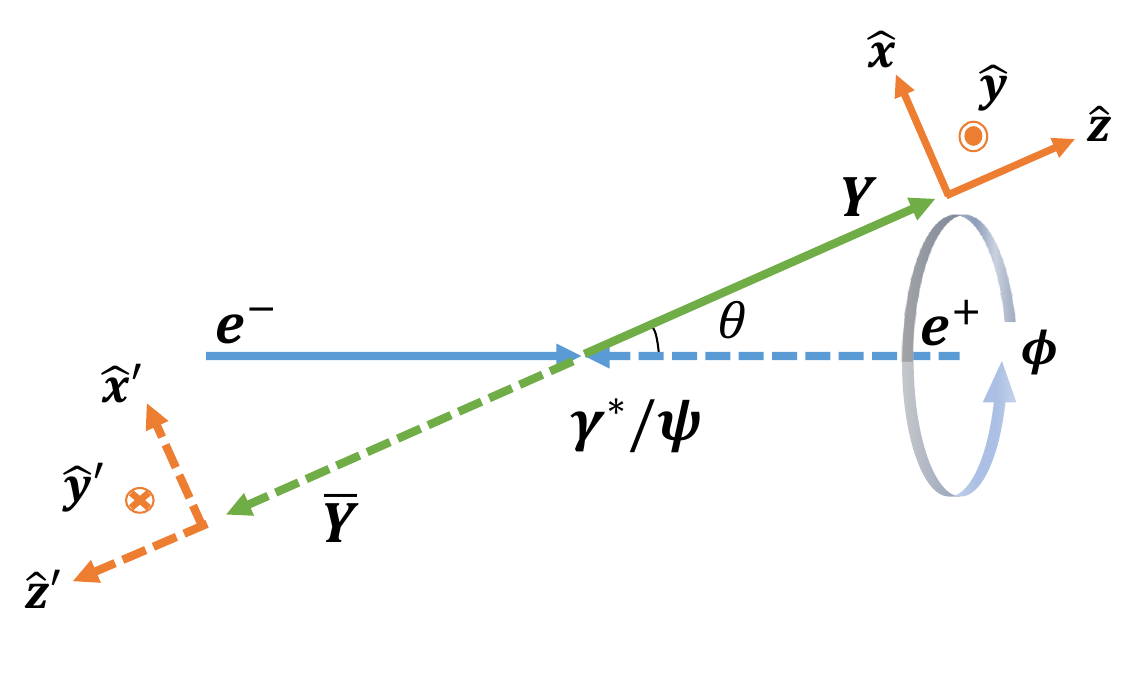}
  \caption{The process of $e^{+}e^{-} \rightarrow \gamma^* /\psi \rightarrow Y\bar{Y}$ in the c.m. frame. The coordinate system of $Y$ and $\bar{Y}$ are chosen to be of the same chirality with $\{\hat{\mathbf{x}},\hat{\mathbf{y}},\hat{\mathbf{z}}\}$ being three axes in the helicity rest frame of $Y$ and $\{\hat{\mathbf{x}}',\hat{\mathbf{y}}',\hat{\mathbf{z}}'\}$ of $\bar{Y}$. For hyperon $Y$, $\hat{\mathbf{y}}={(\hat{\mathbf{p}}_{e}\times\hat{\mathbf{p}}_{Y})}/{\left|\hat{\mathbf{p}}_{e}\times\hat{\mathbf{p}}_{Y}\right|},\ \hat{\mathbf{z}}=\hat{\mathbf{p}}_{Y},\ \hat{\mathbf{x}}=\hat{\mathbf{y}}\times\hat{\mathbf{z}}$, where $\hat{\mathbf{p}}_{e}$ and $\hat{\mathbf{p}}_{Y}$ are unit vectors along momentum directions of the electron and hyperon respectively. The axes of antihyperon $\bar{Y}$ hold the transformation: $\{\hat{\mathbf{x}}',\hat{\mathbf{y}}',\hat{\mathbf{z}}'\}=\{\hat{\mathbf{x}},-\hat{\mathbf{y}},-\hat{\mathbf{z}}\}$. }\label{fig:frame}
\end{figure}

A pair of hyperon-antihyperon ($Y\bar{Y}$) both of spin half, constitutes a massive $2$-qubit system, which can be fully described by a density matrix $\rho_{Y\bar{Y}}$.
As a $(1,1)$-type tensor on the Hilbert space $\mathcal{H}=\mathcal{H}_{Y}\otimes\mathcal{H}_{\bar{Y}}$, its Bloch-Fano representation reads \cite{Fano:1983zz}
\begin{align}
\rho_{Y\bar{Y}}= \;& \frac{1}{4} \bigg(I^{(4)}+\vec{B}^{+}\cdot\boldsymbol{\sigma}\otimes I^{(2)}+I^{(2)}\otimes\vec{B}^{-}\cdot\boldsymbol{\sigma}\nonumber\\
 & +\sum_{i,j}C_{ij}\sigma_{i}\otimes\sigma_{j} \bigg), \quad i,j=1,2,3,\label{eq:two_qubit}
\end{align}
where $\vec{B}^{\pm}$ are polarization or Bloch vectors
of hyperon and antihyperon, $\boldsymbol{\sigma}$ the vector operator combined by three
Pauli operators $\{\sigma_x,\sigma_y,\sigma_z\}$, and $C_{ij} (i,j=1,2,3)$ correlation tensor.
Here $I^{(2)}$ and $I^{(4)}$ are $2\times2$ and $4\times4$ identity matrices, respectively.
Under the basis $\{\sigma_{\mu}\otimes\sigma_{\nu}\}$ with $\sigma_{0} = I^{(2)}$, the Eq.~\eqref{eq:two_qubit} also can be organized into a more compact form:
\begin{equation}
\rho_{Y\bar{Y}}=\frac{1}{4}\Theta_{\mu\nu}\sigma_{\mu}\otimes\sigma_{\nu},\quad\mu,\nu=0,1,2,3,\label{eq:2_qubit}    \end{equation}
with $\Theta_{00}=1$, $\Theta_{i0}=B_{i}^{+}$ and $\Theta_{0j}=B_{j}^{-}$, and $\Theta_{ij}=C_{ij}$ satisfied $\Theta_{\mu\nu}=\langle\sigma_{\mu}\otimes\sigma_{\nu}\rangle=\mathrm{Tr}(\rho_{Y\bar{Y}}\sigma_{\mu}\otimes\sigma_{\nu})$.

In $e^{+}e^{-} \rightarrow \gamma^* /\psi \rightarrow Y\bar{Y}$ process, the produced hyperon and antihyperon move in opposite directions within the center-of-mass (c.m.) frame due to momentum conservation, and the chosen coordinate system is illustrated in Fig.~\ref{fig:frame}.
In the case of a longitudinally polarized electron beam, or when both beams are transversely polarized,
the non-normalized $\Theta_{\mu\nu}$ reads as \cite{Perotti:2018wxm,Batozskaya:2023rek,Salone:2022lpt,Zeng:2023wqw,Cao:2024tvz,Zhang:2024rbl,Zhao:2025cbd}
\begin{widetext}
\begin{equation}
\begin{aligned}
\Theta_{\mu\nu} \propto&\left(\begin{matrix}
1 + \alpha_{\psi} \cos^2 \theta & 0 & \beta_{\psi} \sin \theta \cos \theta & 0 \\
0 & \sin^2 \theta & 0 & \gamma_{\psi} \sin \theta \cos \theta \\
-\beta_{\psi} \sin \theta \cos \theta & 0 & \alpha_{\psi} \sin^2 \theta & 0 \\
0 & -\gamma_{\psi} \sin \theta \cos \theta & 0 & - \alpha_{\psi} - \cos^2 \theta
\end{matrix}\right)
\\&+P_T^2\left(\begin{matrix}
\alpha_{\psi} \sin^2 \theta \cos 2 \phi & -\beta_{\psi} \sin \theta \sin 2 \phi & -\beta_{\psi} \sin \theta \cos \theta \cos 2 \phi & 0 \\
-\beta_{\psi} \sin \theta \sin 2 \phi & (\alpha_{\psi} + \cos^2 \theta) \cos 2 \phi & - (1 + \alpha_{\psi}) \cos \theta \sin 2 \phi & -\gamma_{\psi} \sin \theta \cos \theta \cos 2 \phi \\
\beta_{\psi} \sin \theta \cos \theta \cos 2 \phi & (1 + \alpha_{\psi}) \cos \theta \sin 2 \phi & (1 + \alpha_{\psi} \cos^2 \theta) \cos 2 \phi & -\gamma_{\psi} \sin \theta \sin 2 \phi \\
0 & \gamma_{\psi} \sin \theta \cos \theta \cos 2 \phi & -\gamma_{\psi} \sin \theta \sin 2 \phi & -\sin^2 \theta \cos 2 \phi
\end{matrix}\right)
\\&+P_L\left(\begin{matrix}
0 & \gamma_{\psi} \sin \theta & 0 & (1 + \alpha_{\psi}) \cos \theta \\
\gamma_{\psi} \sin \theta & 0 & 0 & 0 \\
0 & 0 & 0 & -\beta_{\psi} \sin \theta \\
- (1 + \alpha_{\psi}) \cos \theta & 0 & -\beta_{\psi} \sin \theta & 0
\end{matrix}\right),
\end{aligned}
\label{eq:4by4}
\end{equation}
\end{widetext}
where $\theta$ is the scattering angle between the  momenta of incoming electron and
outgoing hyperon, e.g. $\cos\theta=\hat{\mathbf{p}}_{e}\cdot\hat{\mathbf{p}}_{Y}$ in Fig.~\ref{fig:frame}.
The $\alpha_{\psi}\in[-1,1]$ is the decay parameter of the vector charmonium into hyperon and antihyperon pair, $\beta_{\psi}=\sqrt{1-{\alpha_{\psi}}^{2}}\sin\Delta\Phi$ and $\gamma_{\psi}=\sqrt{1-{\alpha_{\psi}}^{2}}\cos\Delta\Phi$,
where $\Delta\Phi\in(-\pi,\pi]$ is the relative phase between electro- and magnetic form factor. The parameters of $ J/\psi\rightarrow Y\bar{Y}$ measured by BESIII collaboration are listed in Table~\ref{tab:decay_parameters}.
The $\rho_{Y\bar{Y}}$ determined by the above is of $\mathrm{rank}\text{-}2$.

Throughout our paper the longitudinal and transverse polarization of beams are separately considered.
The Bloch vector $\vec{B}^{\pm}$ can be read from the coefficient matrix in Eq.~\eqref{eq:4by4}.
For longitudinally polarized beam, $\vec{B}_{L}^{+}$ and $\vec{B}_{L}^{-}$ are given by
\begin{equation}
\begin{aligned}
&\vec{B}_{L}^{+}=\frac{1}{\chi_L}\left(\begin{array}{c}
P_L\gamma_\psi\sin\theta\\-\beta_{\psi}\sin\theta\cos\theta\\-P_L(1+\alpha_{\psi})\cos\theta
\end{array}\right),\\
&\vec{B}_{L}^{-}=\frac{1}{\chi_L}\left(\begin{array}{c}
P_L\gamma_\psi\sin\theta\\\beta_{\psi}\sin\theta\cos\theta\\P_L(1+\alpha_{\psi})\cos\theta
\end{array}\right), \label{eq:BL}
\end{aligned}
\end{equation}
with normalization coefficient $\chi_L =1+\alpha_{\psi}\cos^{2}\theta$, and the correlation tensor $\vec{C}_L$ is
\begin{equation}
  \begin{aligned}
    &\vec{C}_L=\frac{1}{\chi_L}
\left(
\begin{matrix}
\sin^2\theta & 0 &\gamma_{\psi}\cos\theta\sin\theta\\
 0 & \alpha_{\psi}\sin^2\theta & -P_L\beta_{\psi}\sin\theta\\
-\gamma_{\psi}\cos\theta\sin\theta & -P_L\beta_{\psi}\sin\theta & -\alpha_{\psi}-\cos^2\theta
\end{matrix}
\right). \label{eq:CL}
  \end{aligned}
\end{equation}
For transversely polarized beams, $\vec{B}_{T}^{+}$ and $\vec{B}_{T}^{-}$ are
\begin{equation}
\begin{aligned}
\vec{B}_{T}^{+}&=\frac{1}{\chi_T}\left(\begin{array}{c}
-P_T^2\beta_{\psi}\sin\theta\sin 2 \phi \\
\beta_{\psi}\sin\theta\cos\theta(P_T^2\cos 2 \phi-1)\\0
\end{array}\right),\\
\vec{B}_{T}^{-}& =\frac{1}{\chi_T}\left(\begin{array}{c}
-P_T^2\beta_{\psi}\sin\theta\sin 2 \phi, \\
\beta_{\psi}\sin\theta\cos\theta(1-P_T^2\cos 2 \phi)\\0
\end{array}\right), \label{eq:BT}
\end{aligned}
\end{equation}
with $\chi_T =1+\alpha_{\psi}\cos^{2}\theta+P_T^2\alpha_{\psi}\sin^2\theta\cos 2\phi$.
The correlation tensor $\vec{C}_T$ is
\begin{widetext}
\begin{equation}
\vec{C}_T=\frac{1}{\chi_T}
\left(
\begin{matrix}
\sin^2\theta+P_T^2(\alpha_{\psi}+\cos^2\theta)\cos2\phi & -P_T^2(1+\alpha_{\psi})\cos\theta\sin 2 \phi & \gamma_{\psi}\sin\theta\cos\theta(1-P_T^2\cos 2 \phi)\\
P_T^2(1+\alpha_{\psi})\cos\theta\sin 2 \phi&\alpha_{\psi}\sin^2\theta+P_T^2 (1+\alpha_{\psi}\cos^2\theta)\cos 2 \phi & -P_T^2\gamma_{\psi}\sin\theta\sin 2 \phi\\
 -\gamma_{\psi}\sin\theta\cos\theta(1-P_T^2\cos 2 \phi)& -P_T^2\gamma_{\psi}\sin\theta\sin 2 \phi&-\alpha_{\psi}-\cos^2\theta-P_T^2\sin^2\theta\cos 2 \phi
\end{matrix}
\right).\label{eq:CT}
\end{equation}
\end{widetext}
By rotating the coordinate system $\{\hat{\mathbf{x}}', \hat{\mathbf{y}}', \hat{\mathbf{z}}'\} = \{\hat{\mathbf{x}}, \hat{\mathbf{y}}, \hat{\mathbf{z}}\}$, the coefficient matrix $\Theta_{\mu\nu}$ can be symmetrized.
Notably, Eqs.~\eqref{eq:BL} and \eqref{eq:BT} demonstrate that, unlike elementary particles, hyperons acquire transverse polarization even when produced by unpolarized lepton beams \cite{Perotti:2018wxm}.

If the $\vec{B}^{\pm}$ has only one non-zero component, by exchanging the axes corresponding to the non-zero transverse polarization component with the $\hat{z}$-axis, and afterwards diagonalizing the correlation matrix $\vec{C}$, the coefficient matrix $\Theta_{\mu\nu}$ always can be written into a simpler form \cite{Wu:2024asu}:
\begin{equation}
\Theta_{\mu\nu}\rightarrow \Theta'_{\mu\nu}=
\left(
\begin{matrix}
1 & 0 & 0 & B_z \\
0 & \lambda_1 & 0 & 0 \\
0 & 0 & \lambda_2  & 0 \\
B_z & 0 & 0 & \lambda_3
\end{matrix}\right),\label{eq:Theta}
\end{equation}
where $B_z$ is the non-zero polarization component of $\vec{B}^{\pm}$,
and $\lambda_1$, $\lambda_2$, and $\lambda_3$ are the three eigenvalues of the correlation matrix $\vec{C}$ .
With the help of Eq.~\eqref{eq:2_qubit} the density matrix $\rho$ can be expressed in the so-called $X$ state \cite{Yu:2007bwc}:
\begin{equation}
  \begin{aligned}
&\rho\rightarrow\rho^X=\frac{1}{4}\Theta'_{\mu\nu}\sigma_{\mu}\otimes\sigma_{\nu}\\
&=\frac{1}{4}
\left(
\begin{matrix}
1+2 B_z+\lambda_{3} & 0 & 0 & \lambda_{1}-\lambda_{2}\\
0 & 1-\lambda_{3} & \lambda_{1}+\lambda_{2} & 0\\
0 & \lambda_{1}+\lambda_{2} & 1-\lambda_{3} & 0\\
\lambda_{1}-\lambda_{2} & 0 & 0 & 1-2B_z+\lambda_{3}
\end{matrix} \right).
\label{eq:X_state}
\end{aligned}
\end{equation}
The $\rho$ and $\rho^X$ are local unitary equivalent:
\begin{equation}
\begin{aligned}
&\rho^X=(U\otimes V) \;\rho\;(U^\dagger \otimes V^\dagger)\\
&=\frac{1}{4}\biggl(I^{(4)}+B_z\sigma_{z}\otimes I^{(2)}+ I^{(2)}\otimes B_z\sigma_{z}+\sum_{i}\lambda_{i}\sigma_{i}\otimes\sigma_{i}\biggr).\label{eq:uni_trans}
\end{aligned}
\end{equation}
where $U$ and $V$ are local unitary operators acting on the Hilbert spaces of the respective subsystems $Y$ and $\bar{Y}$.
Under such local unitary transformations, the measures of Bell nonlocality and quantum entanglement are strictly invariant \cite{Dur:2000zz}.

For hyperon–antihyperon system $\rho_{Y\bar{Y}}$, the density matrix can be reduced to a $X$-state in the scattering angle of $\theta=0,{\pi}/{2}$ and $\pi$.
Let $\rho^{P_L}_{Y\bar{Y}}$ be the density matrix in the presence of longitudinal beam polarization and $\Theta_L$ its corresponding coefficient matrix. The parameters $\{B_z, \lambda_i (i=1,2,3)\}$ in the symmetric form $\Theta'_L$, obtained via unitary transformation on $\Theta_L$, are given by:
\begin{align}
  \notag
\theta=&\frac{\pi}{2}:\; B_z= P_L \gamma_\psi ,\\
&\qquad\lambda_{1,2}=\pm\sqrt{\alpha_\psi^2+P_L^2\beta_\psi^2},\;\lambda_3=1;&\label{eq:XparaPL1}\\
\theta=&0,\pi:\; B_z= - P_L,\;\lambda_{1,2}=0,\;\lambda_3=1.&\label{eq:XparaPL2}
\end{align}
Similarly, let $\rho^{P_T}_{Y\bar{Y}}$ denote the density matrix for the case of a transversely polarized beam and $\Theta_T$ its corresponding coefficient matrix.
The parameters $\{B_z,\lambda_i (i=1,2,3)\}$ in $\Theta'_T$ matrix are
\begin{align}
  \notag
\theta=&\frac{\pi}{2}:\;  B_z=-\frac{P_T^2 \beta_\psi \sin 2\phi}{1+P_T^2 \alpha_\psi \cos2\phi},\\
\notag
&\qquad\lambda_{1,2}=\pm\frac{\sqrt{\alpha_\psi^2 + 2P_T^2\alpha_\psi \cos2\phi+ P_T^4 \eta}}{1 + P_T^2\alpha_\psi \cos 2\phi},\\
&\qquad\lambda_3=1;\label{eq:XparaPT1}\\
\theta=&0,\pi:\; B_z=0,\;\lambda_{1,2}=\pm P_T^2, \;\lambda_3=1.\label{eq:XparaPT2}
\end{align}
with $\eta =\gamma_\psi^2\sin^2 2\phi + \cos^2 2\phi$.
Eq.~\eqref{eq:XparaPT2} corresponds to a special case of the X-state with vanishing polarization ($\vec{B}^\pm=\vec{0}$), known as a Bell diagonal state (BDS).
Under the case of unpolarized lepton beams Eqs. \eqref{eq:XparaPL1}, \eqref{eq:XparaPL2}, and \eqref{eq:XparaPT1} are also reduced to BDS, as found in literature \cite{Wu:2024asu}.
A local choice of basis allows such state to be expressed as a convex combination of the four maximally entangled Bell states \cite{Bennett:1996gf},
\begin{equation}
\rho_{Y\bar{Y}}^{\mathrm{BDS}}=\frac{1}{4}\biggl(I^{(4)}+\sum_{i}\lambda_{i}\sigma_{i}\otimes\sigma_{i}\biggr).\label{eq:BDS}
\end{equation}

\begin{table*}[htb]
\caption{\label{tab:decay_parameters} The parameters of $J/\psi\rightarrow Y\bar{Y}$ decay
for $Y = \Lambda$, $\Sigma^{+}$, $\Sigma^{0}$, $\Xi^{-}$ and $\Xi^{0}$ measured by BESIII collaboration.}
\begin{ruledtabular}
\begin{tabular}{lccc}
Decay channel & $\alpha_{\psi}$ & $\Delta\Phi/\mathrm{rad}$ & Ref. \tabularnewline
\hline
$J/\psi\rightarrow\Lambda\bar{\Lambda}$ & $0.4748 \pm 0.0022 \pm0.0031$ & $0.748\pm0.006\pm0.004$ & \cite{BESIII:2025wxe,BESIII:2022qax,BESIII:2018cnd,BESIII:2017kqw}\tabularnewline
$J/\psi\rightarrow\Sigma^{+}\bar{\Sigma}^{-}$ & $-0.5047\pm0.0018\pm0.0010$ & $-0.2744\pm0.0033\pm0.0010$ & \cite{BESIII:2025jxt,BES:2008hwe,BESIII:2020fqg}\tabularnewline
$J/\psi\rightarrow\Sigma^{0}\bar{\Sigma}^{0}$ & $-0.4133\pm0.0035\pm0.0077$ & $-0.0828\pm0.0068\pm0.0033$ & \cite{BESIII:2024nif}\tabularnewline
$J/\psi\rightarrow\Xi^{-}\bar{\Xi}^{+}$ & $ 0.5851\pm0.0044\pm0.0034$ & $1.2205\pm0.0159\pm0.0056$ & \cite{BESIII:2026hgj,BESIII:2021ypr,ParticleDataGroup:2022pth}\tabularnewline
$J/\psi\rightarrow\Xi^{0}\bar{\Xi}^{0}$ & $0.514\pm0.006\pm0.015$ & $1.168\pm0.019\pm0.018$ & \cite{BESIII:2016nix,BESIII:2023drj}\tabularnewline
\end{tabular}
\end{ruledtabular}
\end{table*}

\subsection{Bell Nonlocality and Quantum Entanglement} \label{sec:bell}

Bell nonlocality is a fundamental quantum feature that manifests through violation of Bell-type inequalities \cite{Bell:1964kc,Clauser:1978ng}.
Among these, the CHSH inequality provides an experimentally viable and widely adopted framework \cite{Clauser:1969ny}:
\begin{equation}
\left|\left\langle \hat{A}_{1}\hat{B}_{1}\right\rangle + \left\langle \hat{A}_{1}\hat{B}_{2}\right\rangle + \left\langle \hat{A}_{2}\hat{B}_{1}\right\rangle - \left\langle \hat{A}_{2}\hat{B}_{2}\right\rangle \right| \leq 2,
\label{eq:CHSH_inequality}
\end{equation}
where $\hat A_{i}$ and $\hat B_{j}$ ($i,j=1,2$) denote dichotomic observables with possible outcomes $\{0,1\}$ in two subsystems respectively.
Spin operators are defined as
$\hat{A}_{i} = \boldsymbol{\alpha}_{i} \cdot \boldsymbol{\sigma}$ and $\hat{B}_{j} = \boldsymbol{\beta}_{j} \cdot \boldsymbol{\sigma}$,
where $\boldsymbol{\alpha}_{i}$ and $\boldsymbol{\beta}_{j}$ are unit vectors specifying measurement directions for the subsystem $A$ and $B$, respectively.
Then the expectation value for any bipartite state $\rho$ is given by
\begin{equation}
\langle \hat{A}_{i}\hat{B}_{j} \rangle \equiv \operatorname{Tr} \left[ \rho \, (\boldsymbol{\alpha}_{i} \cdot \boldsymbol{\sigma} \otimes \boldsymbol{\beta}_{j} \cdot \boldsymbol{\sigma}) \right].
\label{eq:A_B}
\end{equation}
Since the measurement directions $\boldsymbol{\alpha}_{i}$ and $\boldsymbol{\beta}_{j}$ can be optimized to maximize the violation of CHSH inequality, the CHSH parameter $\mathcal{B}[\rho]$ is defined as the maximal violation of the inequality with respect to the chosen measurement directions. After substituting Eq.~\eqref{eq:A_B} into Eq.~\eqref{eq:CHSH_inequality}, it can be expressed in closed algebraic form:
\begin{align}
\mathcal{B}[\rho] &:= \max_{\boldsymbol{\alpha}_{i},\boldsymbol{\beta}_{j}} \left| \boldsymbol{\alpha}_{1}\cdot \vec{C} \cdot (\boldsymbol{\beta}_{1} + \boldsymbol{\beta}_{2}) + \boldsymbol{\alpha}_{2} \cdot \vec{C} \cdot (\boldsymbol{\beta}_{1} - \boldsymbol{\beta}_{2}) \right| \nonumber \\
&= 2\sqrt{m}.\label{eq:B}
\end{align}
Here $\vec{C}$ denotes the spin correlation matrix,
and $m$ is the sum of two largest eigenvalues of $\vec{C}^{T}\vec{C}$ \cite{Horodecki:1995nsk}, any of which $\leq 1$.
When $\mathcal{B}[\rho]$ lies within $(2, 2\sqrt{2}]$, the bipartite system is Bell nonlocal with $\mathcal{B}=2\sqrt{2}$ representing the maximal Bell nonlocality.

The bipartite states can be always described by a density matrix $\rho \in \mathscr{T}_{\mathcal{H}_A \otimes \mathcal{H}_B}$, the set of all $(1,1)$-type tensor in a Hillbert space ${\mathcal{H}_A \otimes \mathcal{H}_B}$
where $\mathcal{H}_A$ and $\mathcal{H}_B$ are the Hilbert spaces of subsystems $A$ and $B$.
If the density matrix $\rho$ can be expressed as \cite{Werner:1989zz}:
\begin{equation}
\rho = \sum_i c_i \rho_{A}^i \otimes \rho_{B}^i,
\label{eq:general_state}
\end{equation}
with $\sum_i c_i=1,\ \rho_{A}^i \in\mathscr{T}_\mathcal{H_A}$ and $\rho_{B}^i\in\mathscr{T}_\mathcal{H_B}$,
then the state $\rho$ is separable and otherwise it is entangled.
For a two-qubit system, $\rho$ is separable if and only if the partially transposed density matrix of the composite
system with respect to either subsystem $\rho^{T_A}$ or $\rho^{T_B}$ (labeled as $\rho^{\Gamma}$) is positive semidefinite, known as Positive Partial Transpose (PPT) critieron~\cite{Peres:1996dw,Horodecki:1996nc}.
Since $\rho^{\Gamma}$ can have at most one negative eigenvalue for any two-qubit state \cite{Sanpera:1998fb},
a simplified criterion is that $\rho$ is separable if and only if ~\cite{Slater:2005llp,Augusiak:2007xlq}
\begin{equation}
\det(\rho^{\Gamma}) \geq 0 \,. \label{eq:PPTdet}
\end{equation}
Since the partial transpose preserves the trace (Tr$\rho =$ Tr$\rho^{\Gamma} = 1$), the presence of negative eigenvalues in $\rho^{\Gamma}$ necessitates that the sum of the absolute values of the eigenvalues is strictly greater than $1$,
leading to the definition of the negativity as an entanglement measure~\cite{Vidal:2002zz}:
\begin{equation}
\mathcal{N}[\rho] = \frac{\sum_i |t_i| - 1}{2} ,
\end{equation}
where $\{t_i\}$ are the eigenvalues of $\rho^{\Gamma}$.
For qubit systems, the partial transposition on subsystem $A$ or $B$ is equivalent to applying the transformation $\sigma_y\to -\sigma_y$ on $\mathcal{H}_A$ or $\mathcal{H}_B$.
To provide a more comprehensive characterization of quantum correlations alongside negativity, we also consider Wootters' concurrence, defined as \cite{Wootters:1997id}:
\begin{equation}
\mathcal{C}[\rho] = \max\left\{ 0, k_{1} - k_{2} - k_{3} - k_{4} \right\},
\end{equation}
where $k_{i}\ (i=1,2,3,4)$ denote the eigenvalues (sorted in decreasing order) of the Hermitian matrix
$R = \sqrt{\sqrt{\rho} \tilde{\rho} \sqrt{\rho}}$. Here, $\tilde{\rho} = (\sigma_{y} \otimes \sigma_{y}) \rho^{*} (\sigma_{y} \otimes \sigma_{y})$ represents the spin-flipped state of $\rho$, corresponding to the transformation $\boldsymbol{\sigma}\to-\boldsymbol{\sigma}$ on $\mathcal{H}_A$ and $\mathcal{H}_B$.

Both negativity and concurrence are reliable entanglement measures for 2-qubit systems given that:
\begin{itemize}
  \item[i.] $\mathcal{C}[\rho] = 0$, $\mathcal{N}[\rho] = 0$ indicate a separable state;
  \item[ii.] $\mathcal{C}[\rho] > 0$, $\mathcal{N}[\rho] > 0$ signifie an entangled state;
  \item[iii.] $\mathcal{C}[\rho] = 1$, $\mathcal{N}[\rho] = \frac{1}{2}$ correspond to maximal entanglement.
\end{itemize}
Generally no explicit analytical relation links them. For a special case of the $\mathrm{rank}\text{-}1$ density matrix $\rho= |\psi\rangle \langle\psi|$,  $\mathcal{C}(|\psi\rangle) = 2\mathcal{N}(|\psi\rangle)$ constitutes one of the necessary conditions for the system to be in a pure state.
If $|P_L| = 1$ or $|P_T| =1$, the density matrix $\rho_{Y\bar{Y}}$ in Eq.~\eqref{eq:two_qubit} reduces to a $\mathrm{rank}\text{-}1$ matrix.

Furthermore, the Peres–Horodecki criterion states that a $X$ state in Eq.~\eqref{eq:X_state} is entangled if and only if either
$\rho_{22}^{X}\rho_{33}^{X} < |\rho_{14}^{X}|^{2}$ or $\rho_{11}^{X}\rho_{44}^{X} < |\rho_{23}^{X}|^{2}$ holds~\cite{Azuma:2010tqy} (note that two conditions cannot be simultaneously satisfied~\cite{Sanpera:1998fb}).
The Wootters' concurrence and negativity of such a $X$-state are given by~\cite{Yu:2007bwc,Wu:2024asu}
\begin{equation}
  \begin{aligned}
&\mathcal{C}[\rho^{X}]\\
&\;=2\max\left\{ 0,|\rho_{14}^{X}|-\sqrt{\rho_{22}^{X}\rho_{33}^{X}},|\rho_{23}^{X}|-\sqrt{\rho_{11}^{X}\rho_{44}^{X}}\right\}, \label{eq:concurrence}
  \end{aligned}
\end{equation}
\begin{equation}
  \begin{aligned}
&\mathcal{N}[\rho^{X}]\\
&\;\;=-\min\left\{ 0,\;\frac{\rho_{11}^{X}+\rho_{44}^{X}}{2}-\sqrt{{\lf(\frac{\rho_{11}^{X}-\rho_{44}^{X}}{2}\rg)}^2+{\rho_{23}^{X}}^2},\right.\\
&\;\;\left.\quad \frac{\rho_{22}^{X}+\rho_{33}^{X}}{2}-\sqrt{{\lf(\frac{\rho_{22}^{X}-\rho_{33}^{X}}{2}\rg)}^2+{\rho_{14}^{X}}^2}\right\}.\label{eq:negativity}
  \end{aligned}
\end{equation}
which yield $\mathcal{C}[\rho^{X}_{Y \bar Y}]= 2 \mathcal{N}[\rho^{X}_{Y \bar Y}] = |\lambda_2|$ for $\theta=0, {\pi}/{2}, \text{and } \pi$, recalling that
the eigenvalues of $\vec{C}^{T}\vec{C}$ derived from Eq.~\eqref{eq:Theta} are $\{ \lambda_1^2, \lambda_2^2, \lambda_3^2 \}$.
It is also worth noting that $\mathcal{B}[\rho^{X}_{Y\bar{Y}}]=2\sqrt{\lambda^{2}_{2}+\lambda^{2}_{3}}$ at $\theta=0, {\pi}/{2}, \text{and } \pi$.

\section{Longitudinal Beam Polarization}
\label{sec:PL}

\begin{figure}[!thb]
   \centering
   \includegraphics[width=1\linewidth]{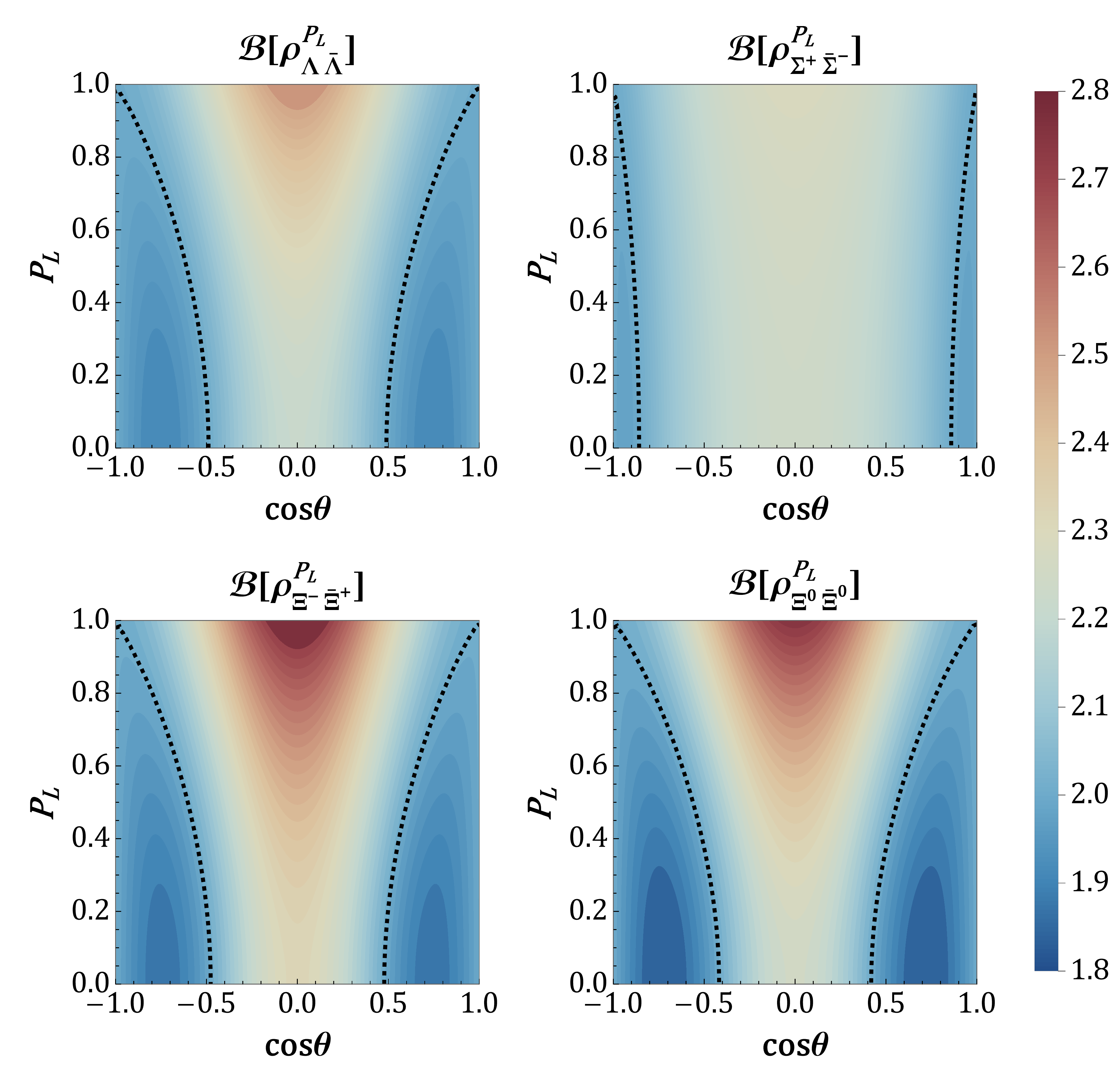}
   \caption{The CHSH parameter $\mathcal{B}[\rho^{P_L}_{Y \bar Y}]$ as a function of $\cos\theta$ ($\theta$ is the scattering angle) and longitudinal beam polarization degree $P_L$ in $J/\psi\to Y{\bar{Y}}$ for $Y = \Lambda$, $\Sigma^{+}$, $\Xi^{-}$ and $\Xi^{0}$. The dashed curve is for $\mathcal{B}[\rho^{P_L}_{Y \bar Y}] = 2$.}
   \label{fig:B_PL}
\end{figure}

\begin{figure}[!t]
  		\includegraphics[width = 0.9\linewidth]{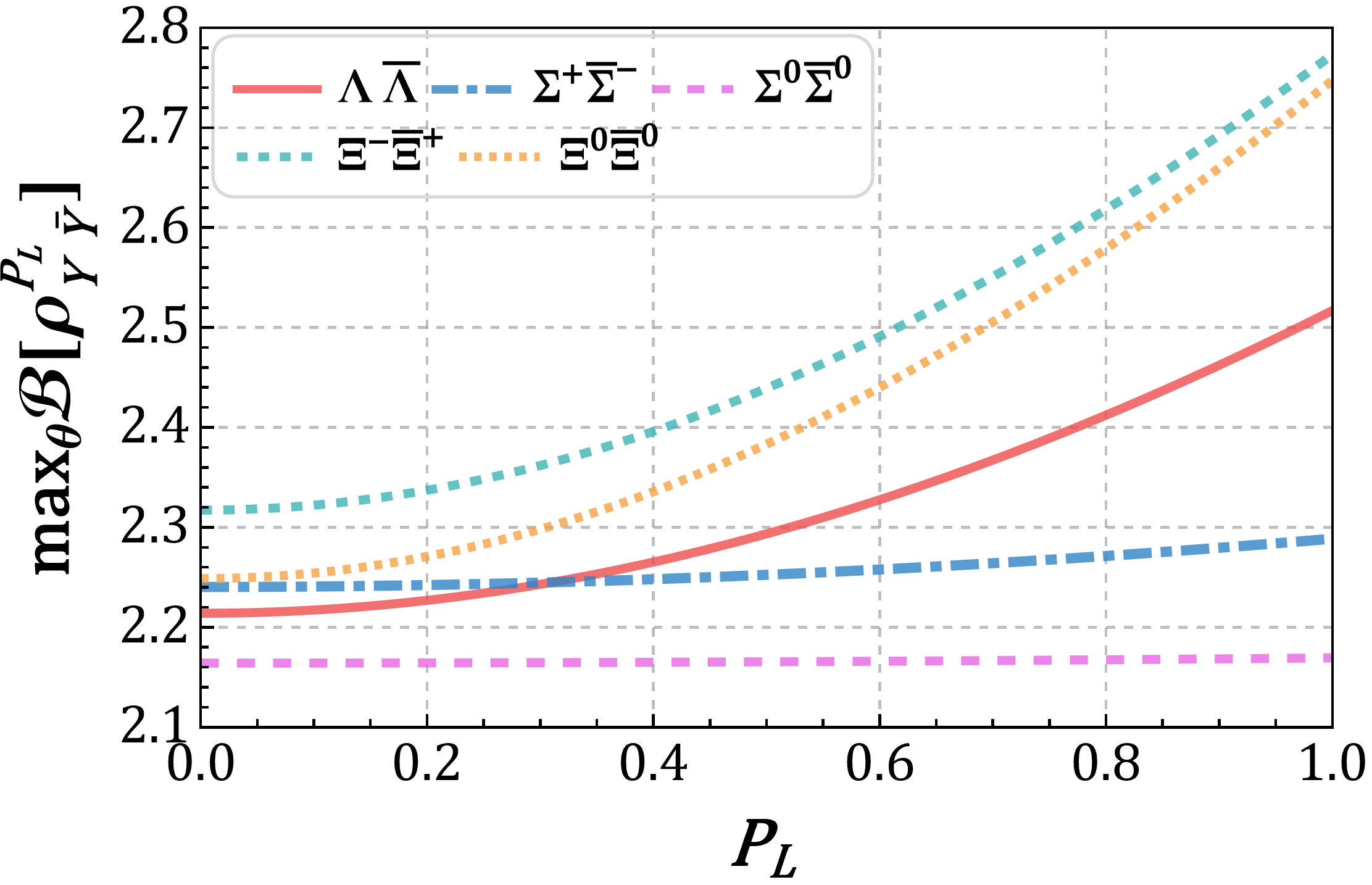}
  		\includegraphics[width = 0.9\linewidth]{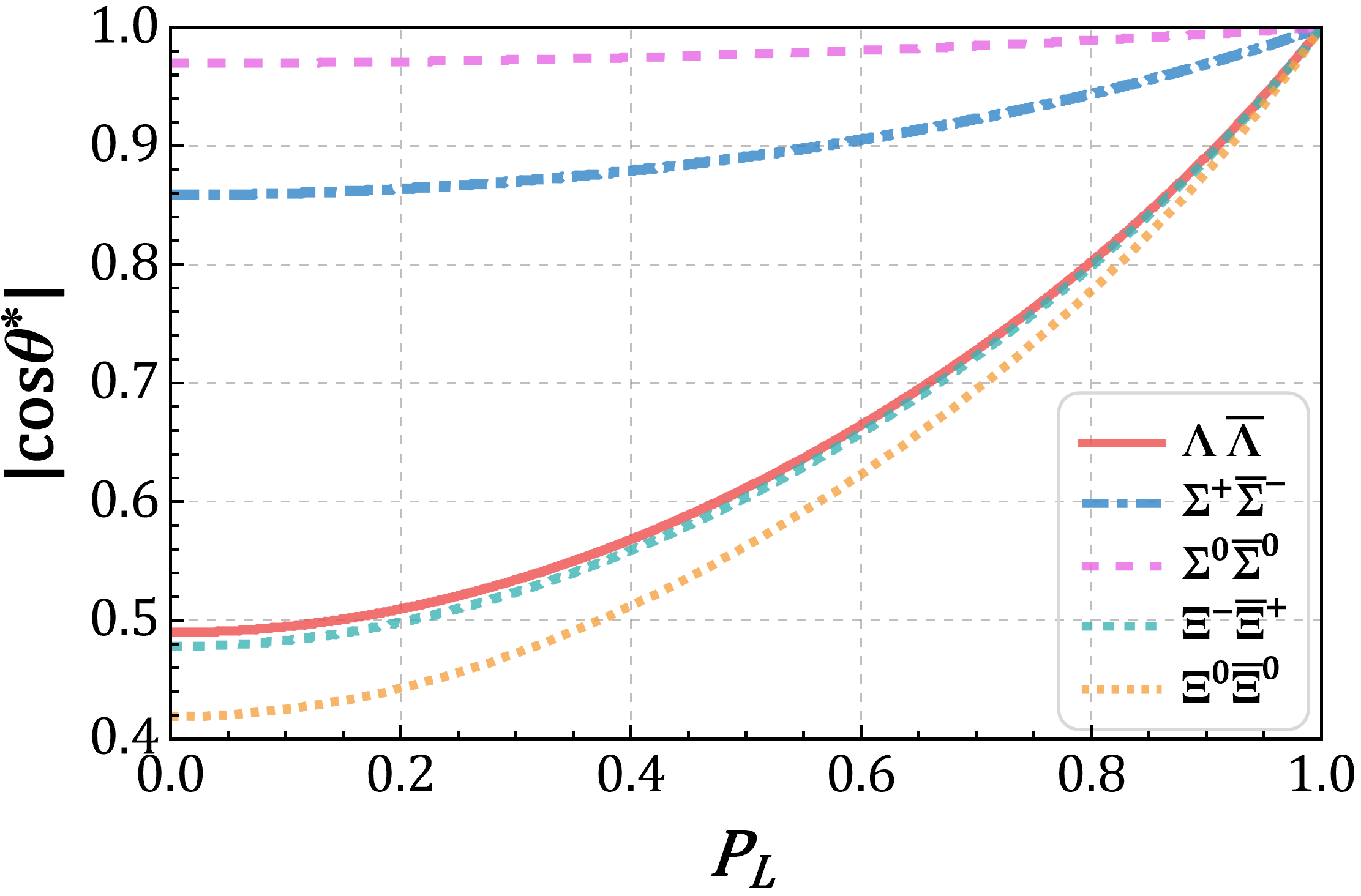}
\caption{The $\max_{\theta}\mathcal{B}[\rho^{P_L}_{Y \bar Y}]$ (upper panel) and $\cos\theta^*$ (lower panel, see main text for defination of $\theta^*$) as functions of $P_L$ in $J/\psi\to Y{\bar{Y}}$ for $Y = \Lambda$, $\Sigma^{+}$, $\Xi^{-}$ and $\Xi^{0}$.
} \label{fig:max_B_PL}
\end{figure}

The section will discuss the impact of longitudinal polarization of beams on Bell-nonlocality and quantum entanglement.
Only the results of $J/\psi \rightarrow Y\bar{Y}$ decay for $Y = \Lambda$, $\Sigma^{+}$, $\Xi^{-}$, and $\Xi^{0}$ are shown, see Appendix.~\ref{sec:Supplpsi} for $J/\psi\rightarrow \Sigma^{0} \bar{\Sigma^{0}}$ and $\psi(3686) \rightarrow Y\bar{Y}$. However, the conclusions for $J/\psi\to\Sigma^0\bar{\Sigma}^0$ will also be presented in the main text, and any conclusions that do not hold universally across all channels are explicitly noted.

The $\theta$ and $P_L$ dependence of the CHSH parameter $\mathcal{B}[\rho^{P_L}_{Y \bar Y}]$ are shown in Fig.~\ref{fig:B_PL} with the parameters in Table~\ref{tab:decay_parameters}. Note that $\mathcal{B}(\theta) = \mathcal{B}(\pi - \theta)$ with $\theta \in [0,\pi]$, and $\mathcal{B}(P_L) = \mathcal{B}(-P_L)$ with $P_L \in [-1,1]$.
It can be seen that increasing $P_L$ enhances Bell-nonlocality across the full range of $\theta$ with an exception of $\theta = 0$ and $\pi$, under which angles $\mathcal{B} \equiv 2$ is independent on $P_L$ (see Eq.~\eqref{eq:XparaPL2}).
In Fig.~\ref{fig:B_PL}, the area between two dashed lines in each panel represents Bell-nonlocally $\mathcal{B}\in (2, \max_{\theta}\mathcal{B}]$ with $\max_{\theta}\mathcal{B}$ being the largest CHSH parameter. As can be seen in Eq.~\eqref{eq:XparaPL1}
the $\max_{\theta}\mathcal{B}$ is present at $\theta = \pi/2$ at any $P_L$ for all channels:
\begin{equation}
\max_{\theta}\mathcal{B} = \mathcal{B}(\theta=\frac{\pi}{2}) = 2 \sqrt{1 + \alpha_\psi^2 + P_L^2\beta_\psi^2 } < 2\sqrt{2}
\end{equation}
which is displayed in the upper panel of Fig.~\ref{fig:max_B_PL}.
Above relation indicates that the impact of $P_L$ on $\max_{\theta}\mathcal{B}$ is mediated by the parameter $\beta_\psi$. For the $\Sigma^{+}$ and $\Sigma^{0}$ channels, $\beta_\psi$ is extremely small---a direct consequence of their small $\Delta\Phi$ (see Table \ref{tab:decay_parameters}). Therefore, the contribution from the lepton beam polarization is strongly suppressed, rendering $\max_{\theta}\mathcal{B}$ largely insensitive to $P_L$ for these two channels.

Increasing $P_L$ expands the angular coverage of Bell nonlocality as can be also seen in Fig.~\ref{fig:B_PL}.
The corresponding angle for $\mathcal{B}[\rho^{P_L}_{Y \bar Y}] = 2$ is defined as the critical angle $\theta^*$, shown in the lower panel of Fig.~\ref{fig:max_B_PL}.
The $|\cos\theta^*|$ moves to 1 when $P_L$ is approaching the maximum degree, so that the system of hyperon pairs is always Bell-nonlocal across the range $\theta \in (0,\pi)$ at $P_L = 1$.
It is noted that for $\Sigma^{+}$ and $\Sigma^{0}$ channels $\theta^*$ is already close to zero in the case of unpolarized beams.

\begin{figure}[!t]
   \centering
   \includegraphics[width=1\linewidth]{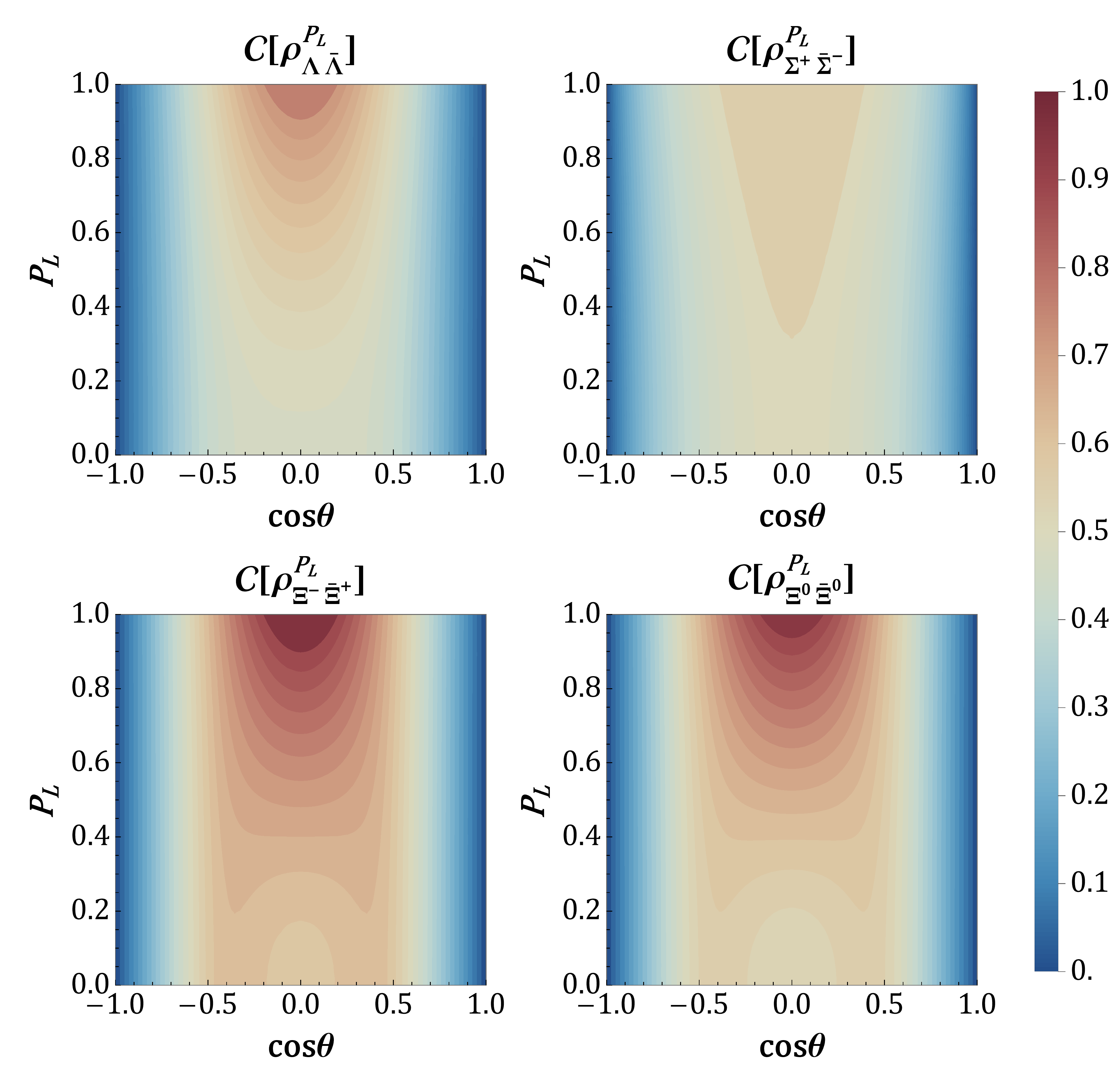}
   \caption{
  The concurrence $\mathcal{C}[\rho^{P_L}_{Y \bar Y}]$ as a function of $\cos\theta$ and
   $P_L$ in $J/\psi\to Y{\bar{Y}}$ for $Y=\Lambda$, $\Sigma^{+}$, $\Xi^{-}$ and $\Xi^{0}$.}
   \label{fig:C_PL}
\end{figure}

\begin{figure}[!t]
   \centering
   \includegraphics[width=0.8\linewidth]{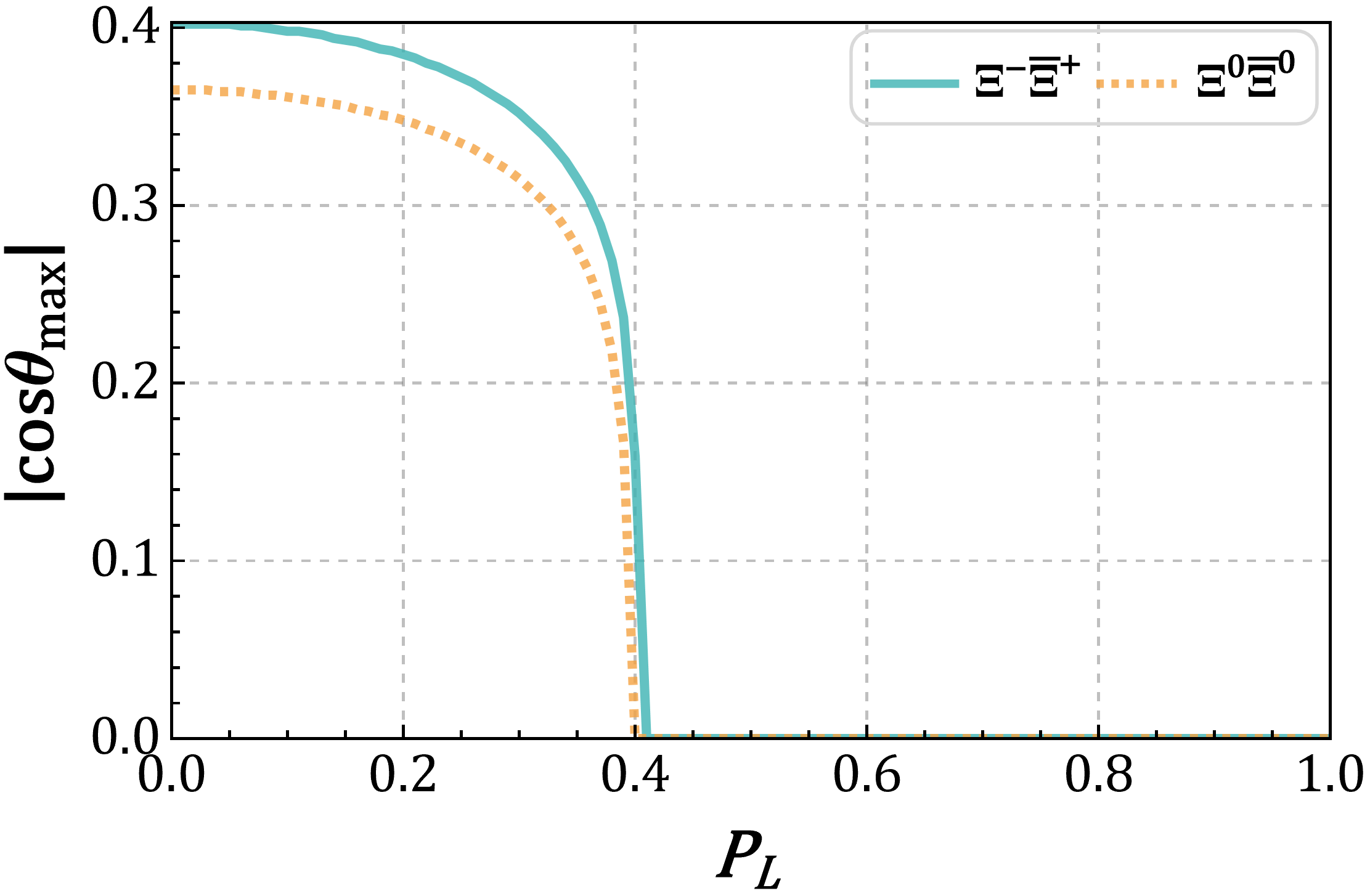}
   \caption{
  The $\cos \theta_{\max}$ (see main text for defination of $\theta_{\max}$) as a function of $P_L$ in $J/\psi\to Y{\bar{Y}}$ for $Y=\Xi^{-}$ and $\Xi^{0}$.
   } \label{fig:max_angle_PL}
\end{figure}

\begin{figure}[!t]
   \centering
   \includegraphics[width=0.9\linewidth]{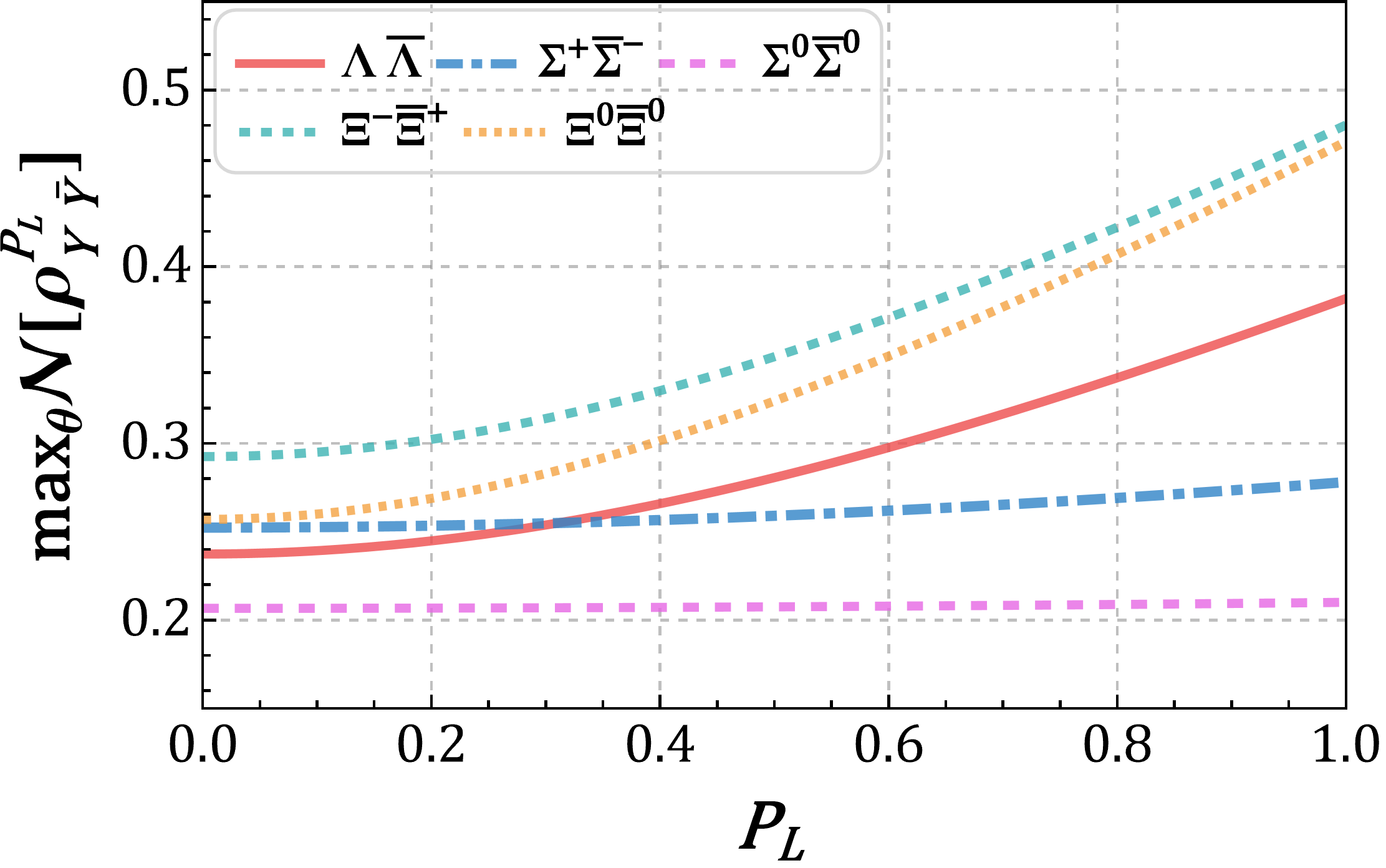}
   \includegraphics[width=0.9\linewidth]{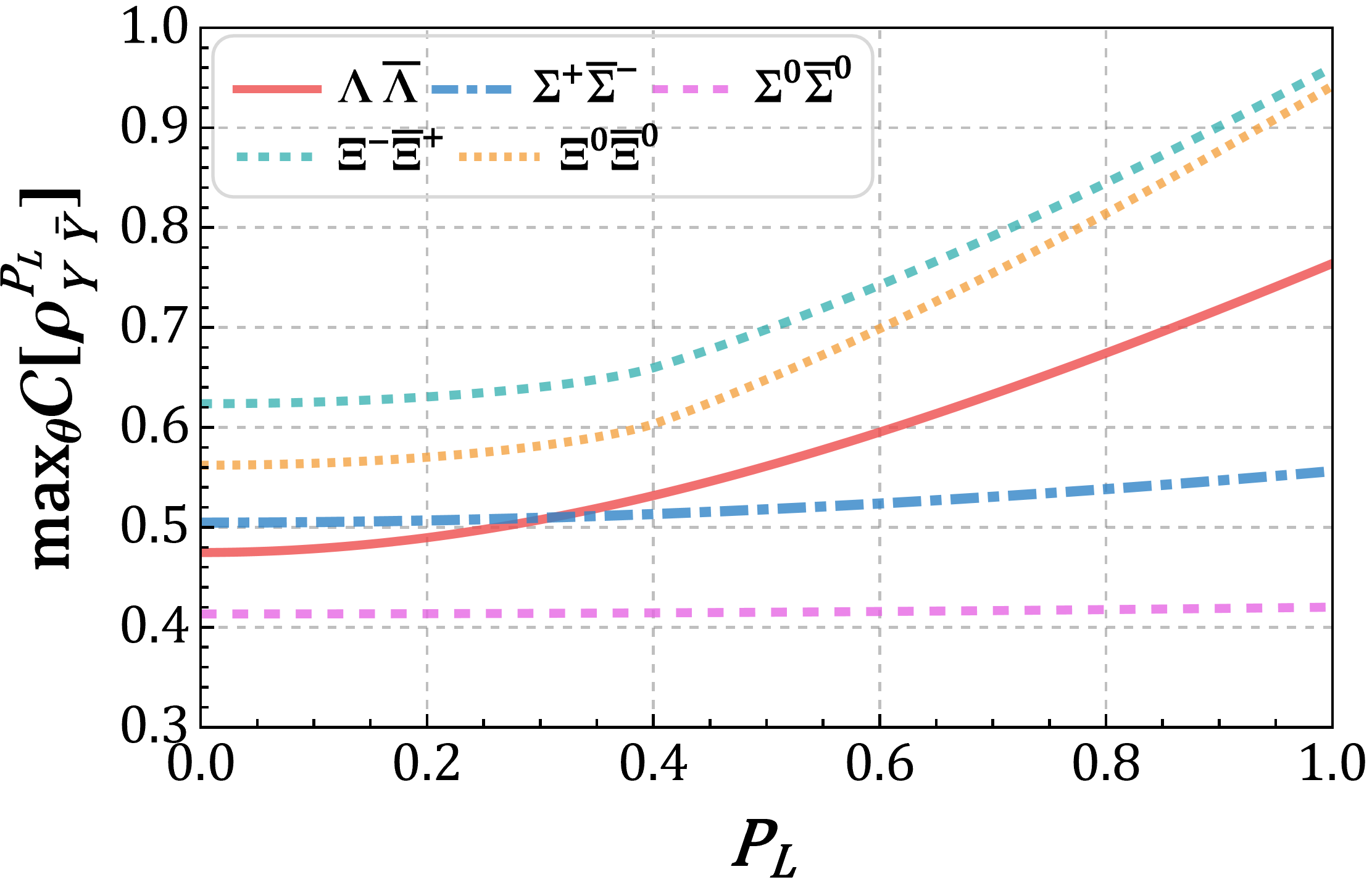}
   \caption{
  The $\max_{\theta}\mathcal{N}[\rho^{P_L}_{Y\bar{Y}}]$ (upper panel) and $\max_{\theta}\mathcal{C}[\rho^{P_L}_{Y\bar{Y}}]$ (lower panel) as functions of $P_L$ in $J/\psi\to Y{\bar{Y}}$ for $Y = \Lambda$, $\Sigma^{+}$, $\Sigma^{0}$,  $\Xi^{-}$ and $\Xi^{0}$.
   }
   \label{fig:max_C_PL}
\end{figure}

Fig.~\ref{fig:C_PL} shows that the $\theta$ and $P_L$ dependence of the concurrence $\mathcal{C}[\rho^{P_L}_{Y \bar Y}]$.
The negativity $\mathcal{N}[\rho^{P_L}_{Y \bar Y}]$ is shown by Fig.~\ref{fig:negJpsi} in Appendix.~\ref{sec:Supplpsi}.
It can be seen that a higher degree of longitudinal polarization leads to enhanced concurrence $\mathcal{C}$ and negativity $\mathcal{N}$ over the full $\theta$ range with a exception of $\theta=0$ and $\pi$, from Eqs.~\eqref{eq:XparaPL2}, ~\eqref{eq:concurrence} and \eqref{eq:negativity}, under which angles $\mathcal{C}=2\mathcal{N}\equiv0$ means that the states are separable and is independent on $P_L$.

The largest negativity with respect to $\theta$, labeled as $\max_{\theta}\mathcal{N}$, locates at $\theta = \pi/2$ across the full $P_L$ range for all channels.
The largest concurrence with respect to $\theta$, denoted as $\max_{\theta}\mathcal{C}$, lies:
\begin{itemize}
  \item[a.] at $\theta = \pi/2$ across the full $P_L$ range for $\Lambda$, $\Sigma^{+}$ and $\Sigma^{0}$ channels;
  \item[b.] at the angles of $\theta = \theta_{\max}$ and $\pi-\theta_{\max}$ when $P_L$ is small, and moves to $\theta = \pi/2$ as the $P_L$ is increasing to certain value for $\Xi^{-}$ and $\Xi^{0}$ channels.
\end{itemize}
For the later case the $|\cos \theta_{\max}|$ is plotted as a function of $P_L$ in Fig.~\ref{fig:max_angle_PL}.
It can be found that at $P_L\gtrsim 0.4$ the $\max_{\theta}\mathcal{C}$ is shifted to $\theta = \pi/2$ for both channels. From Eqs.~\eqref{eq:XparaPL1}, ~\eqref{eq:concurrence} and \eqref{eq:negativity}, one has
\begin{align}
    \mathcal{C}(\theta=\frac{\pi}{2})=2 \mathcal{N}(\theta=\frac{\pi}{2})=\sqrt{\alpha_\psi^2 + P_L^2  \beta_\psi^2},\label{PLCN1}
\end{align}
As shown in Fig.~\ref{fig:max_C_PL}, $\max_{\theta}\mathcal{N}$ and $\max_{\theta}\mathcal{C}$ increases monotonically as the increase of $P_L$ for all channels.
They both reach the largest extrema at $P_L = 1$, but not maximal entangled since $\max_\theta \mathcal{C} < 1$ persists.
For the exact same reason, the $\max_{\theta}\mathcal{N}$ and $\max_{\theta}\mathcal{C}$ of $\Sigma^{+}$ and $\Sigma^{0}$ channels are insensitive to $P_L$, as is the case for $\max_{\theta}\mathcal{B}$.

\section{Transverse Beam Polarization}
\label{sec:PT}
\begin{figure}[!hbt]
  		\includegraphics[width = 1 \linewidth]{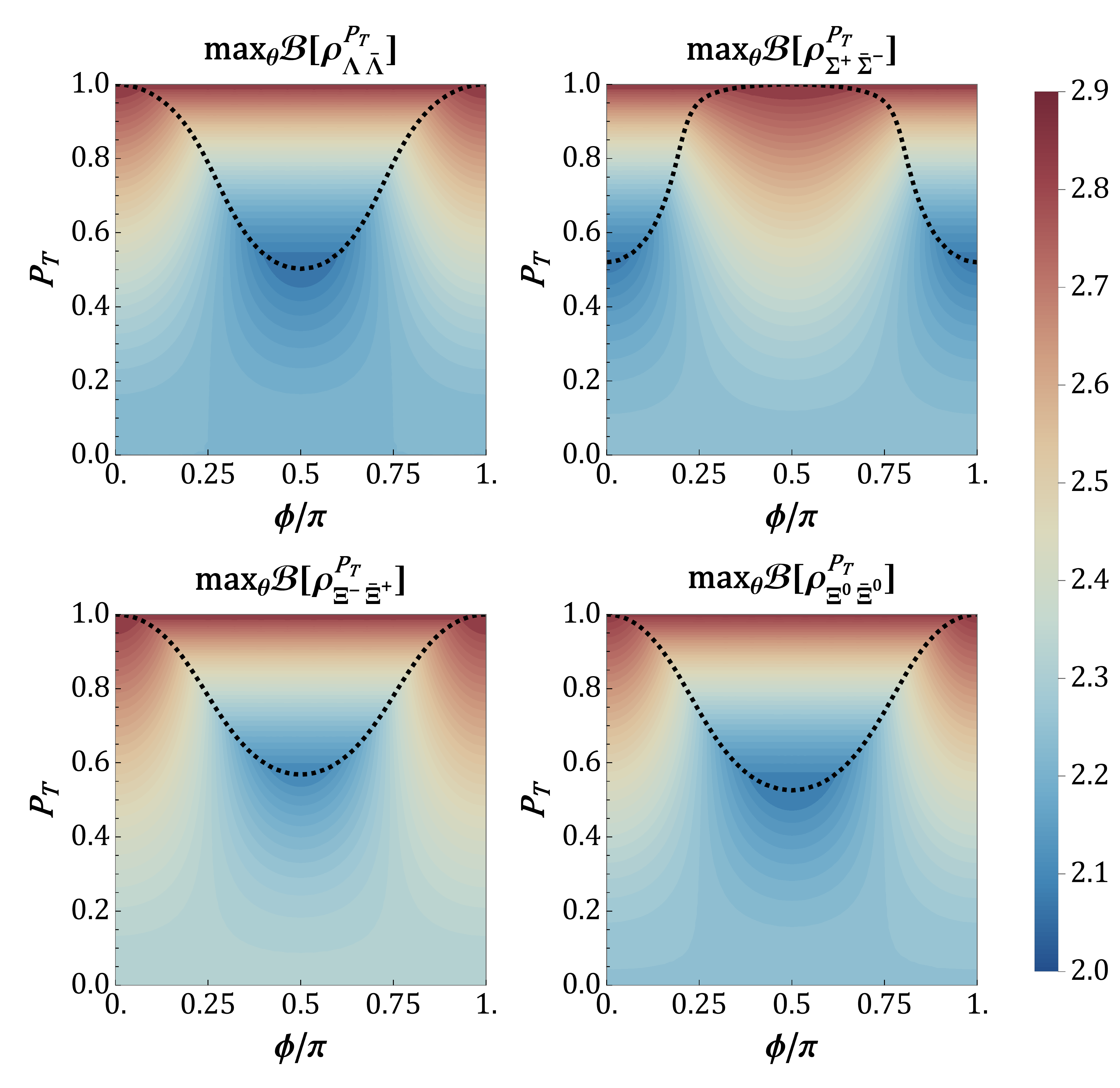}
\caption{The $\max_{\theta}{B}[\rho^{P_T}_{Y \bar Y}]$ as a function of azimuthal angle $\phi$ and transverse beams polarization degree $P_T$ in $J/\psi\to Y{\bar{Y}}$ for $Y = \Lambda$, $\Sigma^{+}$, $\Xi^{-}$ and $\Xi^{0}$. The dashed curve is for $\mathcal{B}(\theta=0,\pi)=\mathcal{B}(\theta=\pi/2)$.} \label{fig:max_B_PT}
\end{figure}

\begin{figure}[!hbt]
    \includegraphics[width = 0.9 \linewidth]{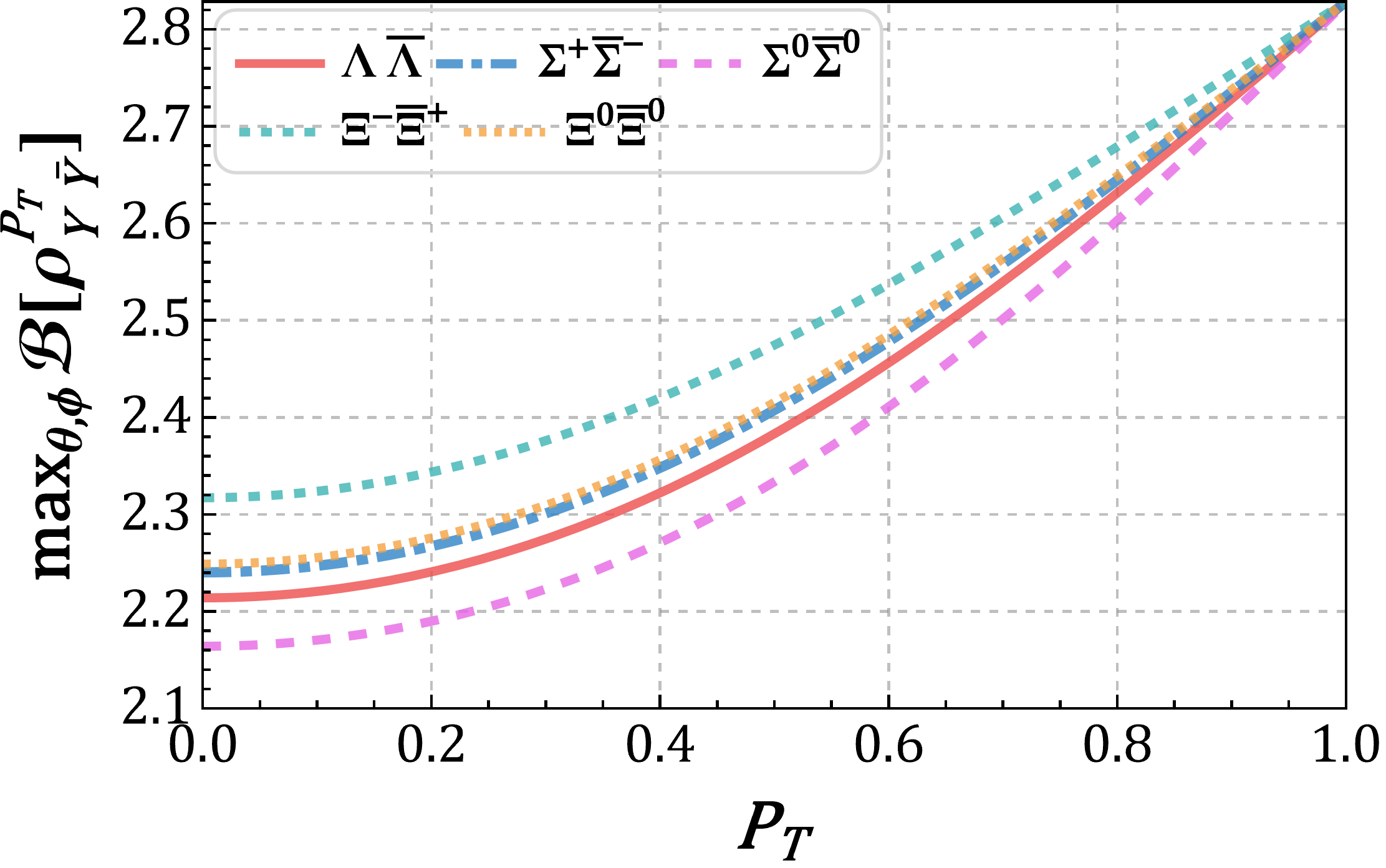}
    \includegraphics[width = 0.9 \linewidth]{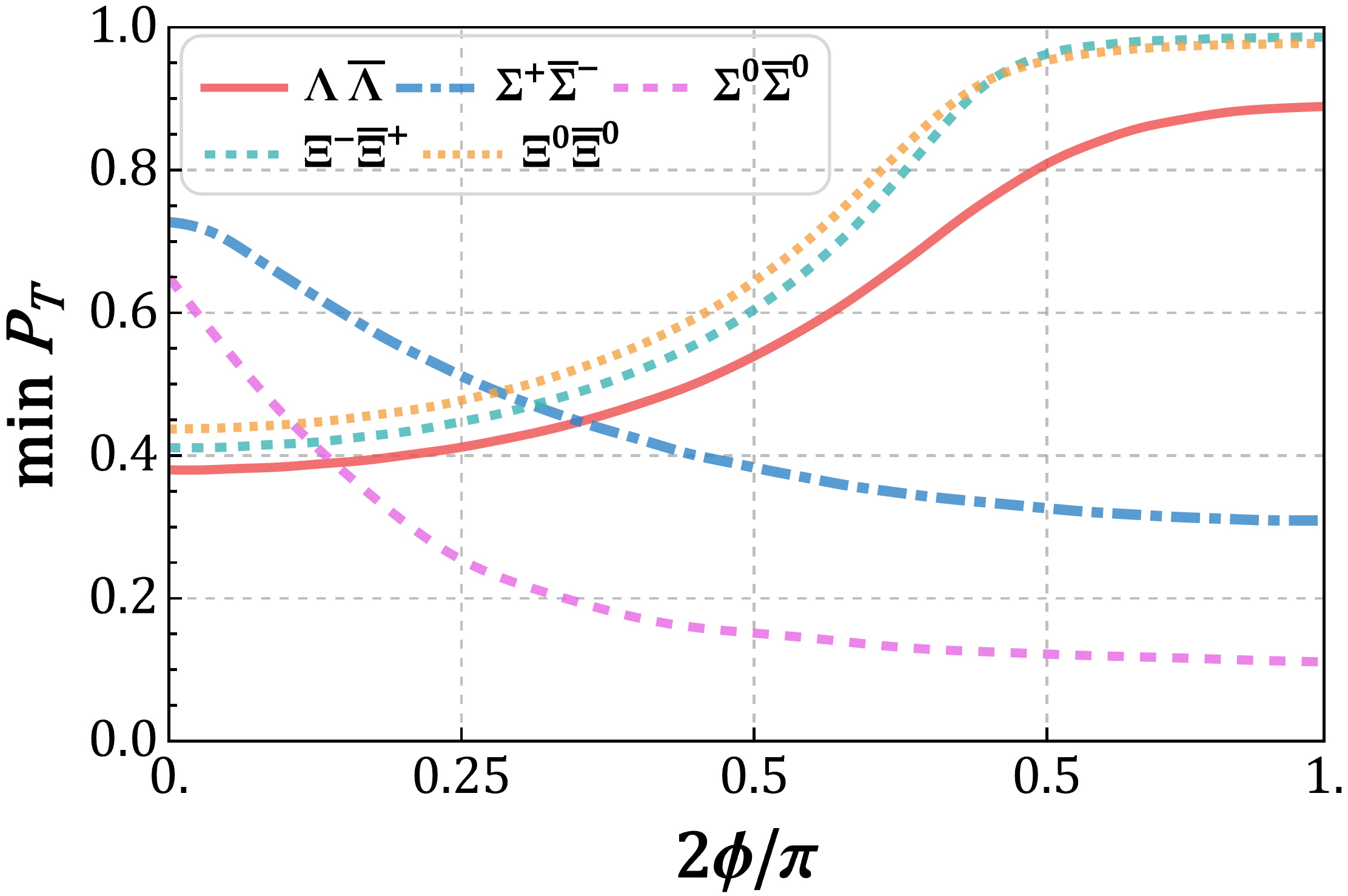}
\caption{The $\max_{\theta,\phi}\mathcal{B}[\rho^{P_T}_{Y \bar Y}]$ as a function of $P_T$(upper panel), and the minimal transverse polarization $\min P_T{(\phi)}$ required to achieve $\mathcal{B}[\rho^{P_T}_{Y \bar Y}] = 2$ as a function of $\phi$ (lower panel) for all five channels. \label{fig:B_complete}
}
\end{figure}

The situation for transverse polarization is more complicated considering that it introduces the dependence on the azimuthal angle $\phi$ with respect to the scattering plane.
Fig.~\ref{fig:max_B_PT} displays the $P_T$ and $\phi$ dependence of the largest CHSH parameter with respect to $\theta$, labeled as $\max_{\theta}\mathcal{B}[\rho^{P_T}_{Y \bar Y}]$, with a note of $\mathcal{B}(\theta) = \mathcal{B}(\pi - \theta)$ with $\theta \in [0,\pi]$, $\mathcal{B}(P_T) = \mathcal{B}(-P_T)$ with $P_T \in [-1,1]$ and $\mathcal{B}(\phi) = \mathcal{B}(\pi-\phi)$, $\mathcal{B}(\phi) = \mathcal{B}(\pi+\phi)$ with $\phi \in [0,\pi]$.
The $\max_{\theta}\mathcal{B}$ appears at $\theta=0,\pi$ or $\theta={\pi}/{2}$ when $0\leq P_T<1$, which can be derived from Eq.~\eqref{eq:XparaPT1},
\begin{align}
&\mathcal{B}(\theta = 0,\pi)  = 2 \sqrt{1+P_T^4},\label{eq:B_m1} \\
&\mathcal{B}(\theta = \frac{\pi}{2}) =  \frac{2\sqrt{1+\alpha_\psi^2 + 4P_T^2\alpha_\psi \cos2\phi+P_T^4 \varepsilon}}{1+P_T^2\alpha_\psi \cos 2\phi}, \label{eq:B_0}
\end{align}
with $\varepsilon=(1+\alpha_{\psi}^2)\cos^2 2\phi+\gamma_\psi^2\sin^2 2\phi$.
When \(P_T = 1\), the $\max_{\theta}\mathcal{B}$ occurs at
\begin{itemize}
  \item[a.] $\theta=0,\pi$ and $\theta={\pi}/{2}$ simultaneously if $\phi=\pi/2+k\pi$ $(k=0,1)$,
  \item[b.] full $\theta$ range if $\phi=k\pi$ $(k=0,1,2)$,
  \item[c.] $\theta=0,\pi$ or $\theta={\pi}/{2}$ as determined by Eqs.
  (\ref{eq:B_m1},\ref{eq:B_0}) otherwise.
\end{itemize}
The dashed curve $\mathcal{B}(\theta = 0,\pi)=\mathcal{B}(\theta =\pi/2)$ in Fig.~\ref{fig:max_B_PT} divides each panel into two regions.
For the region above the dashed curve, the $\max_{\theta}\mathcal{B}$ occurs at $\theta = 0,\pi$, and attain maximal violation of CHSH inequality for arbitrary $\phi$ at $P_T = 1$ from Eq.~\eqref{eq:B_m1}.
For the region below the dashed curve, the $\max_{\theta}\mathcal{B}$ occurs at $\theta = {\pi}/{2}$, and hold the maximal violation for $\phi = k\pi/2$ $(k=0,1,2,3,4)$ at $P_T = 1$ from Eq.~\eqref{eq:B_0}.
Futhermore, the maximal violation is attained across the full $\theta$ range for $\phi = k\pi$ $(k=0,1,2)$ at $P_T = 1$.
It is found for both regions $\max_{\theta}\mathcal{B} \geq 2$ across full $\phi$ coverage, and $P_T = 1$ drives $\max_{\theta} \mathcal{B}$ to approach $2\sqrt{2}$ at any fixed $\phi$.

As can be seen in Fig.~\ref{fig:max_B_PT}, the largest value of $\max_{\theta}\mathcal{B}[\rho^{P_T}_{Y \bar Y}]$, here labeled as $\max_{\theta,\phi}\mathcal{B}[\rho^{P_T}_{Y \bar Y}]$, is attained at specific angles $\phi$ with $P_T$ held constant.
As shown in upper panel of Fig.~\ref{fig:B_complete},
the $\max_{\theta,\phi}\mathcal{B}$ is always enhanced as the increase of $P_T$, and occurs at specific angles of
\begin{itemize}
  \item[a.] $\phi=k\pi$ $(k=0,1,2)$ for $\Lambda$, $\Xi^{-}$ and $\Xi^{0}$ channels;
  \item[b.] $\phi=\pi/2+k\pi$ $(k=0,1)$ for $\Sigma^{+}$, $\Sigma^{0}$ channels.
\end{itemize}
The $\max_{\theta,\phi}\mathcal{B}$ is attained at different azimuthal angles $\phi$ depending on the specific decay channel. This is primarily because the $P_T$-dependence of $\max_{\theta}\mathcal{B}$ is dominated by the term $P_T^2\alpha_\psi\cos2\phi$ (see Eq. \eqref{eq:B_0}). Specifically, the parameter $\alpha_\psi$ for the $J/\psi\rightarrow \Sigma^{+} \bar{\Sigma^{+}}$ and $\Sigma^{0} \bar{\Sigma^{0}}$ channels is negative, which is opposite to those of the $\Lambda \bar{\Lambda}$, $\Xi^{-} \bar{\Xi^{-}}$, and $\Xi^{0} \bar{\Xi^{0}}$ channels; this sign difference directly shifts the optimal angle $\phi$.
This distinct behavior is visually reflected by the different trajectories of the dashed curves in Fig.~\ref{fig:max_B_PT}.
Meanwhile, since the absolute magnitude of $\alpha_\psi$ is comparable across all considered channels, their overall sensitivities to $P_T$ remain remarkably similar.

The violation of CHSH inequality across full \(\theta\) range can be achieved by increasing the transverse beams polarization degree \( P_T \).
Therefore a minimal $P_T$ required to achieve $\mathcal{B}[\rho^{P_L}_{Y \bar Y}] \geq 2$ across full $\theta$ range, labeled as $\min P_T$, can be defined as a function of $\phi$,
as shown in lower panel of Fig.~\ref{fig:B_complete}.
The $\min P_T$ increases as the increase of $\phi \in [0,\pi/2]$ for $\Lambda$, $\Xi^{-}$ and $\Xi^{0}$ channels, opposite to the trends of $\Sigma^{+}$ and $\Sigma^{0}$ channels.
Once again, this behavioral divergence traces back to the aforementioned sign flip of $\alpha_\psi$ among channels.

\begin{figure}[!hbt]
  		\includegraphics[width = 1 \linewidth]{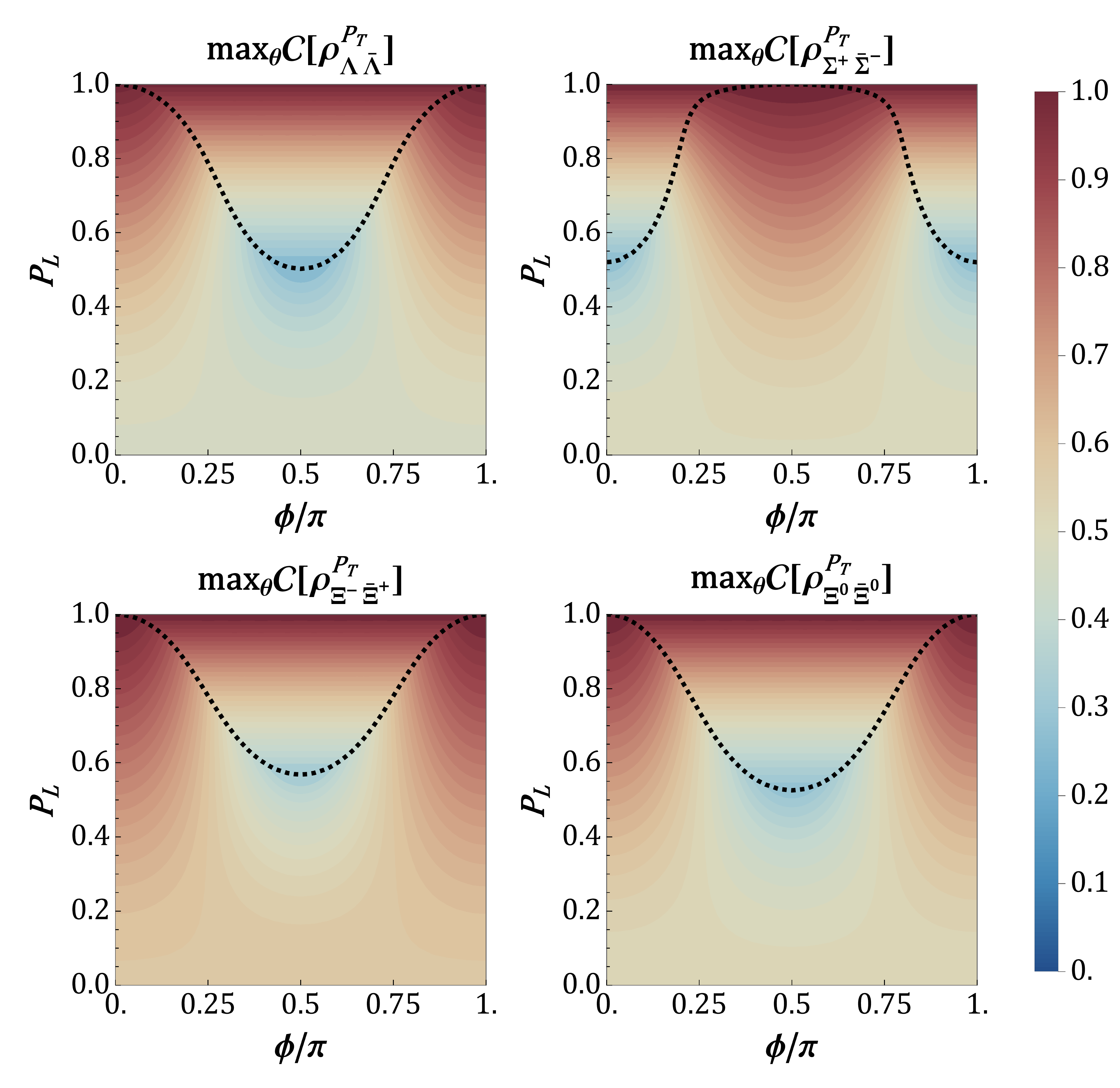}
\caption{
The $\max_{\theta}\mathcal{C}[\rho_{T}]$ as a function of $\phi$ and $P_T$ in $J/\psi\to Y{\bar{Y}}$ for $Y = \Lambda$, $\Sigma^{+}$, $\Xi^{-}$ and $\Xi^{0}$ . See main text for a detailed explanation of the dashed lines.} \label{fig:max_C_PT}
\end{figure}

\begin{figure}[!hbt]
  	\includegraphics[width = 0.9 \linewidth]{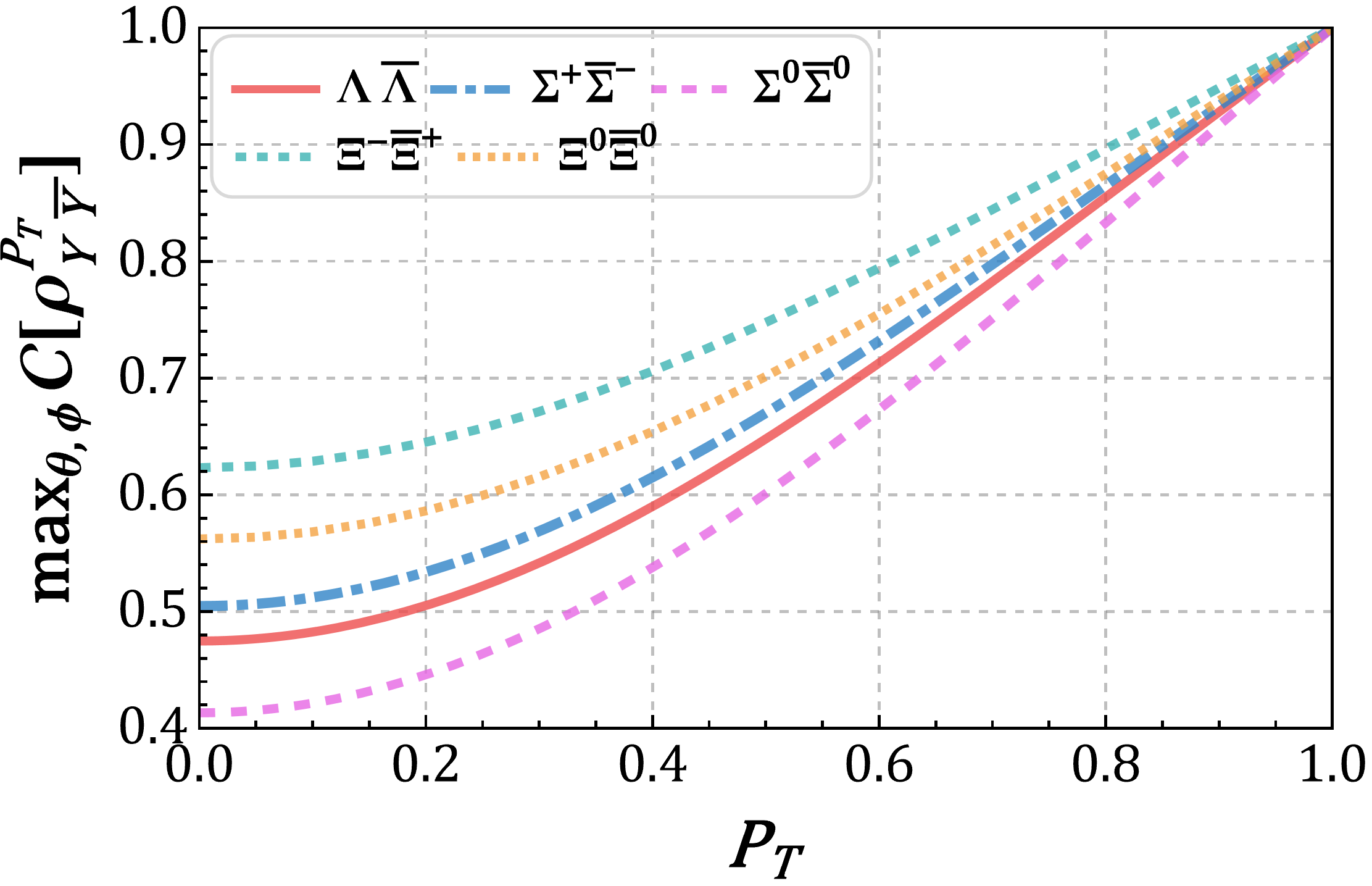}
      \includegraphics[width = 0.9 \linewidth]{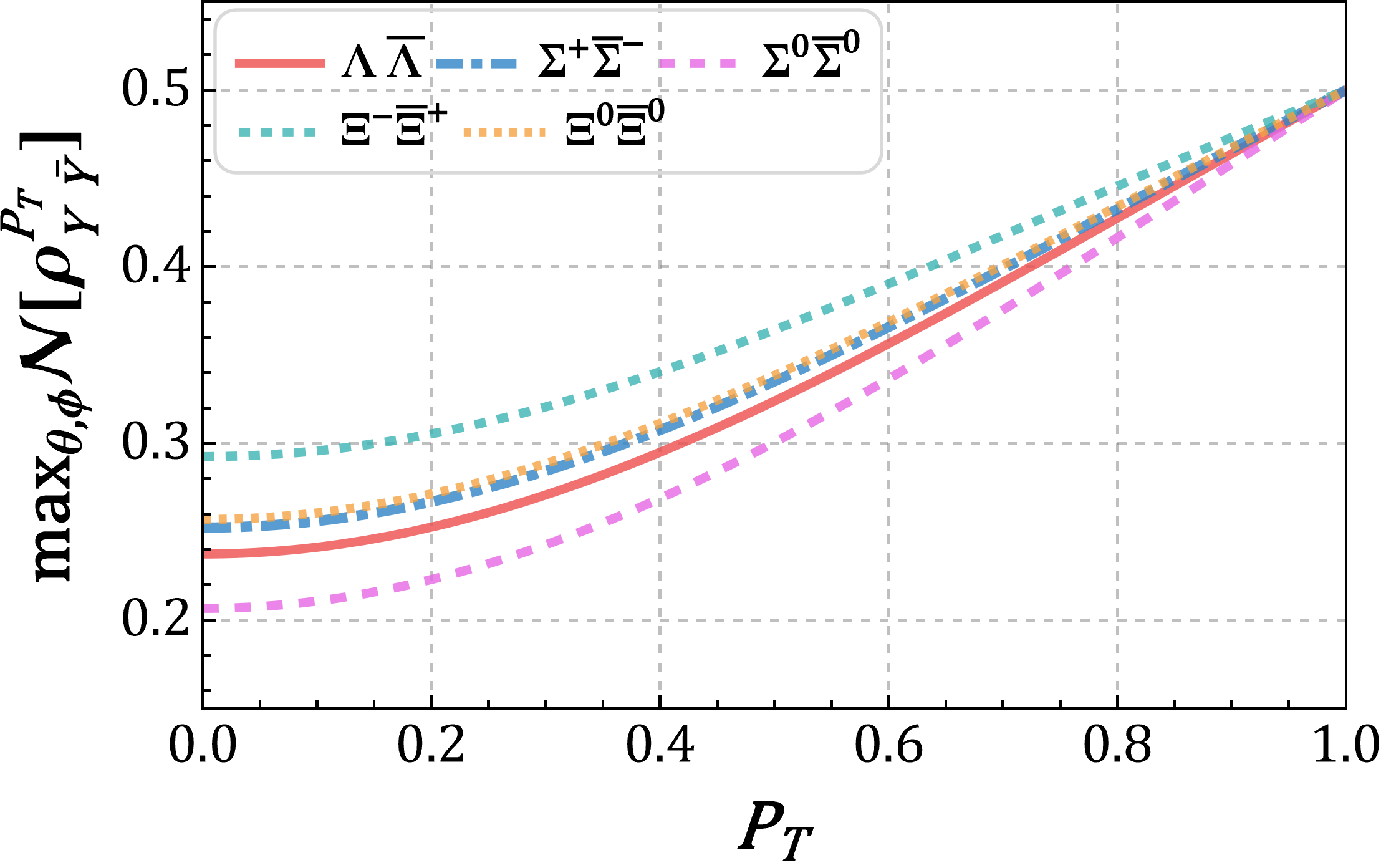}
\caption{
The $\max_{\theta,\phi}\mathcal{C}[\rho^{P_T}_{Y \bar Y}]$ (upper panel) and $\max_{\theta,\phi}\mathcal{N}[\rho^{P_T}_{Y \bar Y}]$ (lower panel) as functions of $P_T$ in $J/\psi\to Y{\bar{Y}}$ for $Y = \Lambda$, $\Sigma^{+}$, $\Sigma^{0}$, $\Xi^{-}$ and $\Xi^{0}$.
} \label{fig:maxphi_C_PT}
\end{figure}

The largest concurrence and negativity with respect to $\theta$, labled as $\max_{\theta}\mathcal{C}[\rho^{P_T}_{Y \bar Y}]$ and $\max_{\theta}\mathcal{N}[\rho^{P_T}_{Y \bar Y}]$, respectively, can be similarly defined.
The behaviour of $\max_{\theta}\mathcal{C}$ in Fig.~\ref{fig:max_C_PT} and $\max_{\theta}\mathcal{N}$ in Appendix.~\ref{sec:Supplpsi} as a function of $P_T$ and $\phi$ are similar to $\max_{\theta}\mathcal{B}$.
When $0 \leq P_T<1$, the position of $\max_{\theta}\mathcal{C}$ is found at:
\begin{itemize}
  \item[a.] for $\Lambda$, $\Sigma^{-}$ and $\Sigma^{0}$ channels: either at $\theta = 0,\pi$ for the region above the dashed curve, or at $\theta = \pi/2$ for the region below the dashed curve.
  \item[b.] for $\Xi^{-}$ and $\Xi^{0}$ channels: either at $\theta = 0,\pi$ for the region above the dashed curve, or $\cos\theta=0.368$ (for $\Xi^{-}$) and $0.403$ (for $\Xi^{0}$) for the region below the dashed curve;
\end{itemize}
For the position of $\max_{\theta}\mathcal{N}$, case-$a$ is applied for all channels.
When \(P_T = 1\), the $\max_{\theta}\mathcal{C}$ and $\max_{\theta}\mathcal{N}$ locate the same position with the situation of $\max_{\theta}\mathcal{B}$.

From Eqs.~\eqref{eq:XparaPT1}, ~\eqref{eq:XparaPT2}, \eqref{eq:concurrence} and \eqref{eq:negativity}, one read as
\begin{align}
  \mathcal{C}(\theta= & 0,\pi)  = 2\mathcal{N}(\theta=0,\pi)=P_T^2,\label{eq:CPT1}\\
  \mathcal{C}(\theta= & \frac{\pi}{2})  =2\mathcal{N}(\theta=\frac{\pi}{2})=\frac{\sqrt{\alpha_\psi^2 + 2P_T^2 \alpha_\psi \cos2\phi+P_T^4 \eta }}{1 + P_T^2\alpha_\psi\cos 2\phi},\label{eq:CPT2}
\end{align}
with $\eta ={\gamma_\psi}^2\sin^2 2\phi + \cos^2 2\phi$.
From Eq.~\eqref{eq:CPT1}, maximal entanglement is attained at $\theta=0$ and $\pi$ for arbitrary $\phi$ if $P_T = 1$.
From Eq.~\eqref{eq:CPT2}, maximal entanglement is attained at $\theta=\pi/2$ for $\phi = k\pi/2$ $(k=0,1,2,3,4)$ if $P_T = 1$.
Furthermore, the maximal entanglement occurs across the full $\theta$ range for $\phi = k\pi$ $(k=0,1,2)$ at $P_T = 1$.

Similar to $\max_{\theta,\phi}\mathcal{B}$, $\max_{\theta,\phi}\mathcal{C}$ and $\max_{\theta,\phi}\mathcal{N}$ can be defined as the largest value of $\max_{\theta}\mathcal{C}$ and $\max_{\theta}\mathcal{N}$ at specific angles $\phi$ with $P_T$ held constant, respectively.
The $\max_{\theta,\phi}\mathcal{C}$ and $\max_{\theta,\phi}\mathcal{N}$, locating at the same azimuthal angle $\phi$ as that of $\max_{\theta,\phi}\mathcal{B}$, increase monotonically as the increase of $P_T$ as shown in Fig.~\ref{fig:maxphi_C_PT}.
Their behavior closely mirror that of $\max_{\theta,\phi}\mathcal{B}$, as their $P_T$-dependence is similarly governed by the aforementioned term $P_T^2\alpha_\psi\cos2\phi$.
For $P_T =1$, the system can always attain maximal entanglement under the proper selection of $\phi$ and $\theta$ angles, as also can be seen in Fig.~\ref{fig:max_C_PT} that $\max_{\theta}\mathcal{C}=2\max_{\theta}\mathcal{N}=1$ at any fixed $\phi$ at $P_T =1$.

\section{Discussions} \label{sec:disc}


\subsection{Hierarchy of Quantum Correlations}
\label{sec:hierarchy}


Bell nonlocality and quantum entanglement satisfy following hierarchy with respect to quantumness:
\begin{equation}  \notag
  \text{Bell nonlocality} \subset \text{Quantum entanglement}.
\end{equation}
Quantitatively, this fundamental hierarchy is manifested through the relationship between their respective measures. For a two-qubit system, the CHSH parameter characterizing Bell nonlocality is strictly upper-bounded by the concurrence~\cite{Verstraete:2001skx}:
\begin{equation}
	\mathcal{B}[\rho]\leq 2 \sqrt{1+\mathcal{C}^2[\rho]}. \label{eq:B_upper}
\end{equation}
of which the equality is saturated iff (if and only if) the singular values of corresponding correlation matrix $\vec{C}$ are $\{1, \mathcal{C}[\rho], \mathcal{C}[\rho]\}$ \cite{Verstraete:2001skx}.
It shall be noted that the singular values of real and symmetric matrix $\vec{C}$ correspond to the absolute values of its eigenvalues.

Furthermore, an upper bound of Wootters’ concurrence is given in terms of the polarization of subsystems \cite{Barr:2024djo,Liu:2026dzv}:
\begin{equation}
	C[\rho]\leq \min \left\{ \sqrt{1-{\left\|\vec{B}^+\right\|^2} }, \sqrt{1-\left\| \vec{B}^-\right\| ^2}\right\},\label{eq:C_upper}
\end{equation}
with $\left\|\vec{B}\right\|=\sqrt{\mathrm{Tr}(\vec{B}^{T}\vec{B})}$.
The bound is saturated iff $\rho$ is a pure state.
The above relation reveals that the maximum achievable entanglement of the system is bounded by the polarization of its subsystem.
Specifically, the upper bound of entanglement of the system decreases as the polarization of its subsystem increases.
By combining Eqs.~\eqref{eq:B_upper} and \eqref{eq:C_upper}, the polarization of the subsystems imposes an upper bound on the CHSH parameter of system:
\begin{equation}
	\mathcal{B}[\rho]\leq \min \left\{ 2\sqrt{2-\left\| \vec{B}^+ \right\|^2}, 2\sqrt{2-\left\| \vec{B}^-\right\|^2 }\right\}\label{eq:B_C_up}.
\end{equation}
The above inequality becomes an equality if and only if Eqs.~\eqref{eq:B_upper} and \eqref{eq:C_upper} are both saturated.
As a special case, Eq.~\eqref{eq:B_C_up} is trivially saturated for elementary fermion pairs, which are always maximally entangled with vanishing polarization ($\| \vec{B}^\pm \| =0$) of subsystems\cite{Altakach:2026fpl,Guo:2026yhz}.

For hyperon-antihyperon pairs without $CP$ violation as considered here, correlation matrix $\vec{C}$ in Eqs. \eqref{eq:CL} and \eqref{eq:CT} are real and symmetric after a rotation of coordinate system.
Using the results from Sec.~\ref{sec:preparation}, one can easily prove that the singular values of the correlation matrix $\vec{C}$ of $\rho^X_{Y\bar{Y}}$ at $\theta=0, {\pi}/{2}\text{ and }{\pi}$ are $\{1, |-\lambda_2|,|\lambda_2|\}=\{1, \mathcal{C}[\rho], \mathcal{C}[\rho]\}$, thus saturating Eq.~\eqref{eq:B_upper} irrespective of the beam polarization, as shown in Sec.~\ref{sec:PL} and Sec. \ref{sec:PT}. Furthermore, from Eqs.~\eqref{eq:BL} and \eqref{eq:BT} the $\left\|\vec{B}^+\right\|=\left\|\vec{B}^-\right\|$ always holds, and:
\begin{align}
  &\left\| \vec{B_L} \right\|=\frac{1}{\chi_L}{\sqrt{\beta_\psi^2 \sin^2 \theta\cos^2\theta + P_L^2 \left( \beta_\psi^2 \cos^2\theta + \gamma_\psi^2 \right)}},\label{eq:PLBupper}\\
  &\left\| \vec{B_T} \right\|=\frac{\beta_\psi \sin\theta}{\chi_T}{\sqrt{\left(1-P_T^2 \cos 2\phi \right)^2\cos^2\theta + P_T^4 \sin^2 2\phi} }.\label{eq:PTBupper}
\end{align}

\begin{figure}[!t]
  	\includegraphics[width = 0.84 \linewidth]{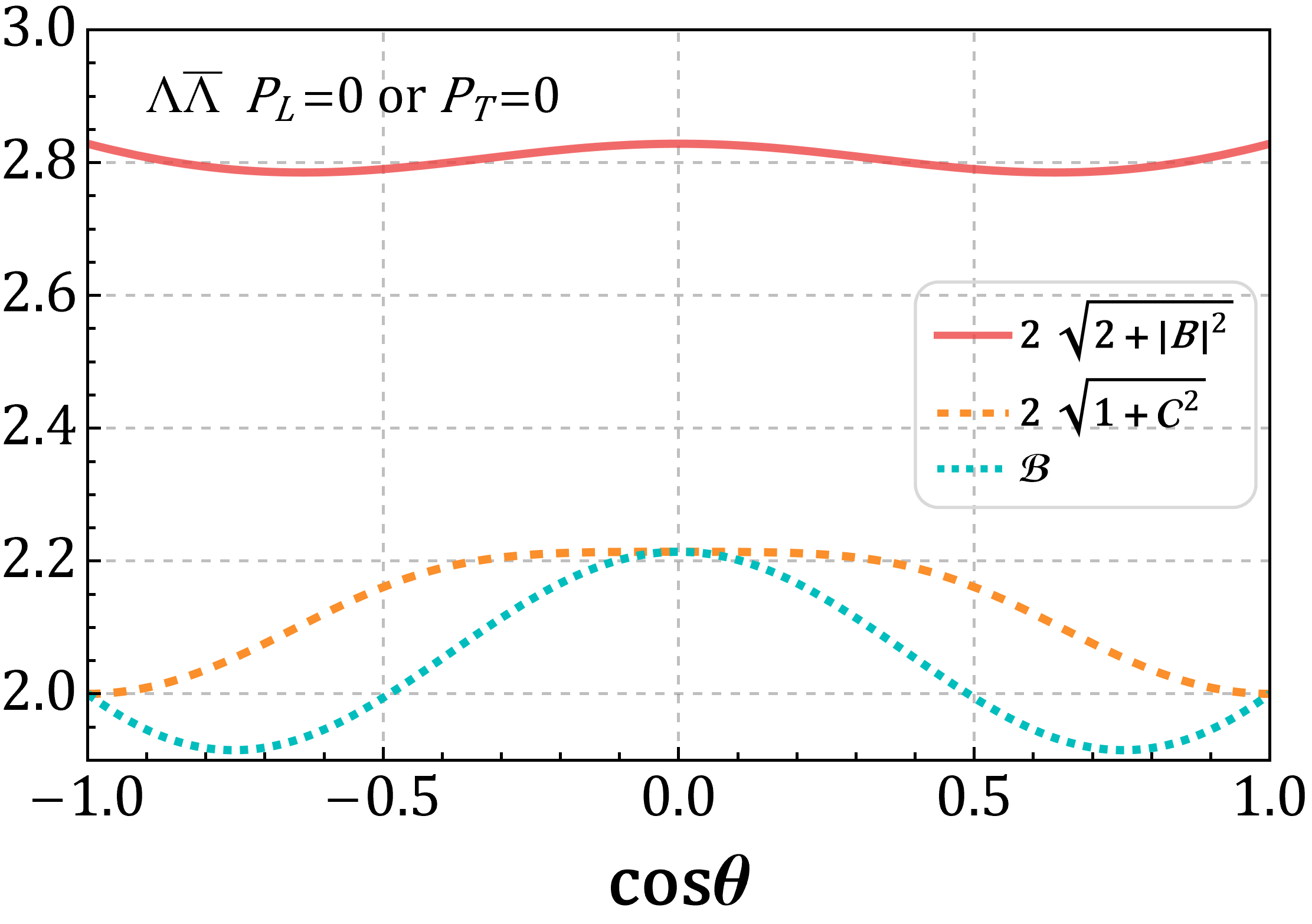}
  		\includegraphics[width = 0.48 \linewidth]{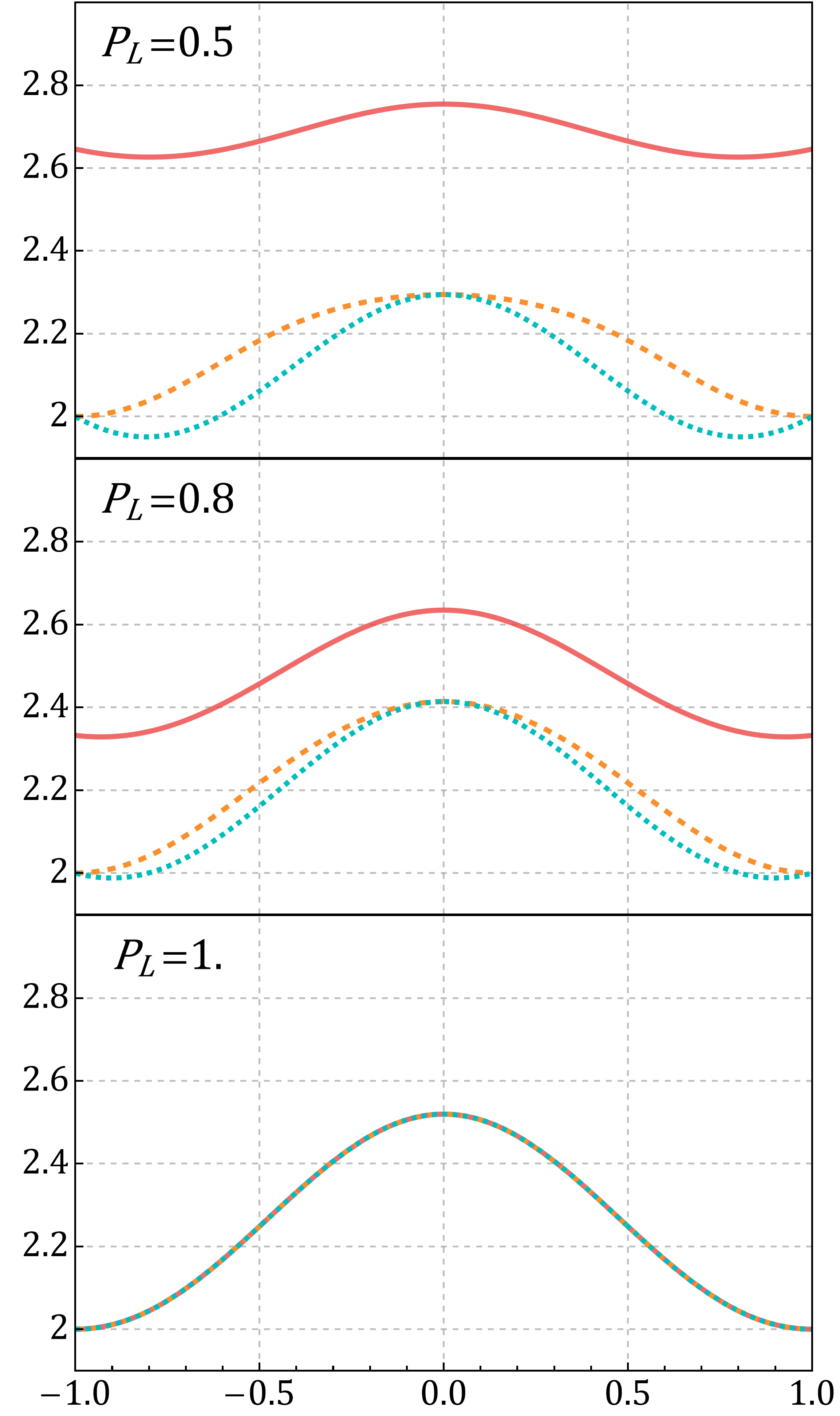}
  		\includegraphics[width = 0.48 \linewidth]{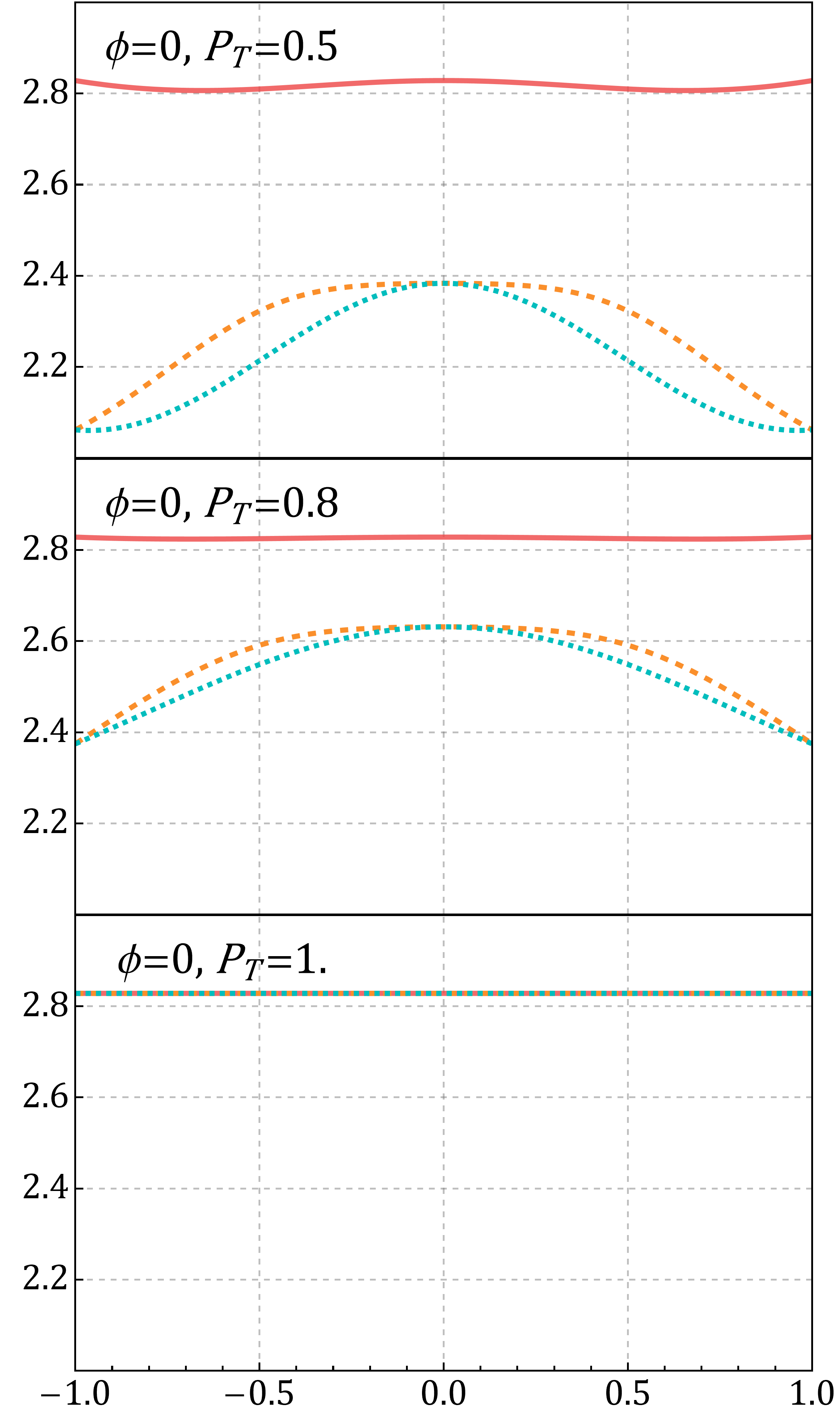}
  \caption{Upper bounds in Eq.~\eqref{eq:B_upper} (orange dashed lines) and Eq.~\eqref{eq:B_C_up} (red solid lines) for $\Lambda$ channel are compared to the CHSH parameters ${\cal B}[\rho]$ (green dotted lines) as a function of $\cos\theta$. Upper panel: unpolarized lepton beams. Lower left panels: longitudinal polarized electron beam. Lower right panels: transversely polarized lepton beams with $\phi = 0$.  The polarization degrees $P_L$ or $P_T$ are indicated in each panel. } \label{fig:kappa_lambda2}
\end{figure}
A necessary condition for the existence of maximal entanglement $(\mathcal{C}=1)$ is that the system is in a pure state.
When $P_L=1$ or $P_T=1$, the $\rho_{Y \bar Y}$ is in a pure state and the Eq.~\eqref{eq:B_C_up} hold as equality.
As can be seen, hyperon polarization remains non-zero ($\|\vec{B^{\pm}_L}\| \neq 0$) across full scattering angles.
Therefore maximal entanglement is unachievable since $\mathcal{C}[\rho^{P_L}_{Y \bar Y}] < 1$ as constrained by Eq.~\eqref{eq:C_upper}.
Even so, the system accesses the largest possible entanglement.
On the other hand, tuning the transverse polarization degree to maximum degree makes it possible to realize vanishing polarization of hyperons by properly choosing $\theta$ and $\phi$ as shown by Eq.~\eqref{eq:PTBupper} and in detail discussed in Sec.~\ref{sec:PT}.

Taking $\Lambda$ channel as an example, Fig.~\ref{fig:kappa_lambda2} shows the evolution of the upper bounds in Eqs.~\eqref{eq:B_upper} and \eqref{eq:B_C_up} as the longitudinal or transverse beam polarization degree increases in comparison of the CHSH parameters $\mathcal{B}[\rho_{\Lambda \bar \Lambda}]$.
The upper bounds in Eq.~\eqref{eq:B_upper} is more stringent than that in Eq.~\eqref{eq:B_C_up}.
When the $P_L$ or $P_T$ approaches the maximum degree, and two upper bounds become close to each other and ultimate equal at any scattering angle when the system is in a pure state.
It can be also seen in Fig.~\ref{fig:kappa_lambda2} that if $P_T =1$, maximal entanglement is attained across full $\theta$ range at particular angles $\phi = k\pi, (k=0, 1, 2)$.

\begin{figure}[!thb]
  		\includegraphics[width = 0.9 \linewidth]{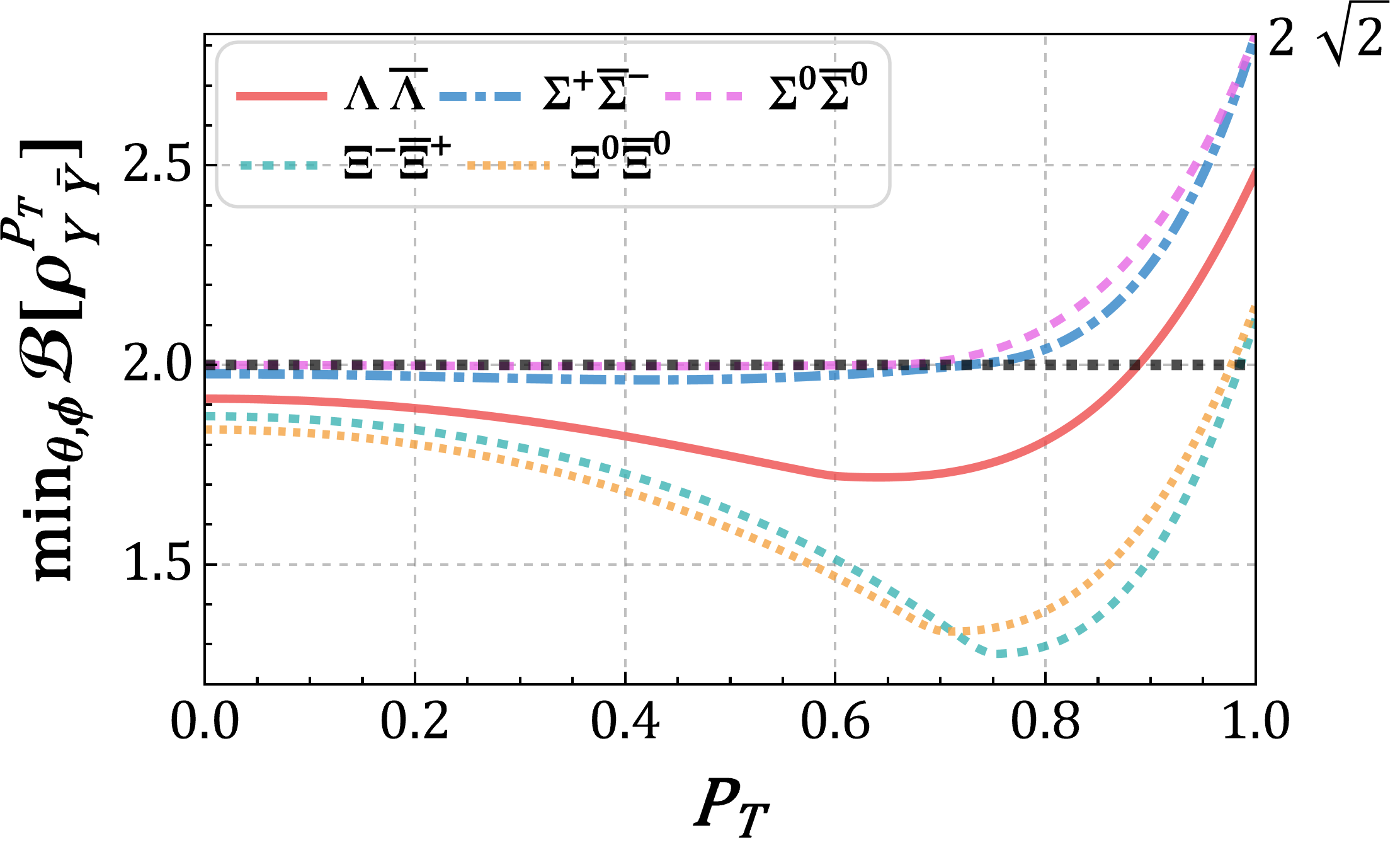}
  		\includegraphics[width = 0.86 \linewidth]{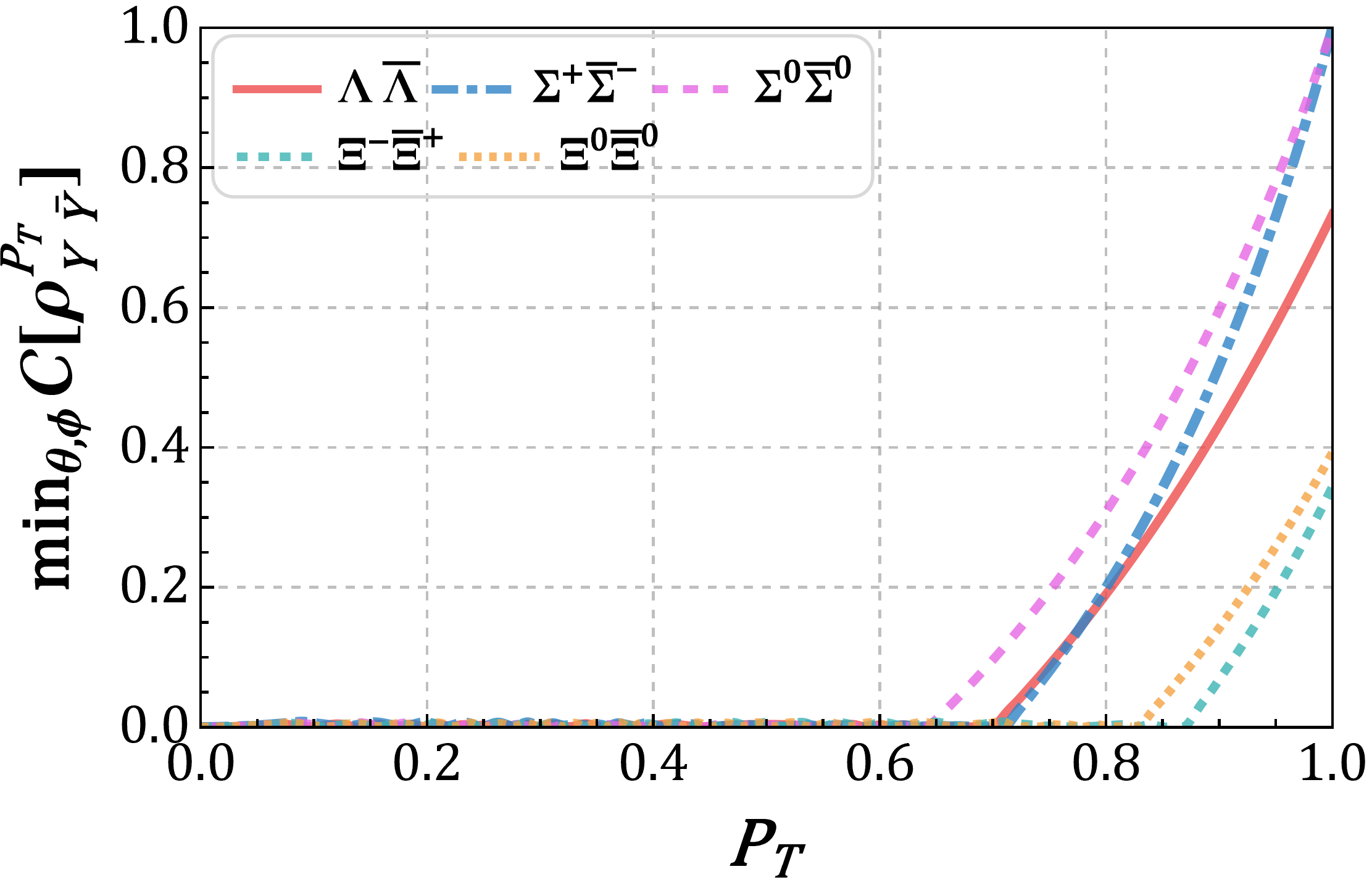}
  \caption{The $\min_{\theta,\phi}\mathcal{B}[\rho^{P_T}_{Y \bar Y}]$ (upper panel) and $\min_{\theta,\phi}\mathcal{C}[\rho^{P_T}_{Y \bar Y}]$ (lower panel) as a function of $P_T$ in $J/\psi\to Y{\bar{Y}}$ for $Y = \Lambda$, $\Sigma^{+}$, $\Sigma^{0}$, $\Xi^{-}$ and $\Xi^{0}$ respectively.
} \label{fig:min_B_C_PT}
\end{figure}

\subsection{Separable state with transversely polarized beams}
\label{sec:separable}

In the case of transverse polarized beams, minimum values of $\mathcal{B}[\rho^{P_T}_{Y \bar Y}]$ and $\mathcal{C}[\rho^{P_T}_{Y \bar Y}]$, labeled as $\min_{\theta,\phi}\mathcal{B}[\rho^{P_T}_{Y \bar Y}]$ and $\min_{\theta,\phi}\mathcal{C}[\rho^{P_T}_{Y \bar Y}]$, can be found across the full range of solid angle, as shown in Fig.~\ref{fig:min_B_C_PT}.
It is observed that over a substantial $P_T$ range the system is of vanishing entanglement, and the CHSH inequality is not violated,
indicating the separability of the system at certain azimuthal angle $\phi^*$:
\begin{itemize}
  \item[a.] $\phi^* = k\pi,(k=0,1,2)$ for $\Sigma^{+}$, $\Sigma^{0}$ channels with $\alpha_\psi<0$;
  \item[b.] $\phi^* = \pi/2+k\pi,(k=0,1)$ for $\Lambda$, $\Xi^{-}$, and $\Xi^{0}$ channels with $\alpha_\psi>0$.
\end{itemize}
At these $\phi^*$, $\vec{B_T^{\pm}}$ in Eq.~\eqref{eq:BT} has only one nonzero component, and the $\rho_{Y\bar{Y}}^{P_T}$ as a $X$-state with the parameters in Eq.~\eqref{eq:Theta} are given by
\begin{align}
B_z &= -\frac{\beta_\psi \left(1\mp P_T^2\right)\sin 2\theta}{2\chi'_T}, \nonumber \\
\lambda_{1} &= \frac{(1\pm P_T^2)(1+\alpha_{\psi}) + (1\mp P_T^2)\sqrt{\zeta}}{2 \chi'_T}, \nonumber \\
\lambda_{2} &= \frac{(1\pm P_T^2)(1+\alpha_{\psi}) - (1\mp P_T^2)\sqrt{\zeta}}{2 \chi'_T}, \label{eq:XparaPTsep12} \\
\lambda_3 &= -\frac{\alpha_{\psi} \sin^2\theta \pm P_T^2 \left(1+ \alpha_{\psi} \cos^2\theta\right)}{\chi'_T }. \nonumber
\end{align} 
with $\chi'_T = 1+\alpha_{\psi}\cos^2\theta \pm P_T^2\alpha_{\psi}\sin^2\theta$ and $\zeta = (\alpha_{\psi}+\cos 2\theta)^2+\gamma_{\psi}^2\sin^2 2\theta$.
The upper and lower sign of ``$\pm$'' and ``$\mp$'' are for case-$a$ and case-$b$ of $\phi^*$, respectively.
The $\min_{\phi}\mathcal{B}[\rho^{P_T}_{Y \bar Y}]$ is given by:
\begin{align}
\min_{\phi}\mathcal{B}[\rho^{P_T}_{Y \bar Y}] &= 2\sqrt{m'}.
\label{eq:CN0}
\end{align}
where $m'$ is the sum of two largest eigenvalues of  $\{\lambda_1^2,\lambda_2^2,\lambda_3^2\}$ in Eq.~\eqref{eq:XparaPTsep12}.
From Eqs. \eqref{eq:concurrence} and \eqref{eq:negativity}, $\min_{\phi}\mathcal{C}[\rho^{P_T}_{Y \bar Y}]$ and $\min_{\phi}\mathcal{N}[\rho^{P_T}_{Y \bar Y}]$ are given by:
\begin{equation}
\begin{aligned}
    \min_{\phi}\mathcal{C}[\rho^{P_T}_{Y \bar Y}]&=\frac{1}{2} \max \bigg\{ 0, \, \lambda_{1} - \lambda_{2} + \lambda_{3}-1, \big.\\
    &\quad\bigg.\lambda_{1} + \lambda_{2} - \sqrt{(1 + \lambda_{3})^{2} - 4B_{z}^{2}} \bigg\},\label{eq:CN1}
\end{aligned}
\end{equation}
\begin{equation}
\begin{aligned}
    \min_{\phi}\mathcal{N}[\rho^{P_T}_{Y \bar Y}]&=\frac{1}{4} \max \bigg\{ 0, \, \lambda_{1} - \lambda_{2} + \lambda_{3}-1, \big.\\
    &\quad\bigg. \sqrt{(\lambda_{1} + \lambda_{2})^{2} + 4B_{z}^{2}} -1 - \lambda_{3} \bigg\}.
    \label{eq:CN2}
\end{aligned}
\end{equation}
For each channel, when the polarization degree $P_T$ is increased to be a critical value, nonseparablity of the system is emergent.
If $P_T$ keeps increasing to another critical value, the Bell nonlocality of the system appears with the CHSH inequality violated.
This is consistent with the hierarchy of quantum correlations mentioned in Sec.~\ref{sec:hierarchy}.
Those critical values are summarized in Table.~\ref{tab:PT_separable}.

\begin{table}[!hb]
\caption{\label{tab:PT_separable}
The critical $P_T$ values of $J/\psi\rightarrow Y\bar{Y}$ for $Y = \Lambda$, $\Sigma^{+}$, $\Sigma^{0}$, $\Xi^{-}$ and $\Xi^{0}$.}
\begin{ruledtabular}
\begin{tabular}{lcc}
$J/\psi\rightarrow$ & Bell nonlocality & Nonseparablity \tabularnewline
\hline
$\Lambda\bar{\Lambda}$ & $0.889$ & $0.699$ \tabularnewline
$\Sigma^{+}\bar{\Sigma}^{-}$ & $0.727$ & $0.711$ \tabularnewline
$\Sigma^{0}\bar{\Sigma}^{0}$ & $0.681$ & $0.609$\tabularnewline
$\Xi^{-}\bar{\Xi}^{+}$ & $0.985$ & $0.869$ \tabularnewline
$\Xi^{0}\bar{\Xi}^{0}$ & $0.971$ & $0.820$ \tabularnewline
\end{tabular}
\end{ruledtabular}
\end{table}

\begin{figure}[!t]
  \centering \includegraphics[width = 0.9 \linewidth]{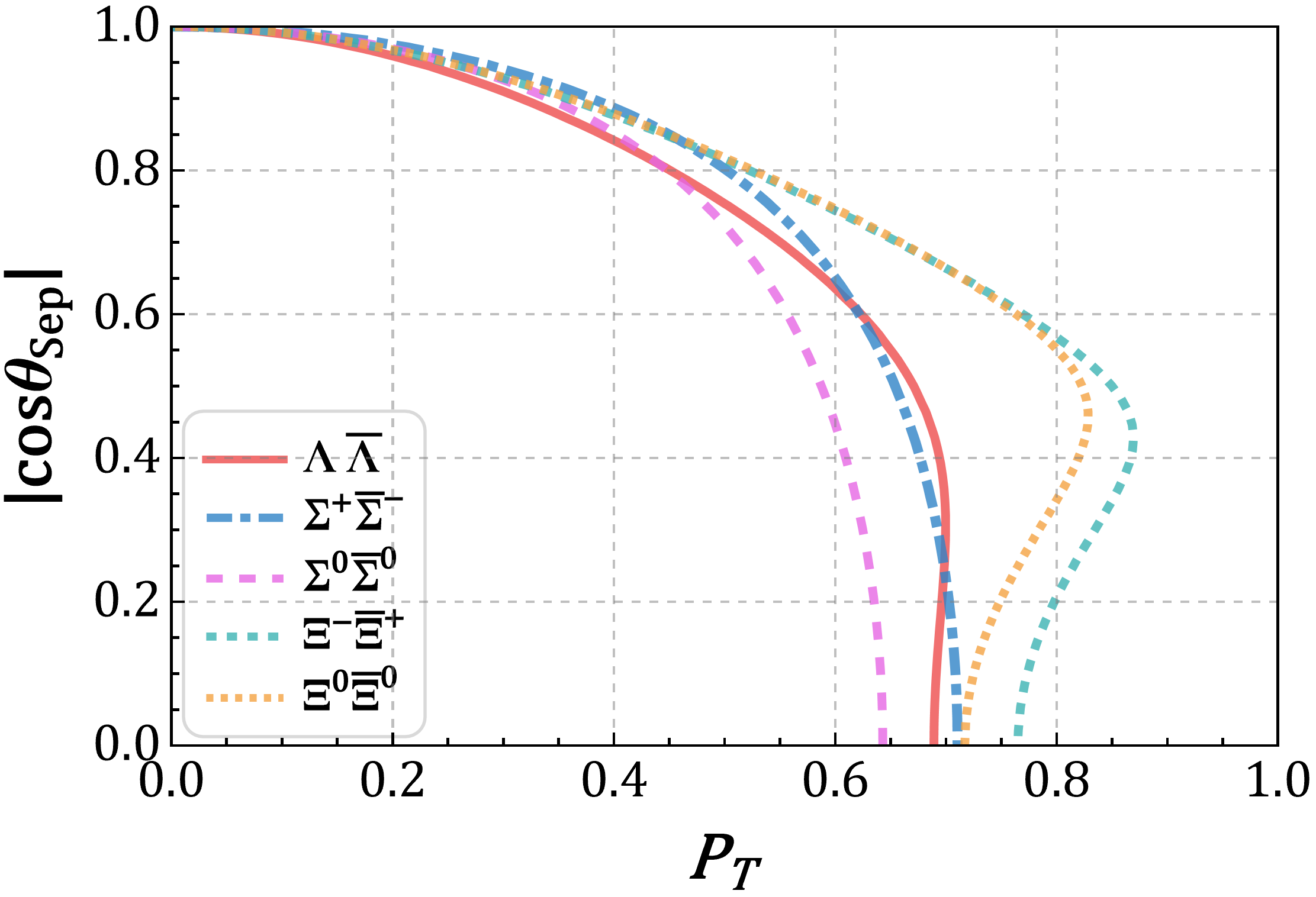}
\caption{
The $\cos\theta_{Sep}$ as a function of the beam transverse polarization degree $P_T$ in $J/\psi\to Y{\bar{Y}}$ for $Y = \Lambda$, $\Sigma^{+}$, $\Sigma^{0}$, $\Xi^{-}$ and $\Xi^{0}$ respectively.\label{fig:theta_Sep}
}
\end{figure}

The system is separable with fixed $P_T$ and $\phi^*$ at scattering angles $\theta_{Sep}$ and $\pi-\theta_{Sep}$, indicating from Eqs.~\eqref{eq:CN1} and \eqref{eq:CN2} that
\begin{equation}
  \min_{\phi}\mathcal{C}(\theta_{Sep})=\min_{\phi}\mathcal{N}(\theta_{Sep})=0.
\end{equation}
Furthermore, a useful theorem states that if a $\mathrm{rank} \text{-}2$ density matrix $\rho$  is separable, its partially transposed density operator $\rho^{\Gamma}$ is also of $\mathrm{rank} \text{-}2$ \cite{Sanpera:1998fb}. Since $\rho_{Y\bar{Y}}$ is always of $\mathrm{rank}\text{-}2$, the $\rho^{\Gamma}_{Y\bar Y}$ must also be $\mathrm{rank} \text{-}2$ within the separable region, implying the relation
\begin{equation}
\text{det}\rho^{\Gamma}_{Y \bar Y}\big|_{\theta_{Sep}} = 0.\label{eq:det0}
\end{equation}
which implies that the separability criterion in Eq.~\eqref{eq:PPTdet} is exactly saturated.
The $\cos{\theta_{Sep}}$ can be plotted as a function of ${P_T}$ as shown in Fig.~\ref{fig:theta_Sep} at fixed $\phi^*$.
For unpolarized lepton beams, the system is separable at $\theta_{Sep}= 0$ and $\pi$.
It can be seen that increasing the transverse polarization degree moves the $\theta_{Sep}$ to the direction of $\pi/2$.
There are two $\theta_{Sep}$ at some $P_T$ range for $\Lambda\bar{\Lambda}$, $\Xi^+\bar{\Xi}^-$ and $\Xi^0\bar{\Xi}^0$ pairs.
When the $P_T$ is big enough, the separate state vanishes for all channels.

\section{Conclusion} \label{sec:conclusion}

This work systematically investigates the influence of longitudinal and transverse beam polarization on quantum entanglement and Bell nonlocality of hyperon-antihyperon pairs produced in electron-positron annihilation.
Based on the explicit spin density matrix,
the CHSH parameter, concurrence, and negativity of the $Y\bar{Y}$ system are quantitatively evaluated.
It is found that both polarization configurations can enhance quantum correlations, though the manifestation differ remarkably.
Longitudinal polarization of beam broadens the angular range of Bell nonlocality and increases monotonically entanglement of the system,
but fails to attain maximal entanglement even at maximum polarization degree.
In contrast, transverse polarization enables the system to be maximal entangled across full range of scattering angle if choosing optimally the azimuthal angle.
The underlying reason is ascribed to different reaction of hyperons polarization to beam polarization modes.
Hierarchy of quantum correlations under two polarization configurations of beams and separable state in the case of transversely polarized beams are thoroughly discussed as well.

Compared to the $\Lambda \bar{\Lambda}$, $\Xi^{-} \bar{\Xi^{-}}$, and $\Xi^{0} \bar{\Xi^{0}}$ channels, the $J/\psi\rightarrow \Sigma^{+} \bar{\Sigma^{+}}$ and $\Sigma^{0} \bar{\Sigma^{0}}$ channels not only possess negative decay parameters, but also exhibit much smaller relative phases between electric and magnetic couplings.
As a result, Bell nonlocality and quantum entanglement in the respective $Y\bar{Y}$ systems exhibit distinctly different sensitivities to the manipulation of the lepton beam polarization. As a byproduct, this provides a novel perspective on the anomalous decay parameters of the $\Sigma$ channels, revealing that these anomalies physically manifest as distinct responses in quantum correlations.

Transversely polarized beams would be already available at BESIII as discussed in literature \cite{Cao:2024tvz}, and longitudinally polarized beams would be feasible at future Super $\tau$-Charm facility \cite{Achasov:2023gey}.
The findings in this paper highlight the potential of exploiting polarized beams as a tool to study quantum correlations of $Y\bar{Y}$ system, and also as a means to explore theoretically the quantum information in other high-energy reactions.
Furthermore, whether the manipulation of beam polarization can be utilized to mitigate fundamental loopholes (such as the detection or locality loopholes using only classical observables ) in Bell tests remains an intriguing open question for particle physics.

\begin{acknowledgments}

This work is supported by the National Key R\&D Program of China under Grant No. 2023YFA1606703, the National Natural Science Foundation of China  (Grant Nos. 12547111 and 12235008), and the Hebei Natural Science Foundation with Grant Nos. A2022201017 and A2023201041, and Natural Science Foundation of Guangxi Autonomous Region with Grant No. 2022GXNSFDA035068.

\end{acknowledgments}

\begin{table*}[!hbt]
\caption{\label{tab:decay_parameters_phi3686}
The decay parameters of $\psi(3686) \rightarrow Y\bar{Y}$
for $Y = \Lambda$, $\Sigma^{+}$, $\Sigma^{0}$, $\Xi^{-}$ and $\Xi^{0}$.}
\begin{ruledtabular}
\begin{tabular}{lccc}
 Decay channels & $\alpha_{\psi}$ & $\Delta\Phi/\mathrm{rad}$ & Ref.\tabularnewline
\hline
$\psi(3686)\to\Lambda\bar{\Lambda}$ & $0.69\pm0.07\pm0.02$ & $0.40\pm0.14\pm0.03$ & \citep{ParticleDataGroup:2022pth,BESIII:2023euh}\tabularnewline
$\psi(3686)\to\Sigma^{+}\bar{\Sigma}^{-}$ & $0.682\pm0.030\pm0.011$ & $0.397\pm0.07\pm0.014$ & \citep{ParticleDataGroup:2022pth,BESIII:2020fqg}\tabularnewline
$\psi(3686)\to\Sigma^{0}\bar{\Sigma}^{0}$ & $0.814\pm0.028\pm0.028$ & $0.512\pm0.085\pm0.034$ & \citep{ParticleDataGroup:2022pth,BESIII:2024nif}\tabularnewline
$\psi(3686)\to\Xi^{-}\bar{\Xi}^{+}$ & $0.693\pm0.048\pm0.049$ & $0.667\pm0.111\pm0.058$ & \citep{ParticleDataGroup:2022pth,BESIII:2022lsz}\tabularnewline
$\psi(3686)\to\Xi^{0}\bar{\Xi}^{0}$ & $0.665\pm0.086\pm0.081$ & $-0.050\pm0.150\pm0.020$ & \citep{ParticleDataGroup:2022pth,BESIII:2023drj}\tabularnewline
\end{tabular}
\end{ruledtabular}
\end{table*}

\appendix

\section{Supplementary figures for $J/\psi$ and $\psi(3686) \rightarrow Y\bar{Y}$}
\label{sec:Supplpsi}

\begin{figure*}[h]
  \centering
  		\includegraphics[width = 0.225 \linewidth]{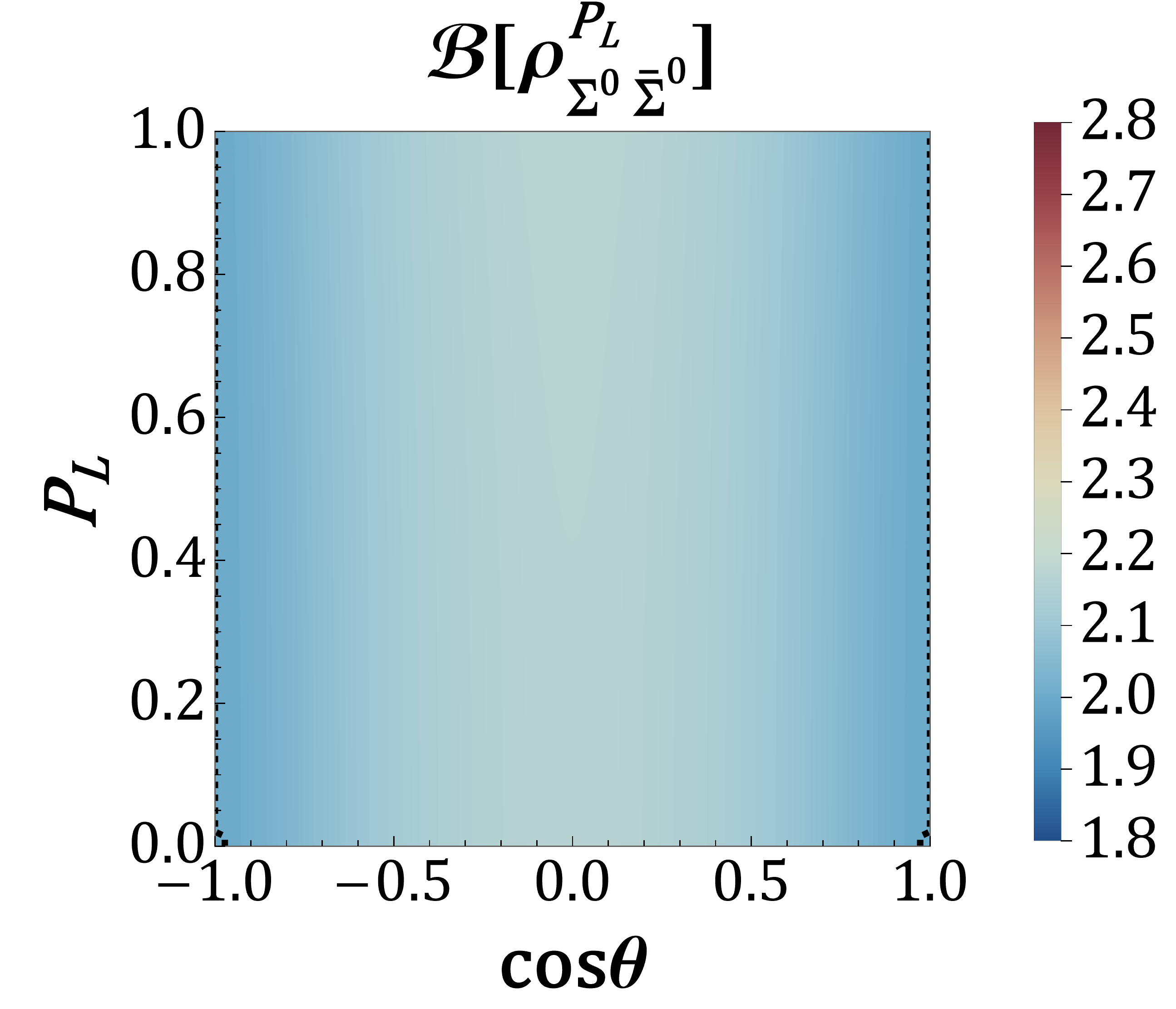}
  		\includegraphics[width = 0.225 \linewidth]{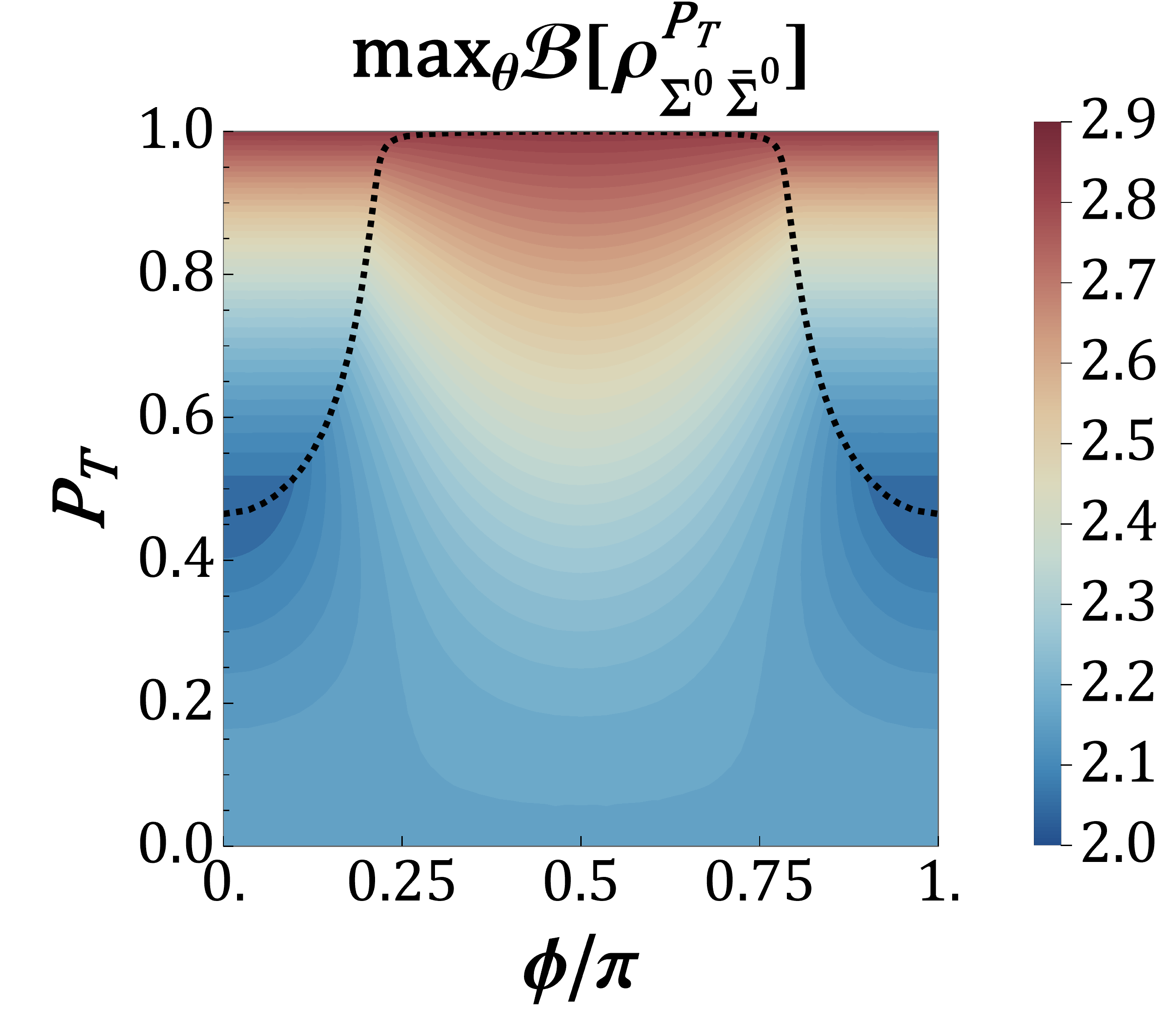}
      \includegraphics[width = 0.225 \linewidth]{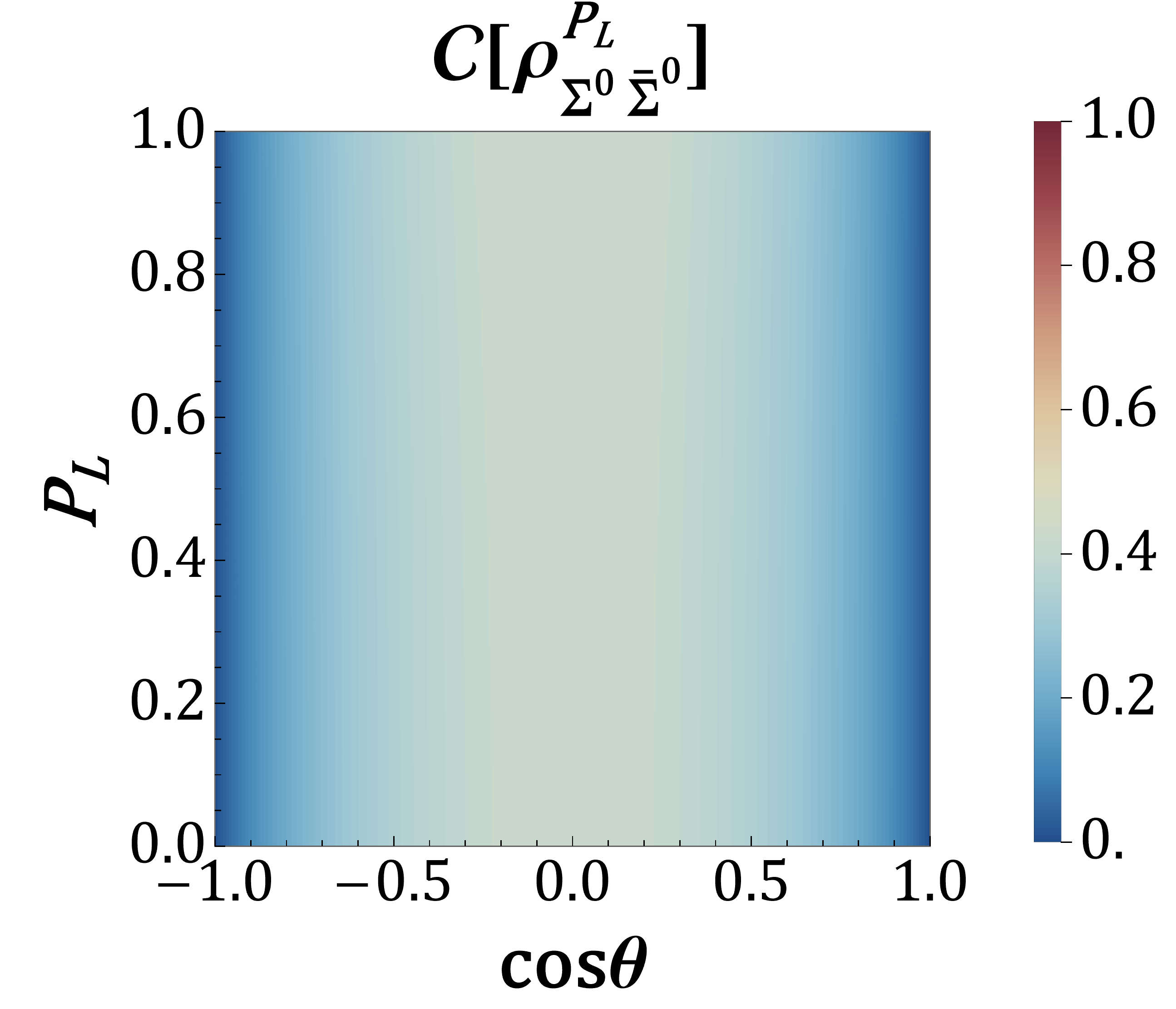}
  		\includegraphics[width = 0.225 \linewidth]{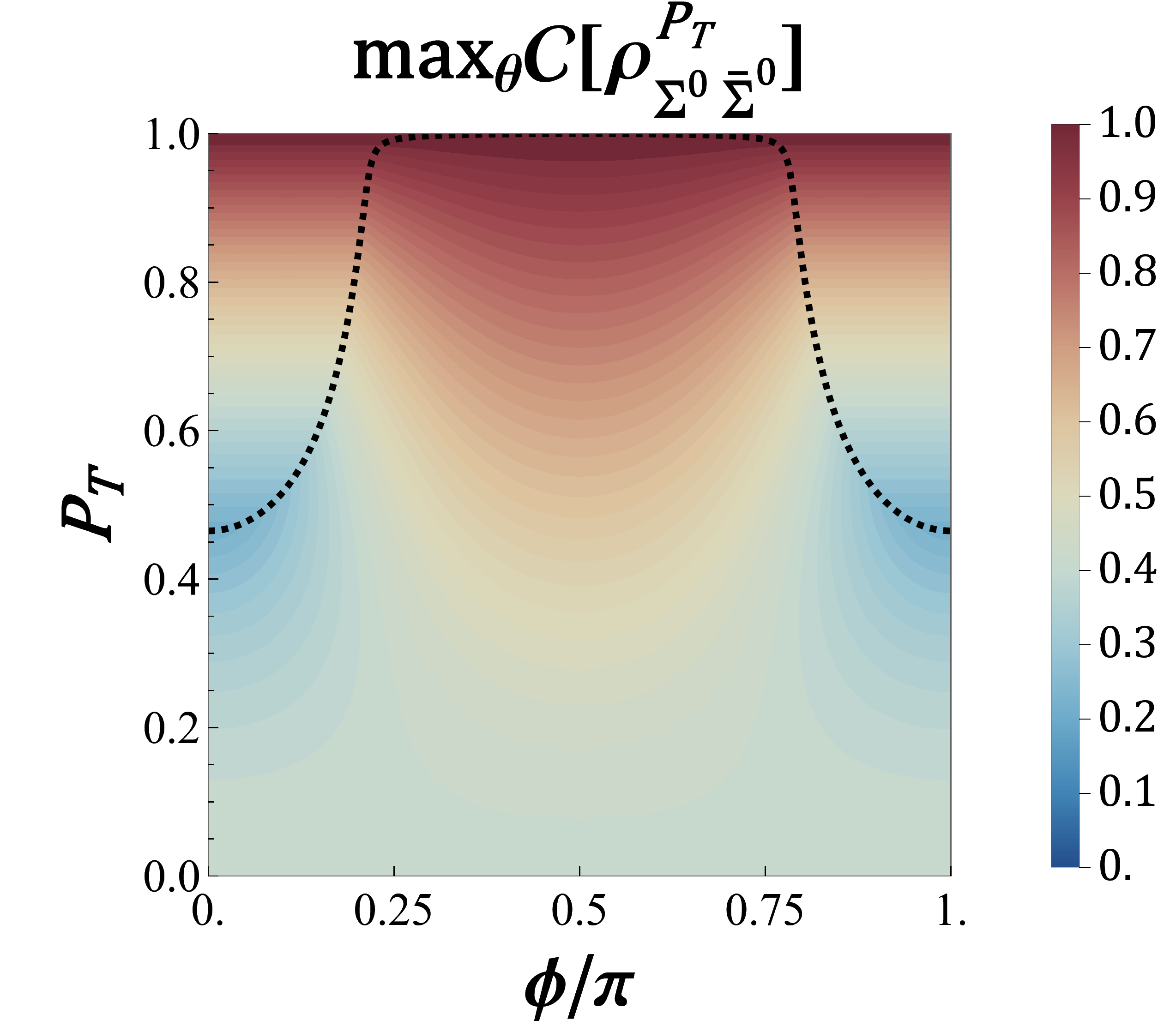}
   \caption{The CHSH parameters and concurrence for $J/\psi \to \Sigma^0{\bar\Sigma}^0$. The dashed curves are for $\mathcal{B}(\theta=0,\pi)=\mathcal{B}(\theta=\pi/2)$ and $\mathcal{C}(\theta=0,\pi)=\mathcal{C}(\theta=\pi/2)$, repsectively. \label{fig:sig0Jpsi}
}
\end{figure*}


\begin{figure*}[h]
  \centering
  		\includegraphics[width = 0.195 \linewidth]{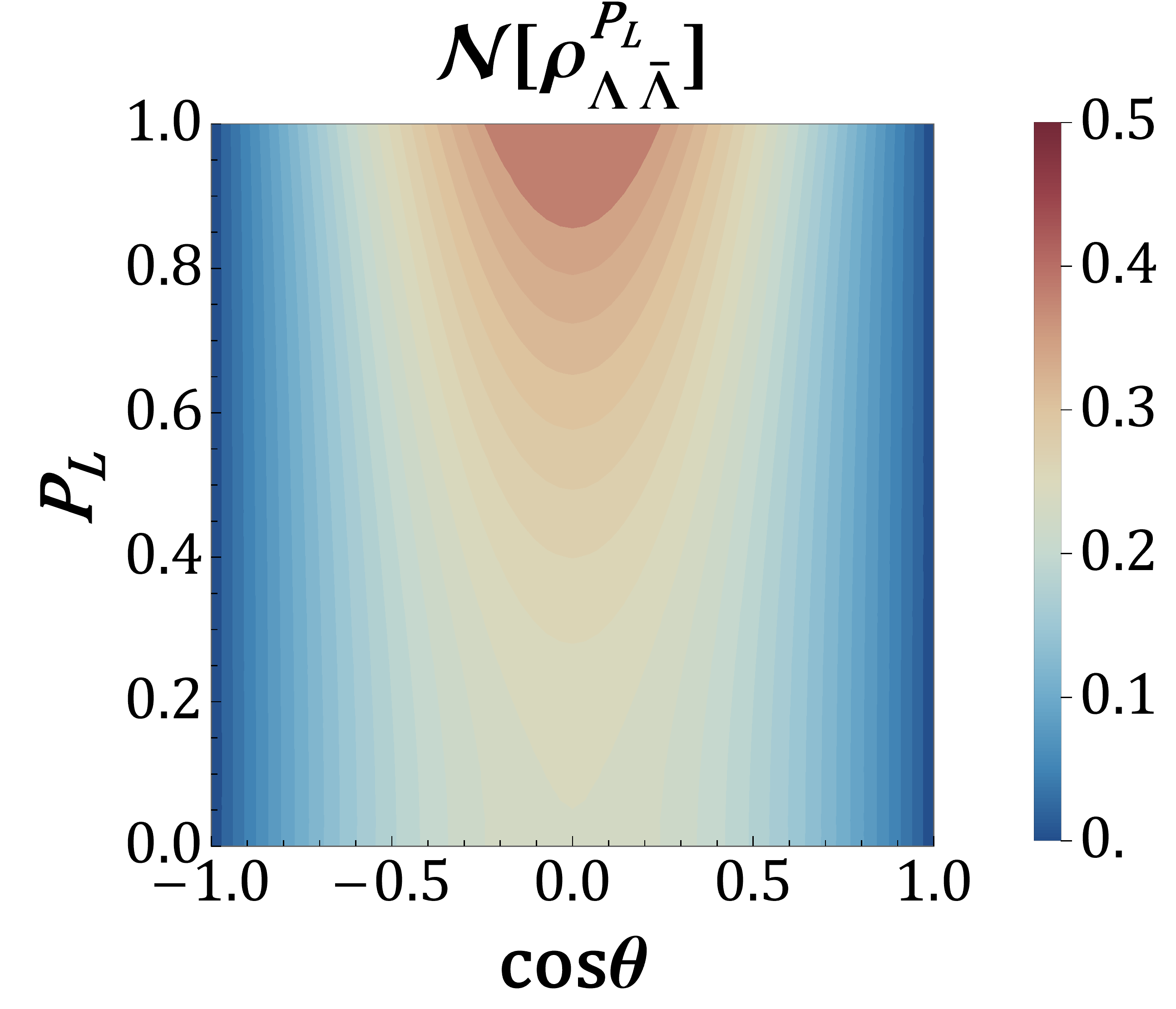}
  		\includegraphics[width = 0.195 \linewidth]{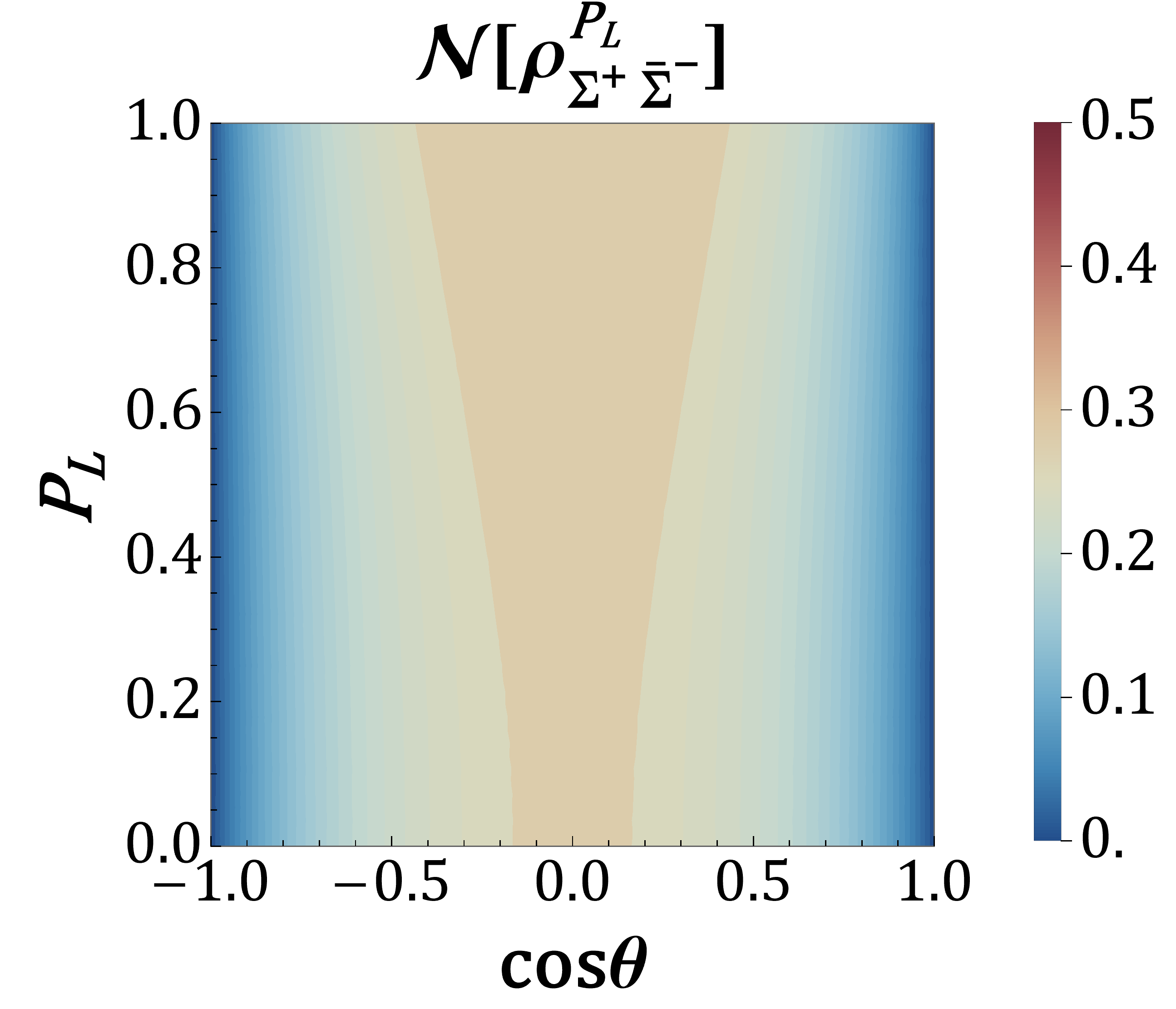}
      \includegraphics[width = 0.195 \linewidth]{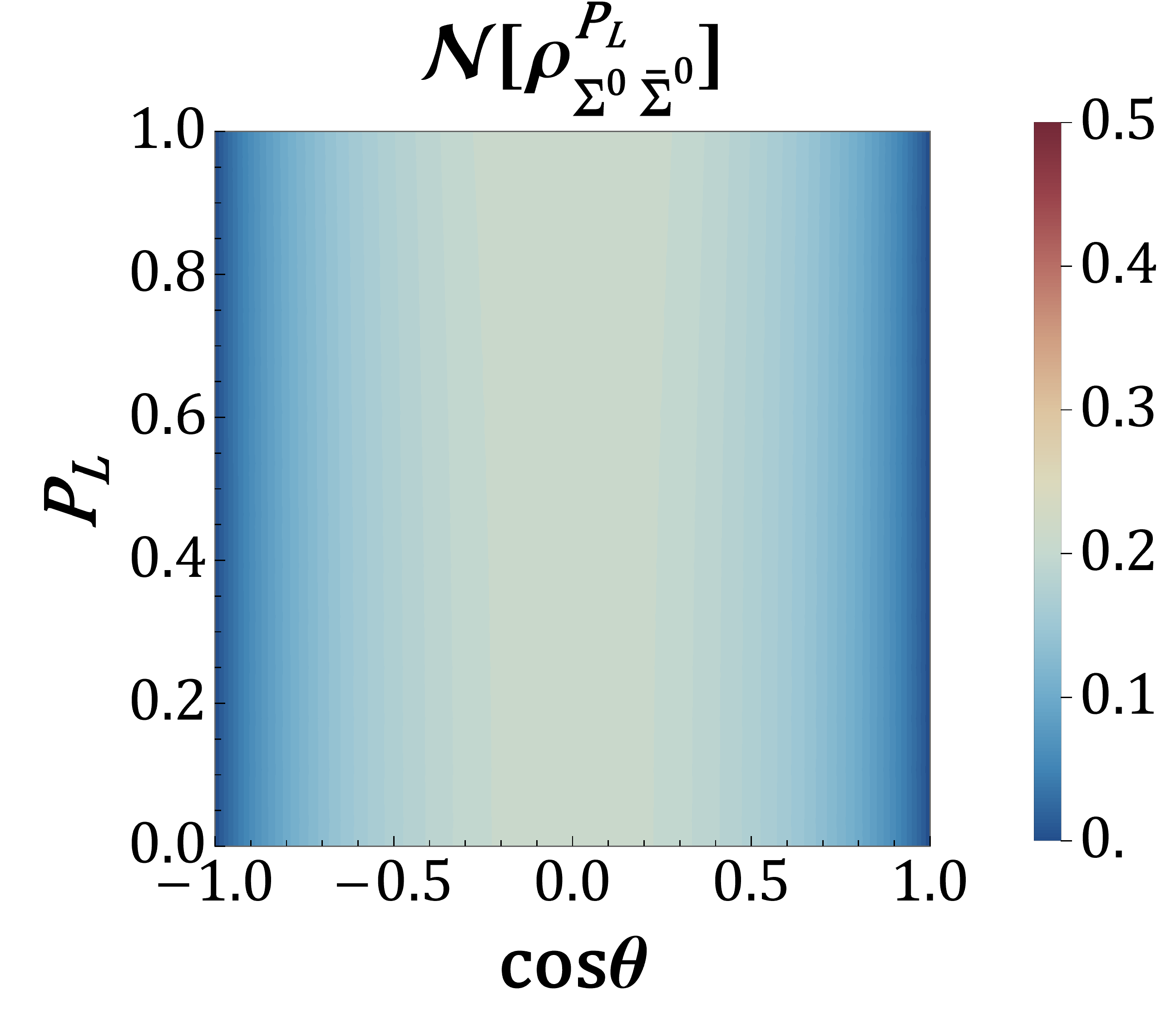}
      \includegraphics[width = 0.195 \linewidth]{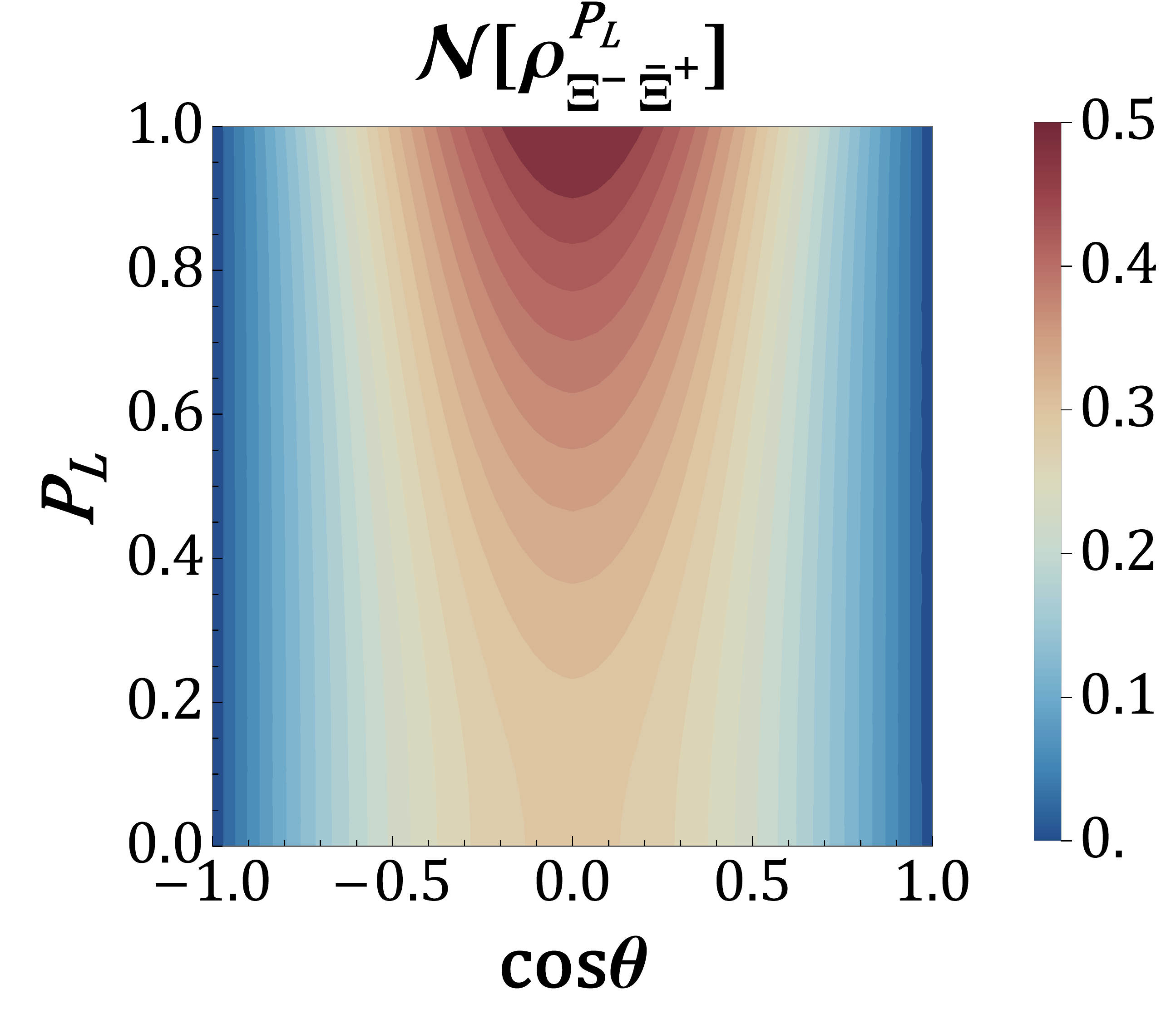}
  		\includegraphics[width = 0.195 \linewidth]{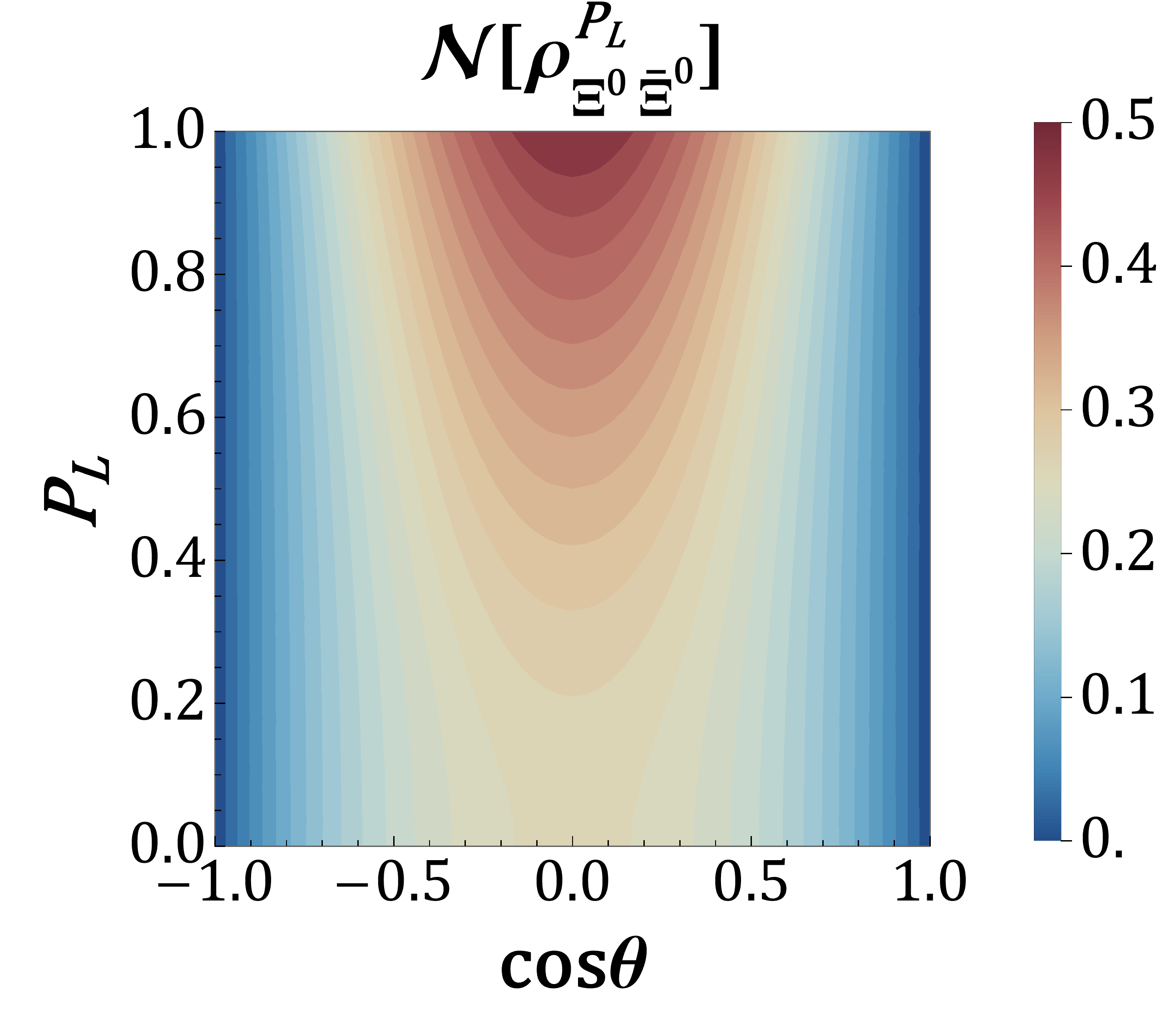}
      \includegraphics[width = 0.195 \linewidth]{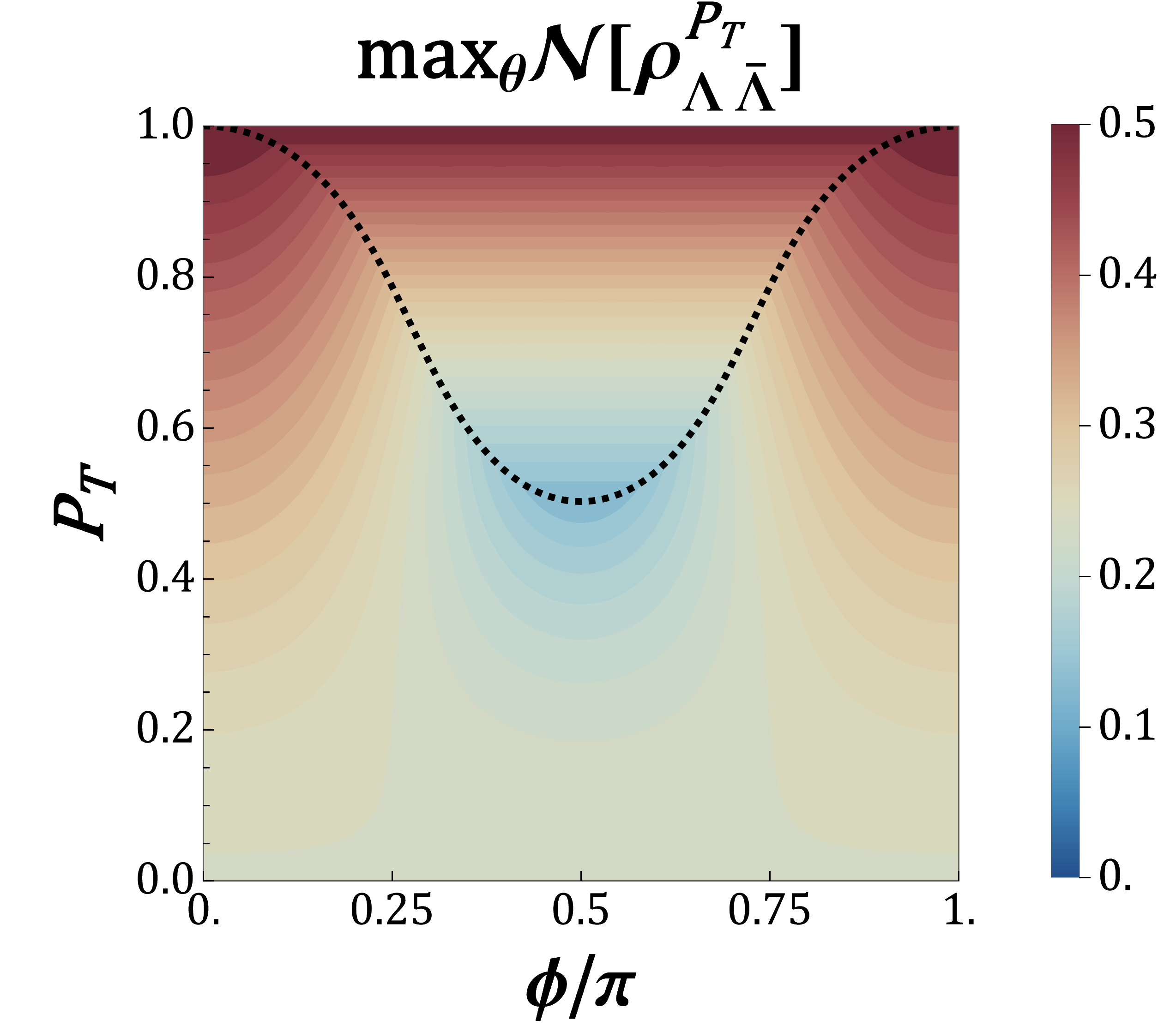}
  		\includegraphics[width = 0.195 \linewidth]{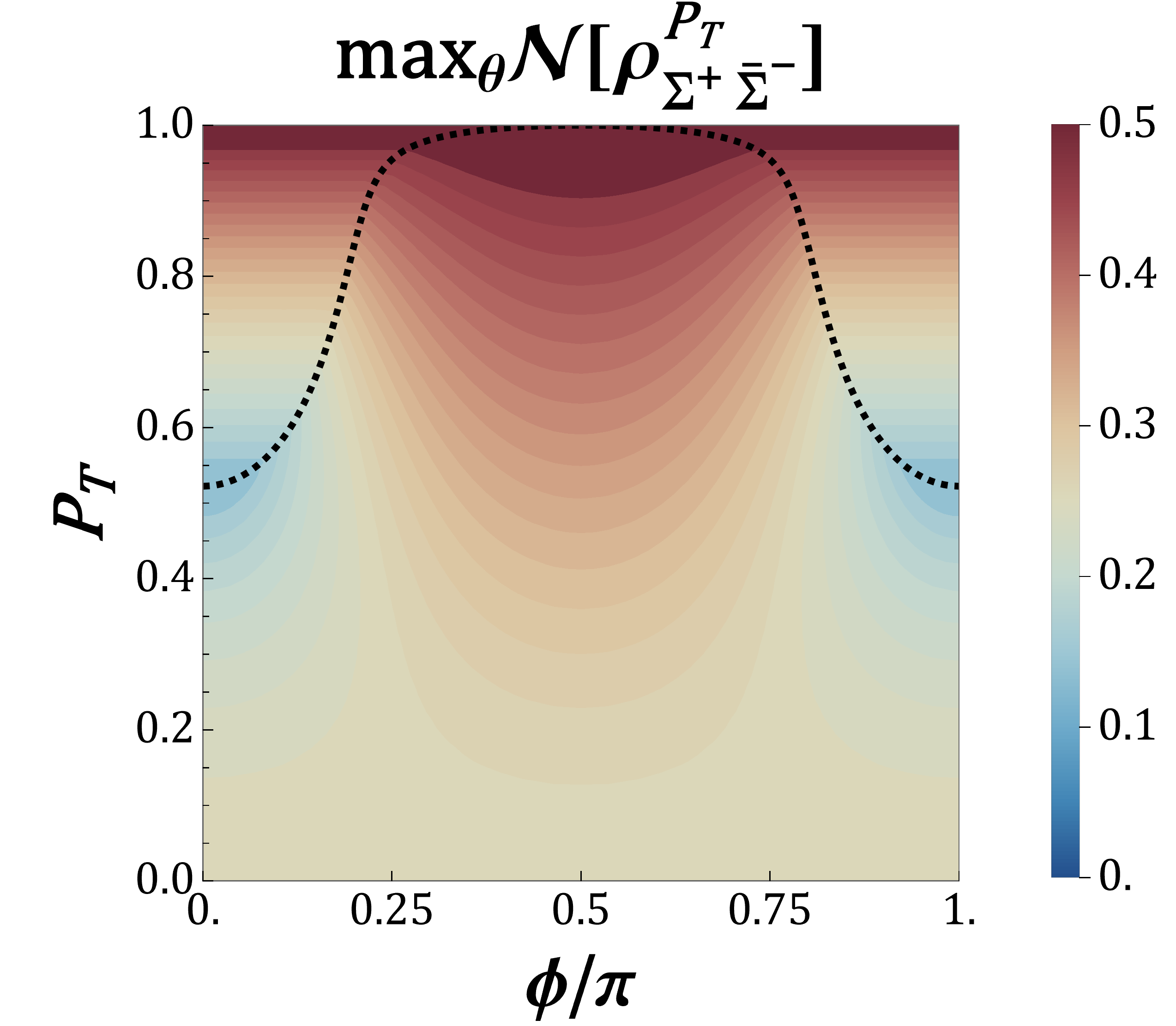}
      \includegraphics[width = 0.195 \linewidth]{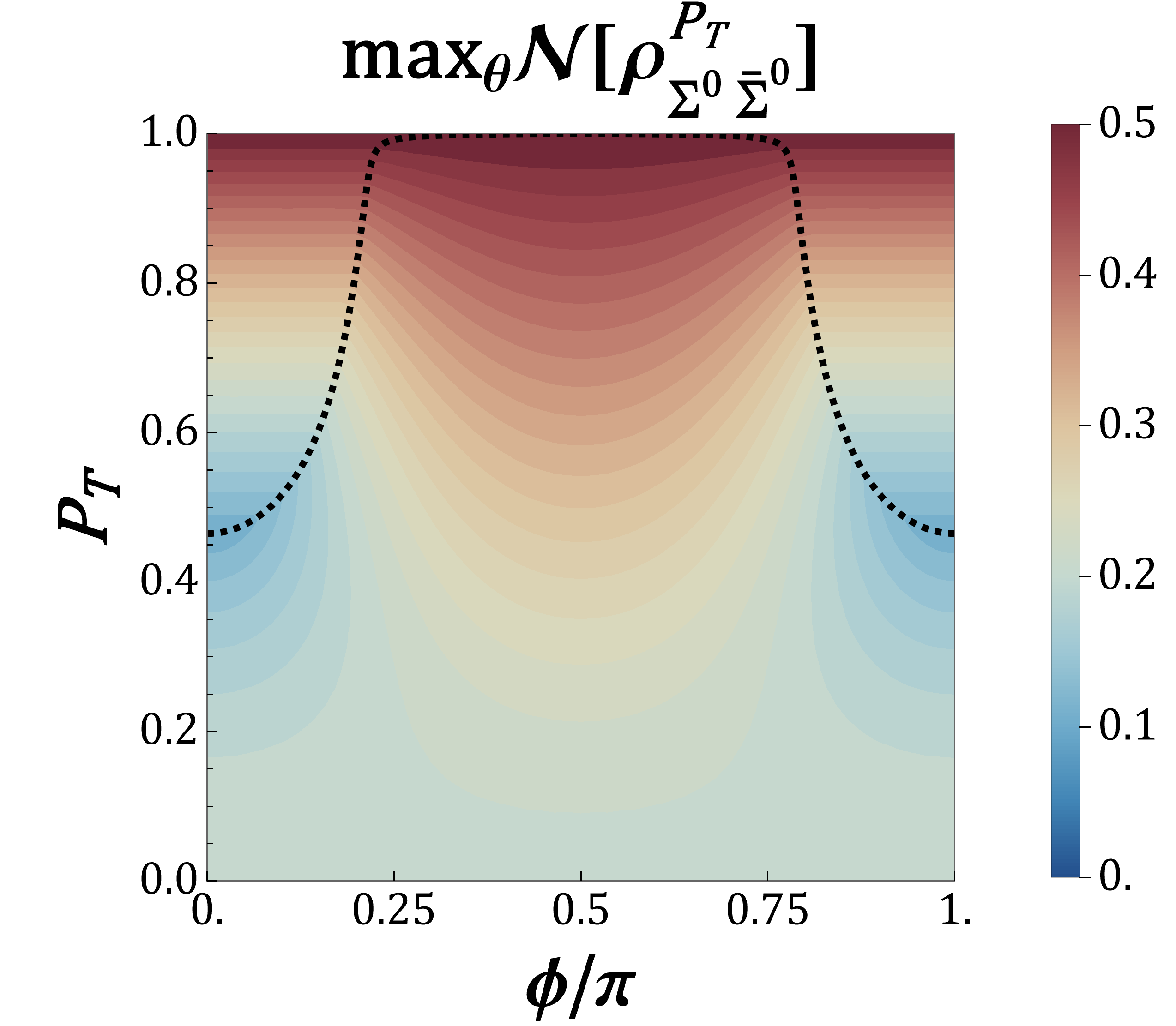}
      \includegraphics[width = 0.195 \linewidth]{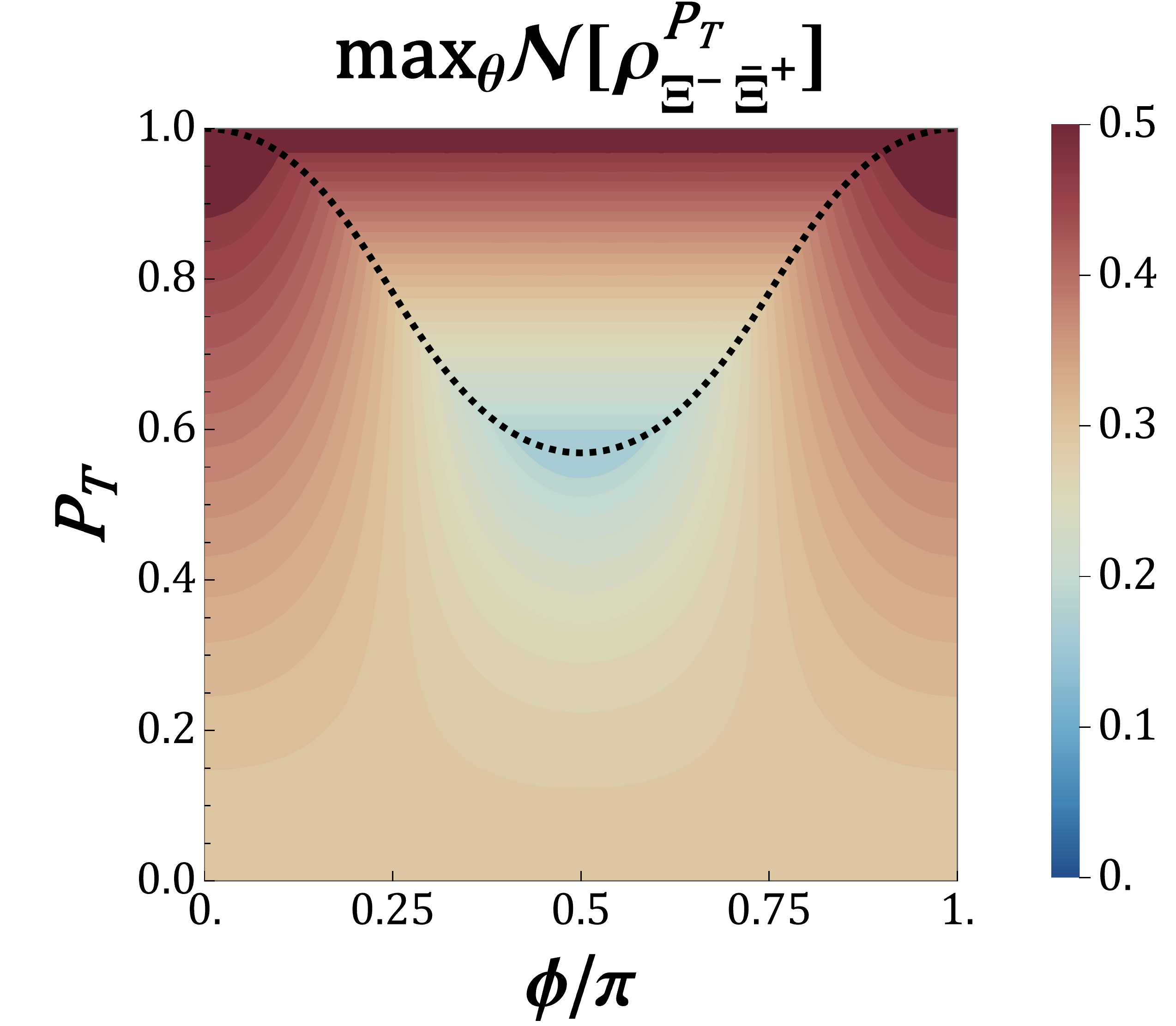}
  		\includegraphics[width = 0.195 \linewidth]{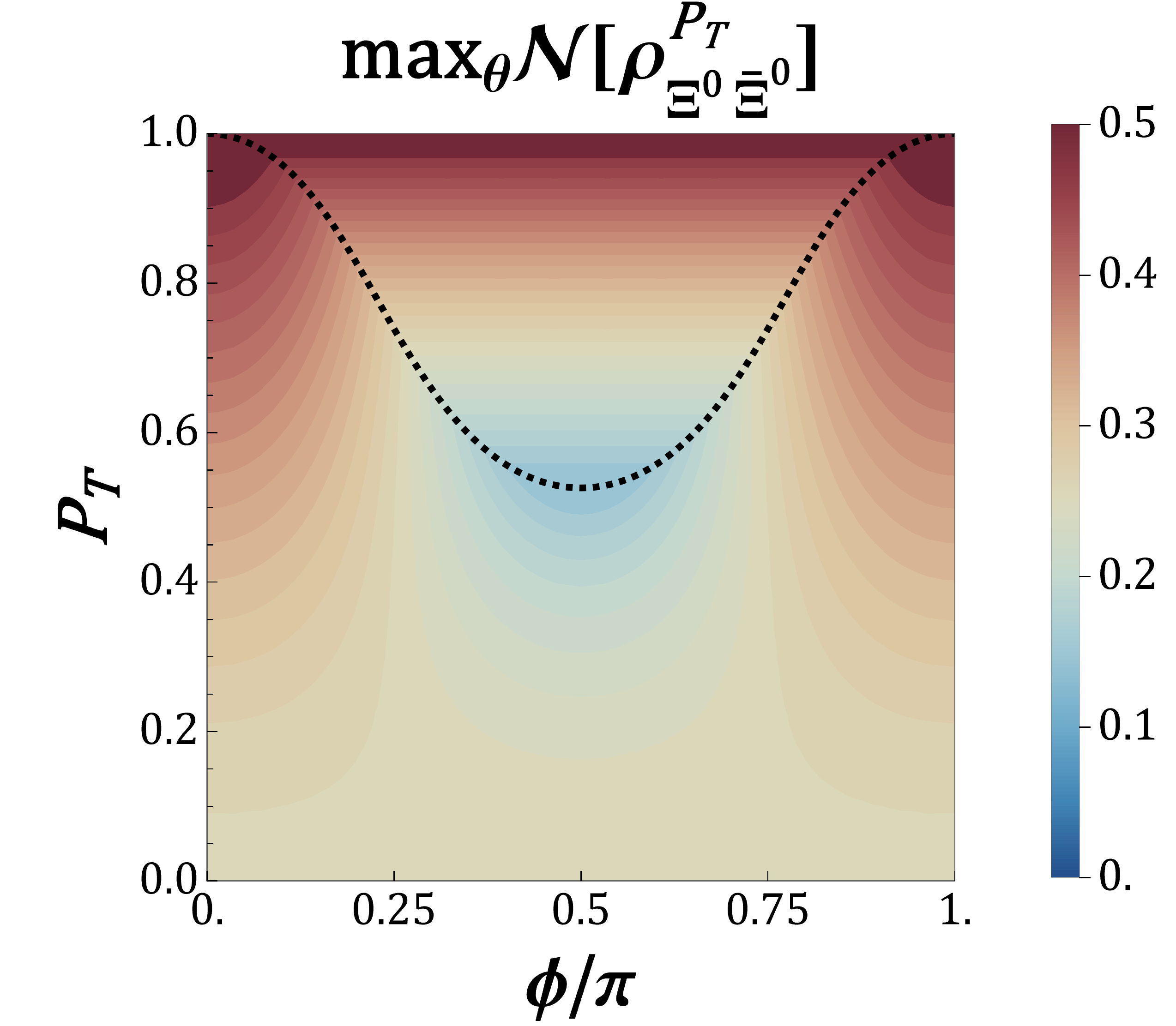}
   \caption{Negativity $\mathcal{N}[\rho^{P_L}_{Y \bar Y}]$ as a function of $P_L$ and $\theta$ (upper panels), and the $\max_{\theta}\mathcal{N}[\rho^{P_T}_{Y \bar Y}]$ as a function of $P_T$ and $\phi$ (lower panels) in in $J/\psi \rightarrow Y\bar{Y}$ for $Y = \Lambda$, $\Sigma^{+}$, $\Sigma^{0}$, $\Xi^{-}$ and $\Xi^{0}$. The dashed curve is $\mathcal{N}(\theta=0,\pi)=\mathcal{N}(\theta=\pi/2)$.\label{fig:negJpsi}
}
\end{figure*}

\begin{figure*}[h]
  \centering
  		\includegraphics[width = 0.195 \linewidth]{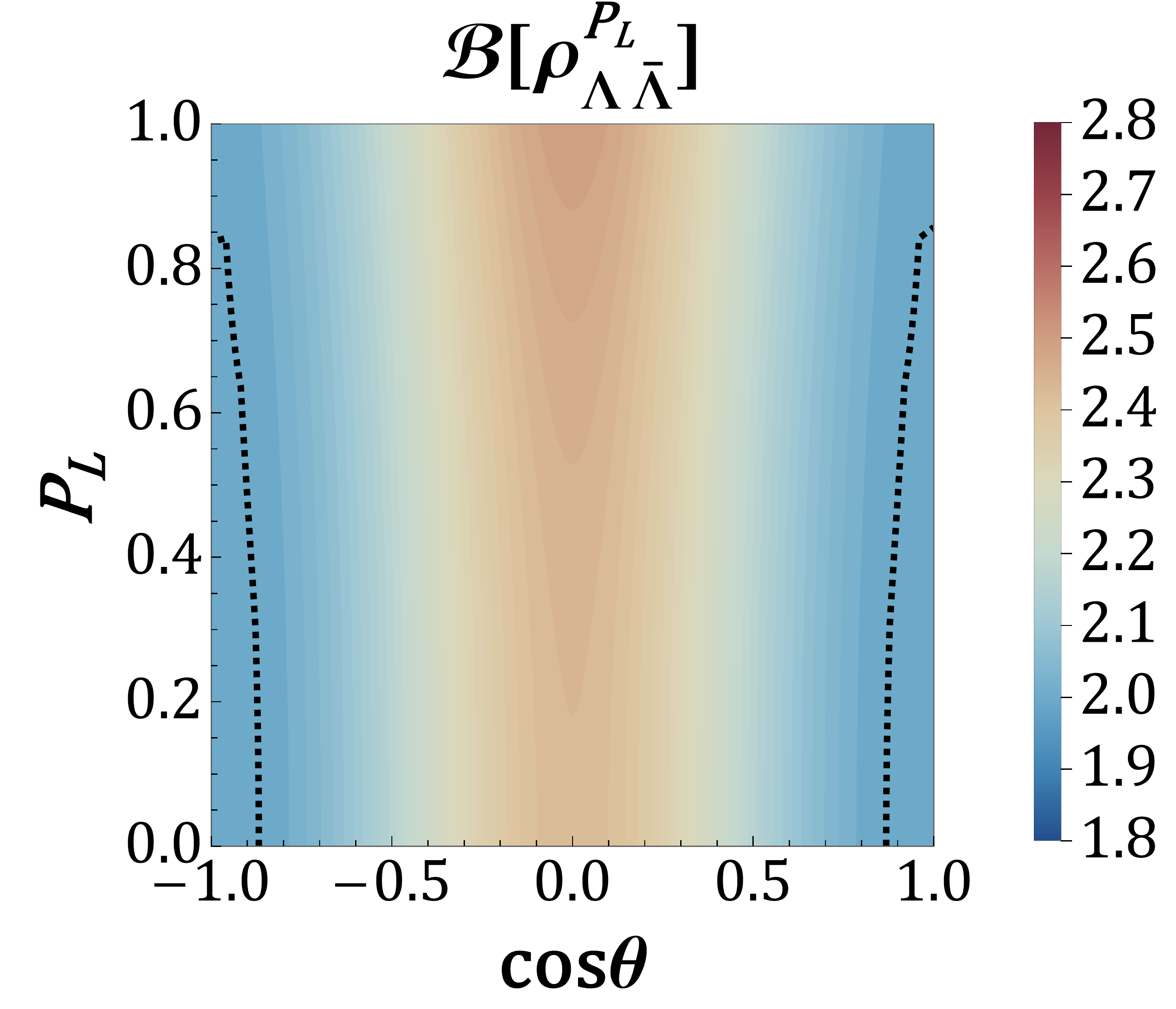}
  		\includegraphics[width = 0.195\linewidth]{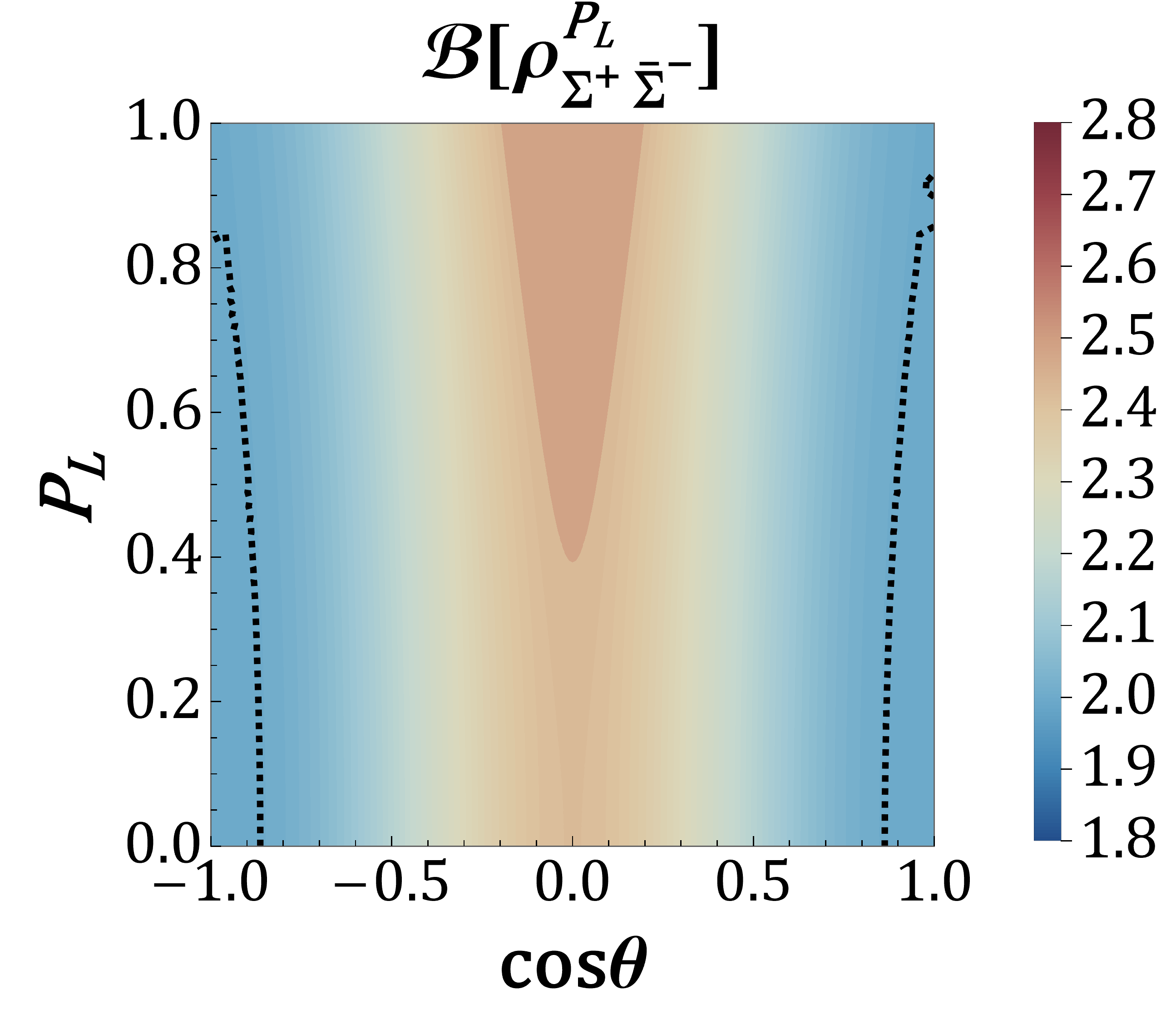}
      \includegraphics[width = 0.195\linewidth]{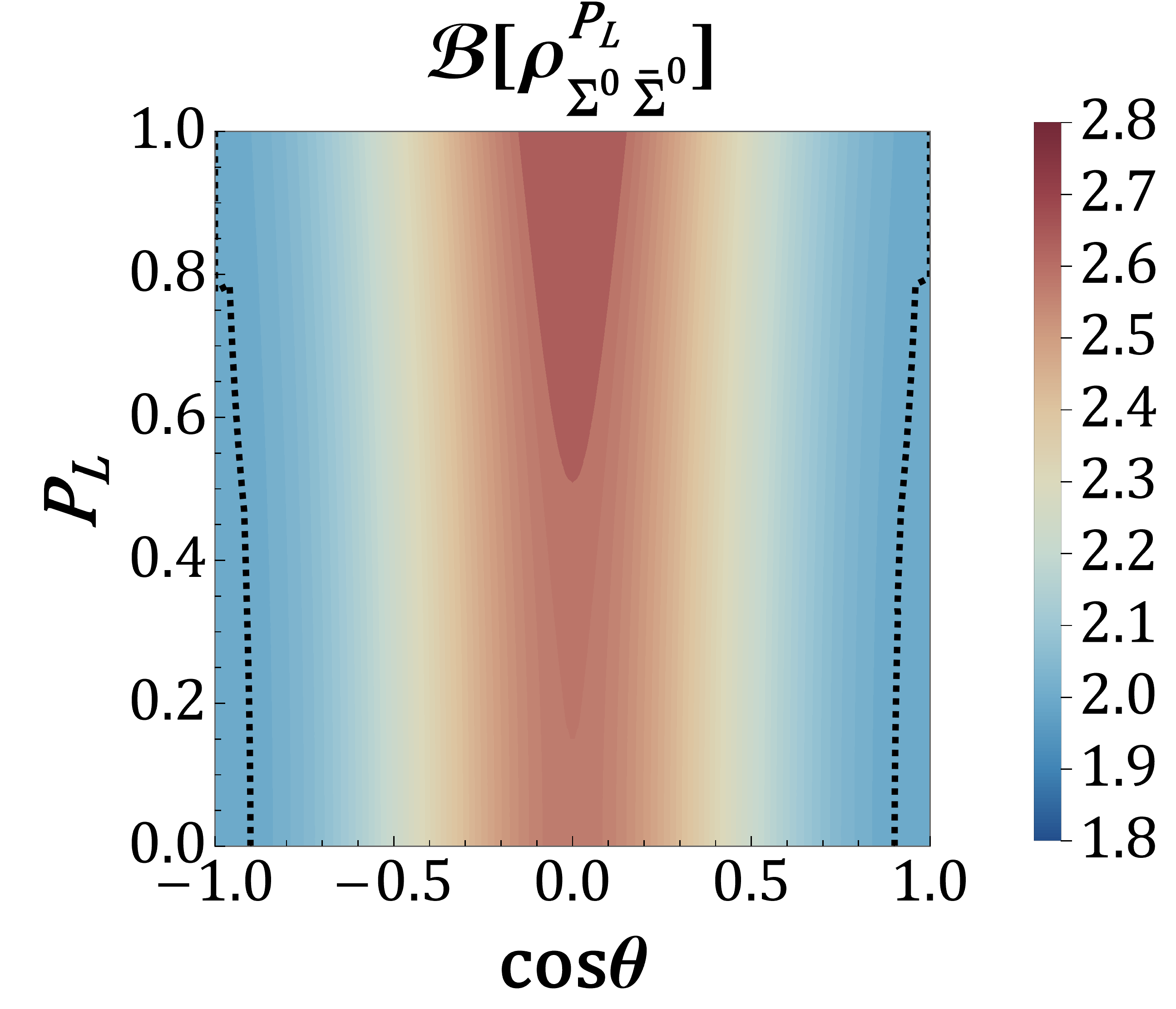}
      \includegraphics[width = 0.195 \linewidth]{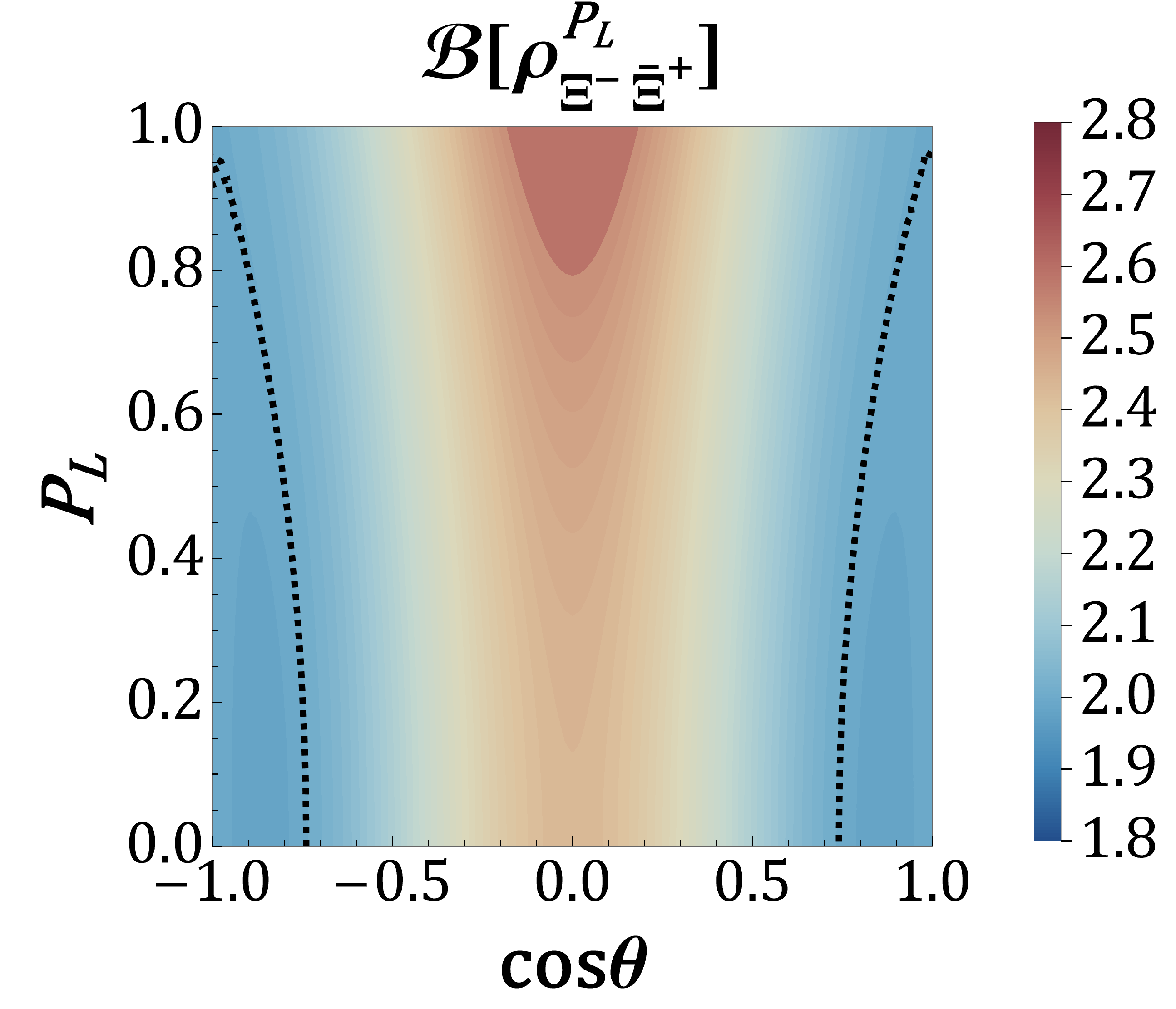}
  		\includegraphics[width = 0.195 \linewidth]{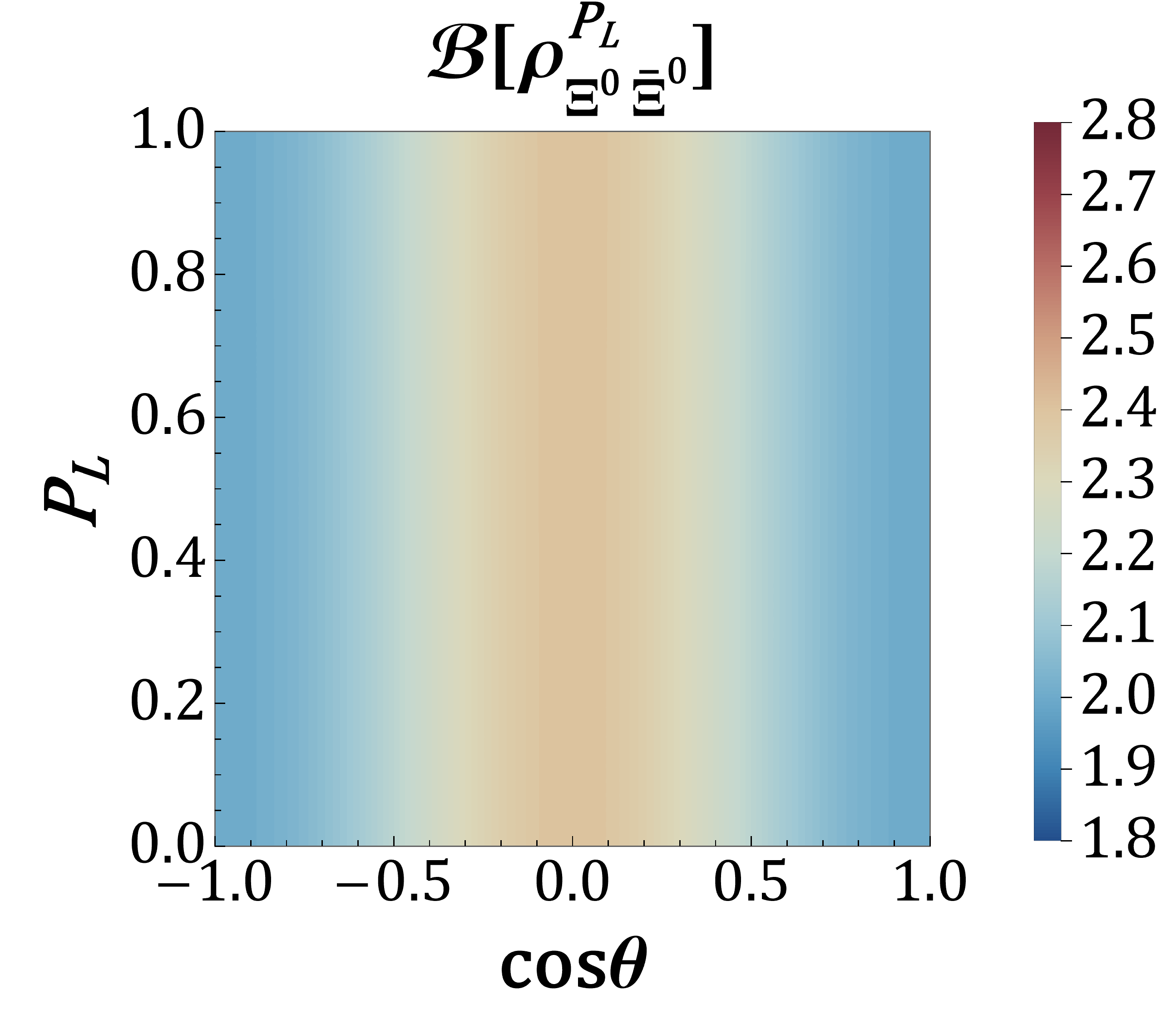}
      \includegraphics[width = 0.195 \linewidth]{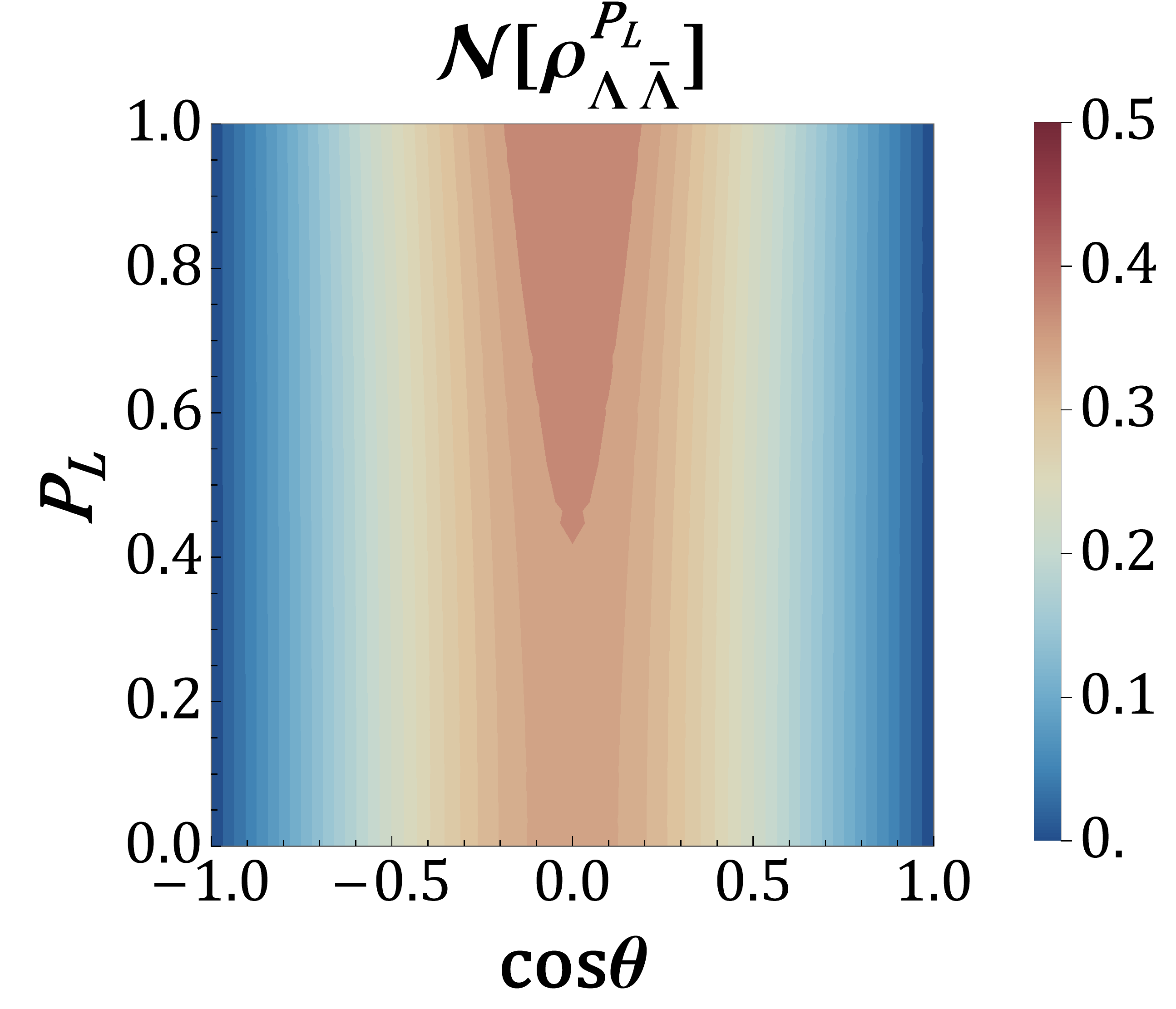}
  		\includegraphics[width = 0.195\linewidth]{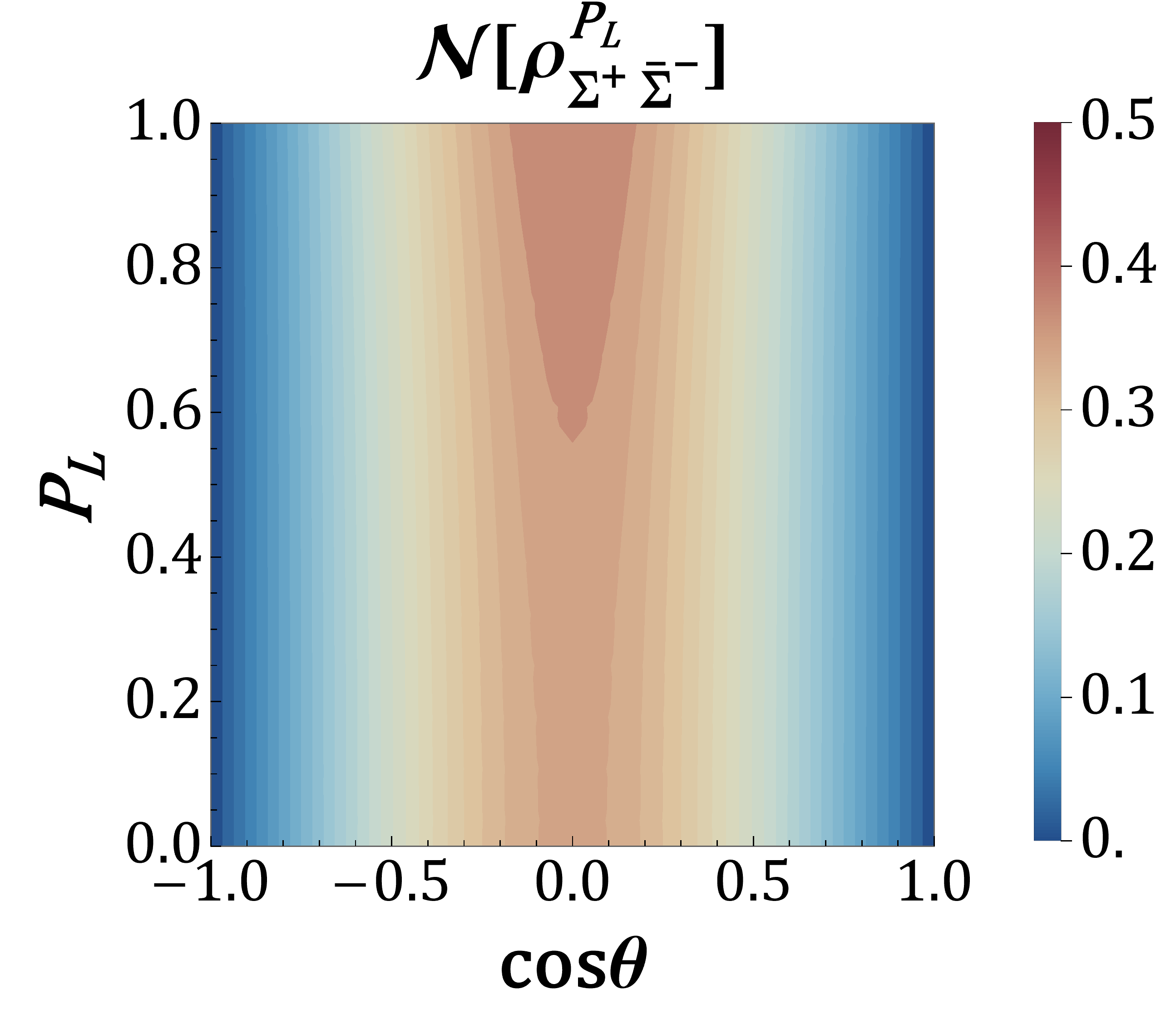}
      \includegraphics[width = 0.195\linewidth]{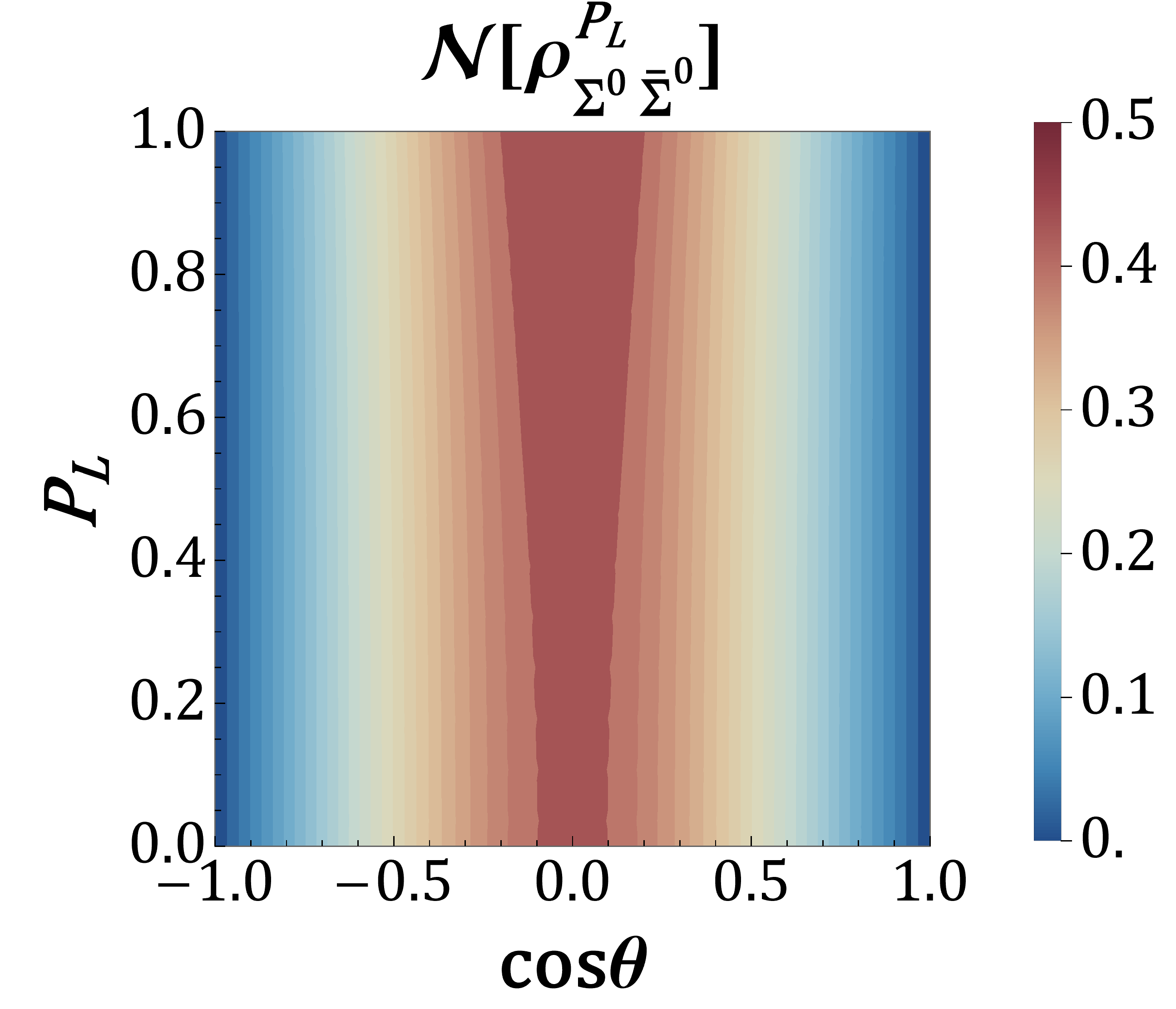}
      \includegraphics[width = 0.195 \linewidth]{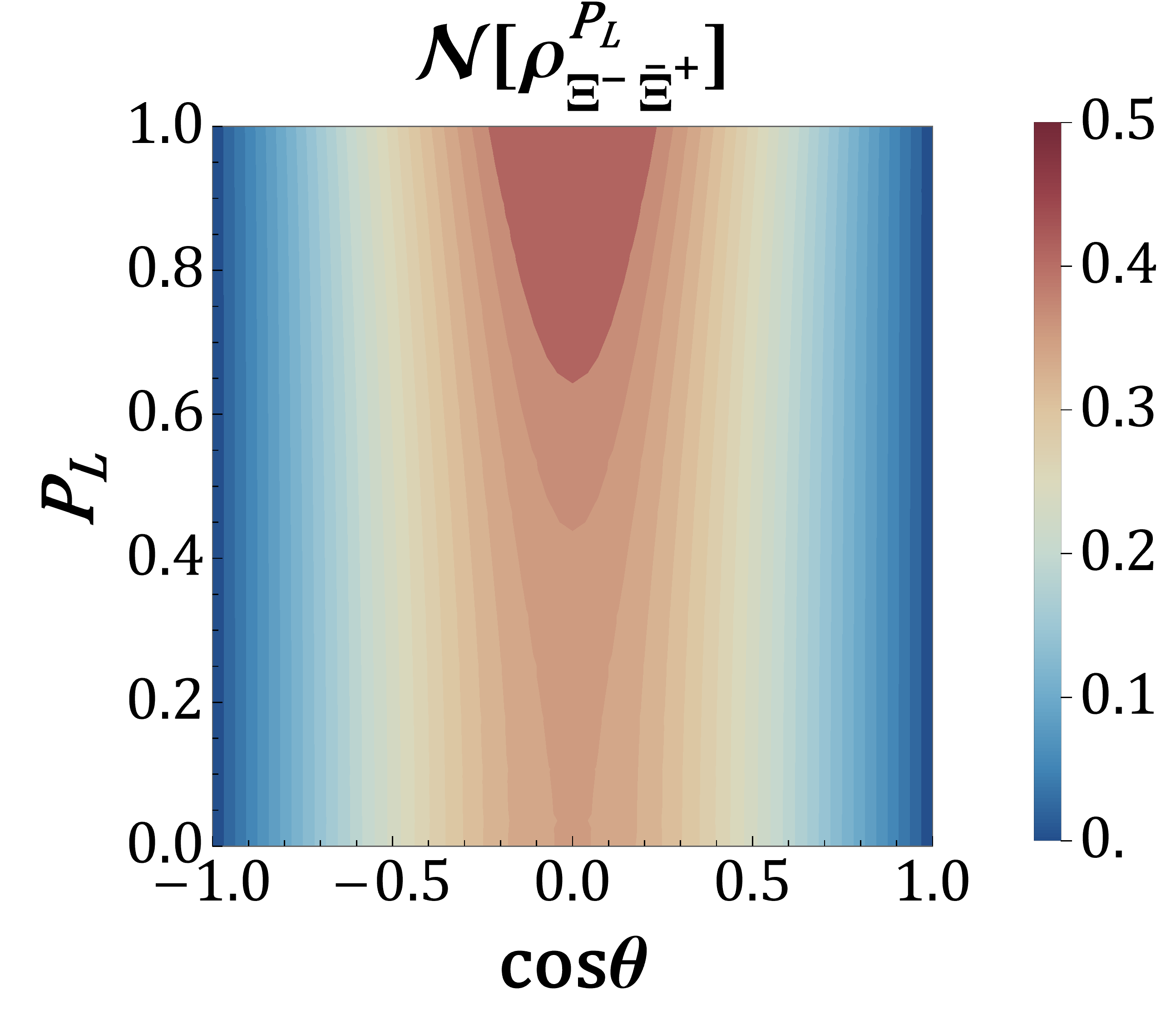}
  		\includegraphics[width = 0.195 \linewidth]{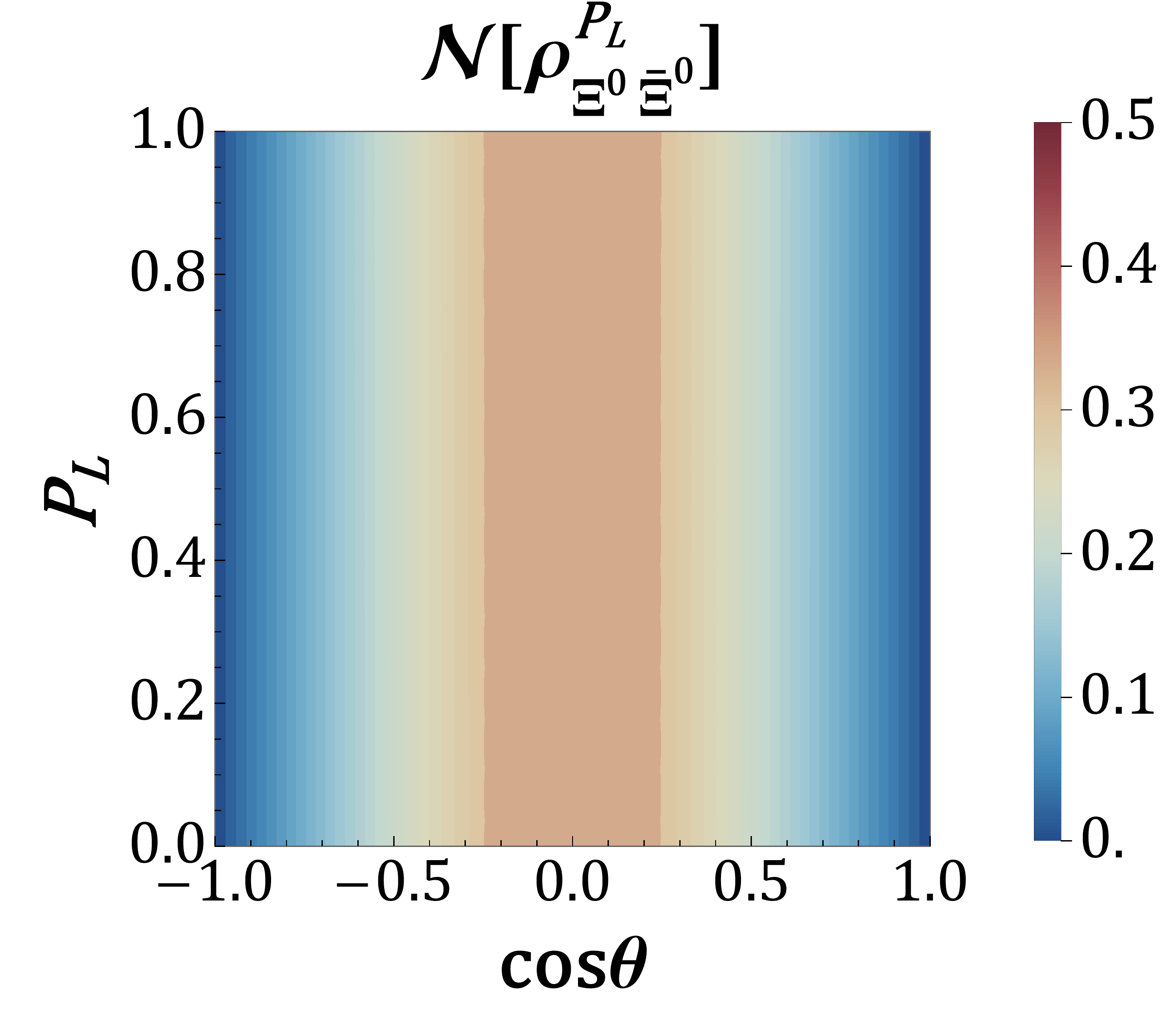}
      \includegraphics[width = 0.195 \linewidth]{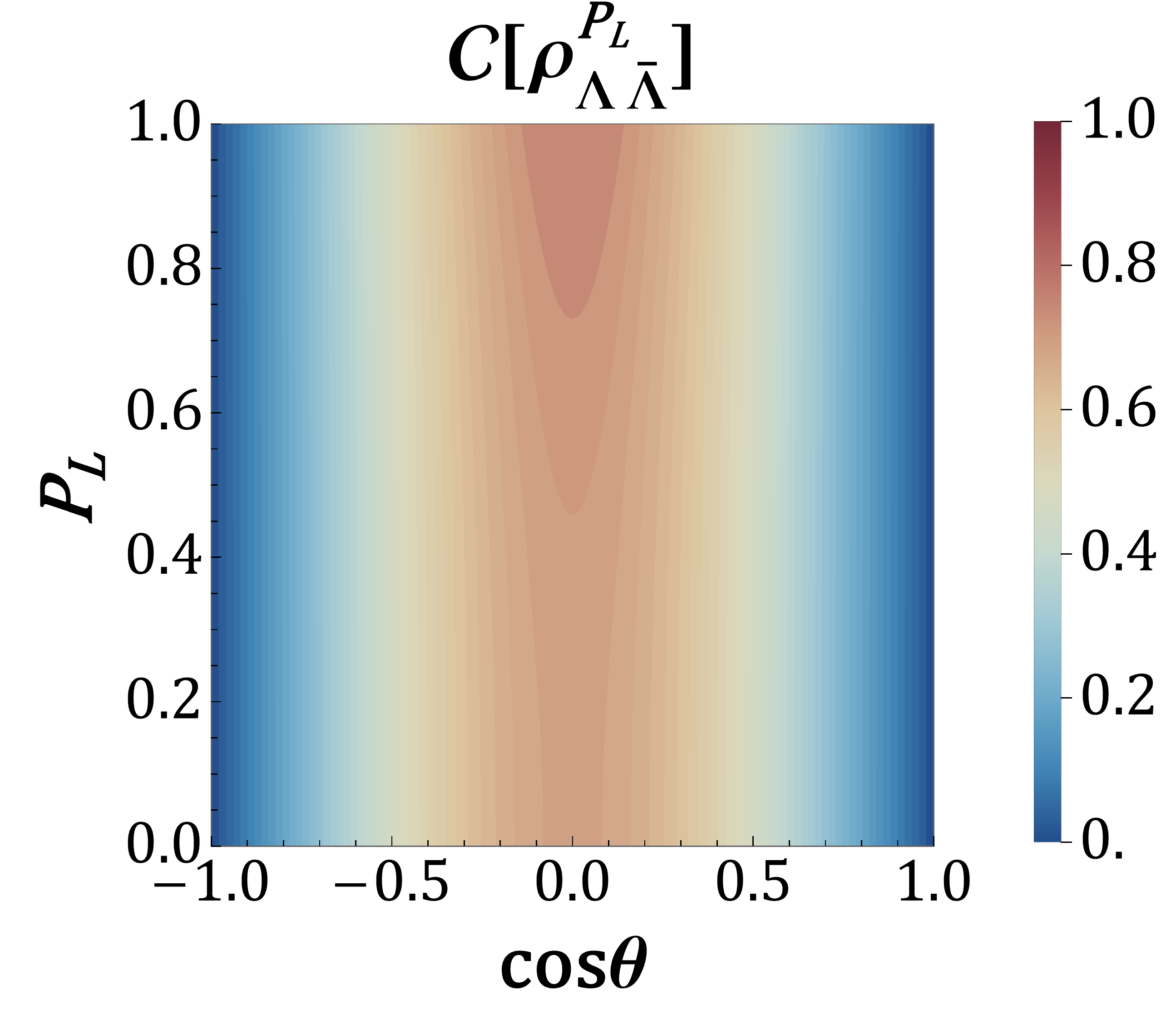}
  		\includegraphics[width = 0.195\linewidth]{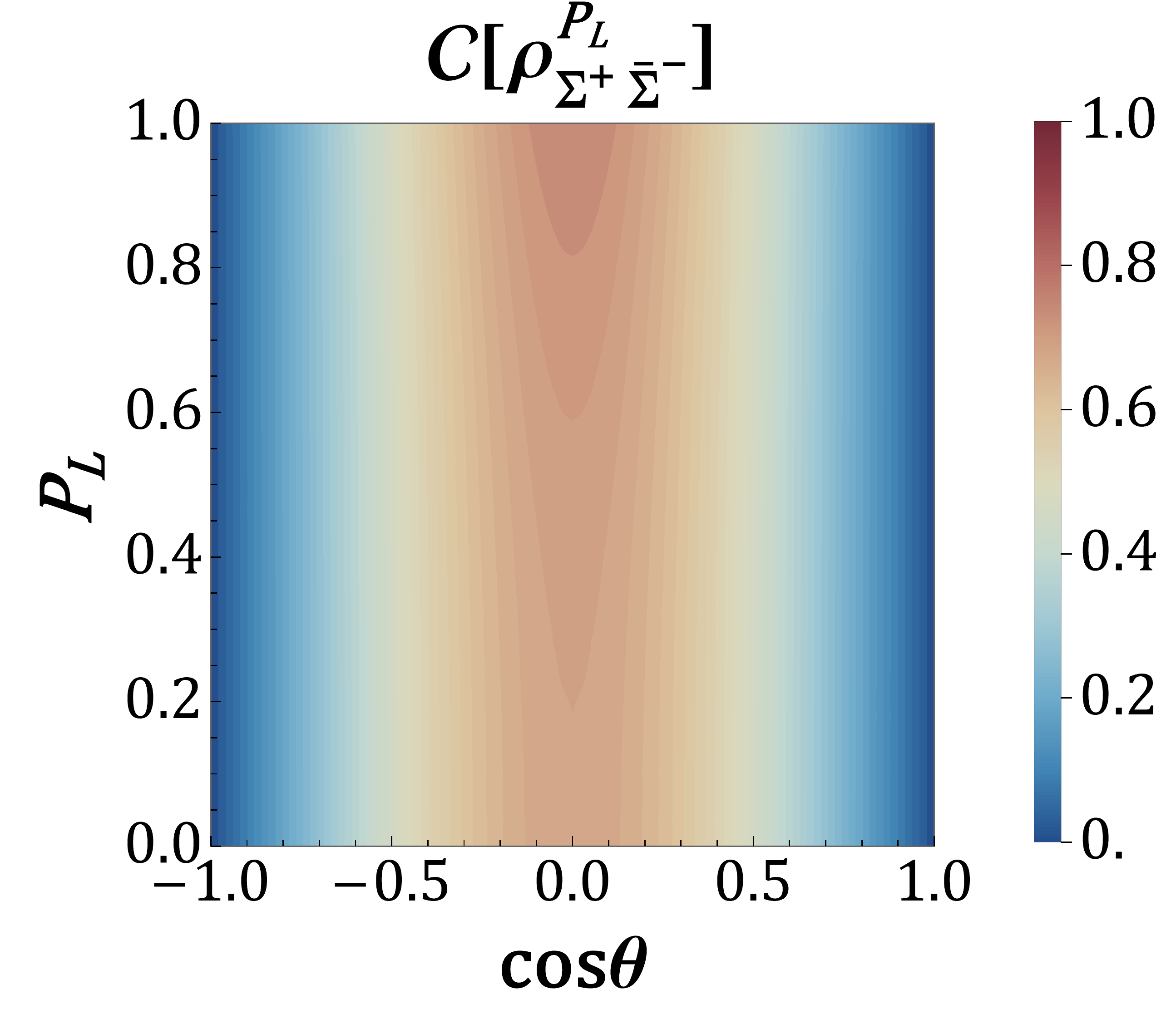}
      \includegraphics[width = 0.195\linewidth]{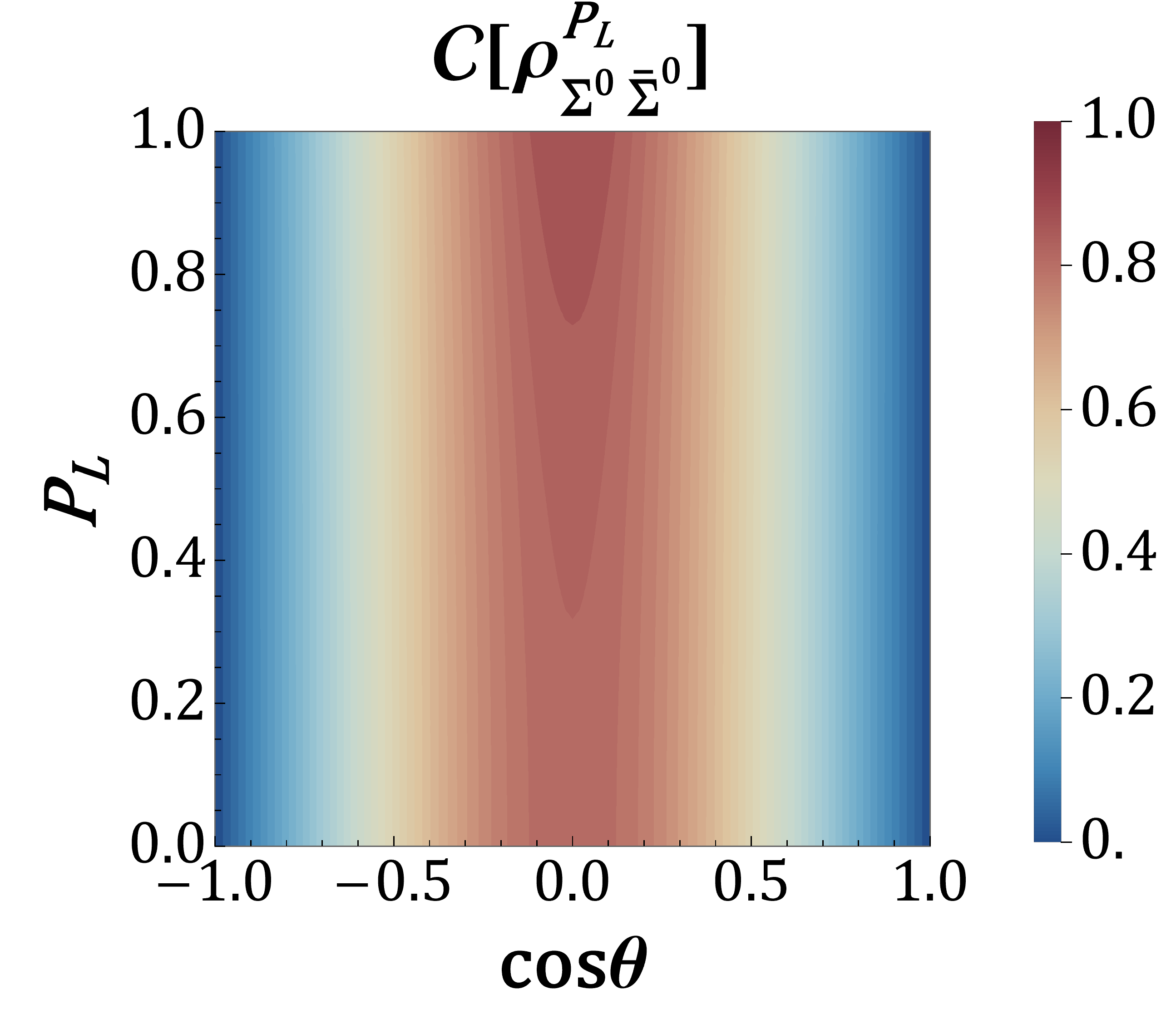}
      \includegraphics[width = 0.195 \linewidth]{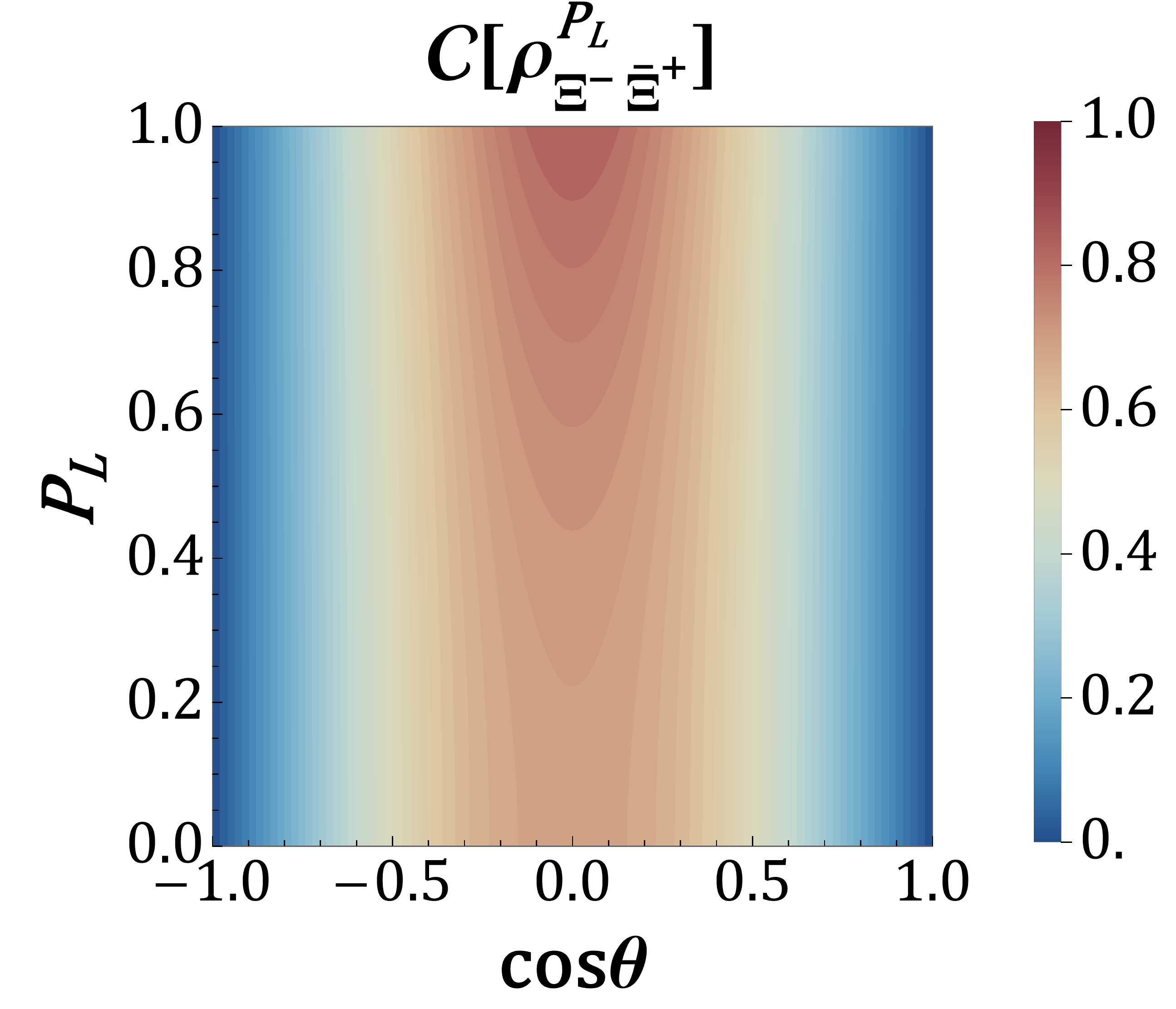}
  		\includegraphics[width = 0.195 \linewidth]{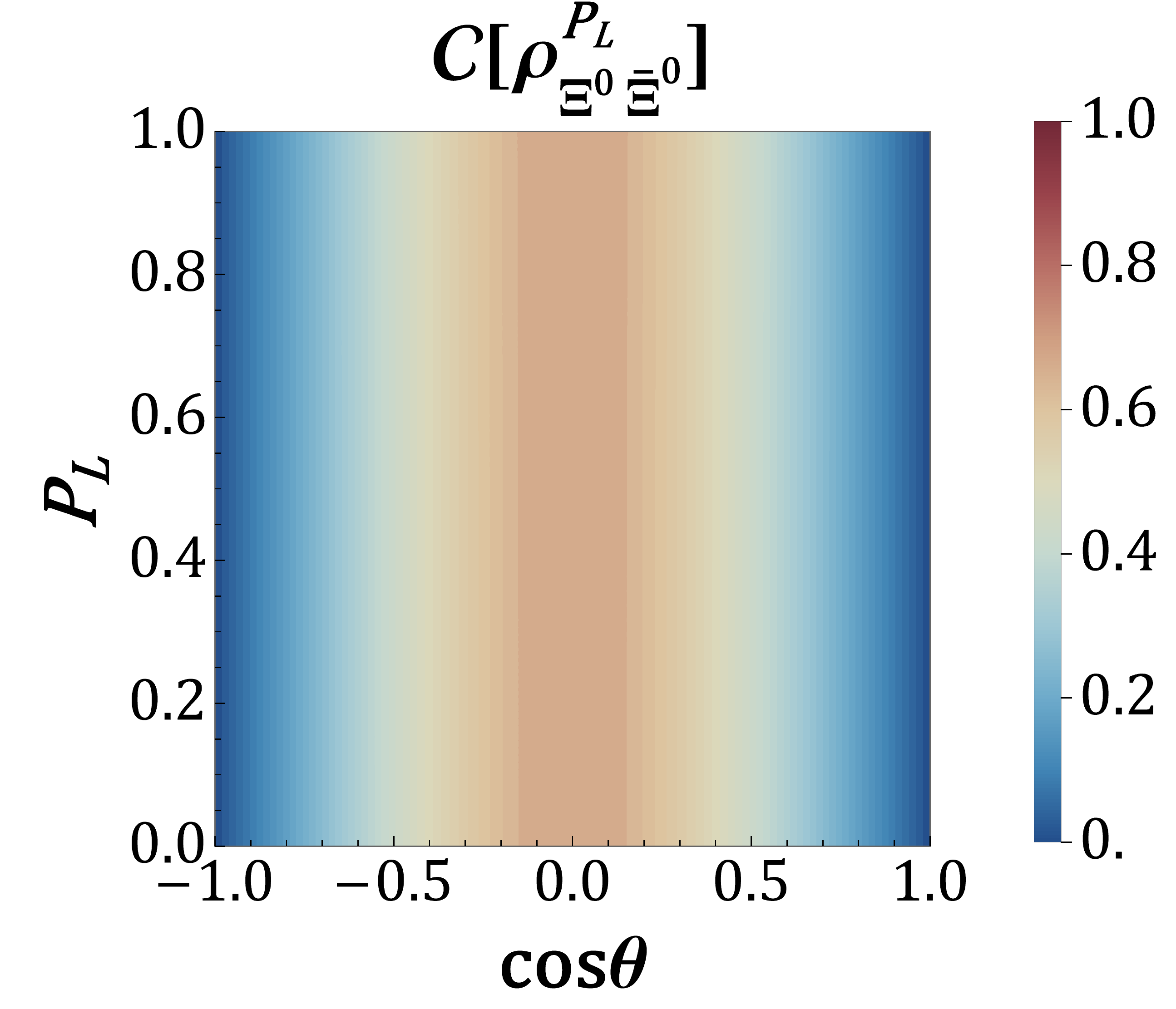}
   \caption{The CHSH parameter $\mathcal{B}[\rho^{P_L}_{Y \bar Y}]$ (upper panels), negativity $\mathcal{N}[\rho^{P_L}_{Y \bar Y}]$ (middle panels) and concurrence $\mathcal{C}[\rho^{P_L}_{Y \bar Y}]$ (lower panels) as a function of $P_L$ and $\theta$ in $\psi(3686) \rightarrow Y\bar{Y}$ for $Y = \Lambda$, $\Sigma^{+}$, $\Sigma^{0}$, $\Xi^{-}$ and $\Xi^{0}$. The dashed line in each of upper panels is $\mathcal{B}=2$.\label{fig:entanglepsi}
}
\end{figure*}

\begin{figure*}[h]
      \includegraphics[width = 0.325 \linewidth]{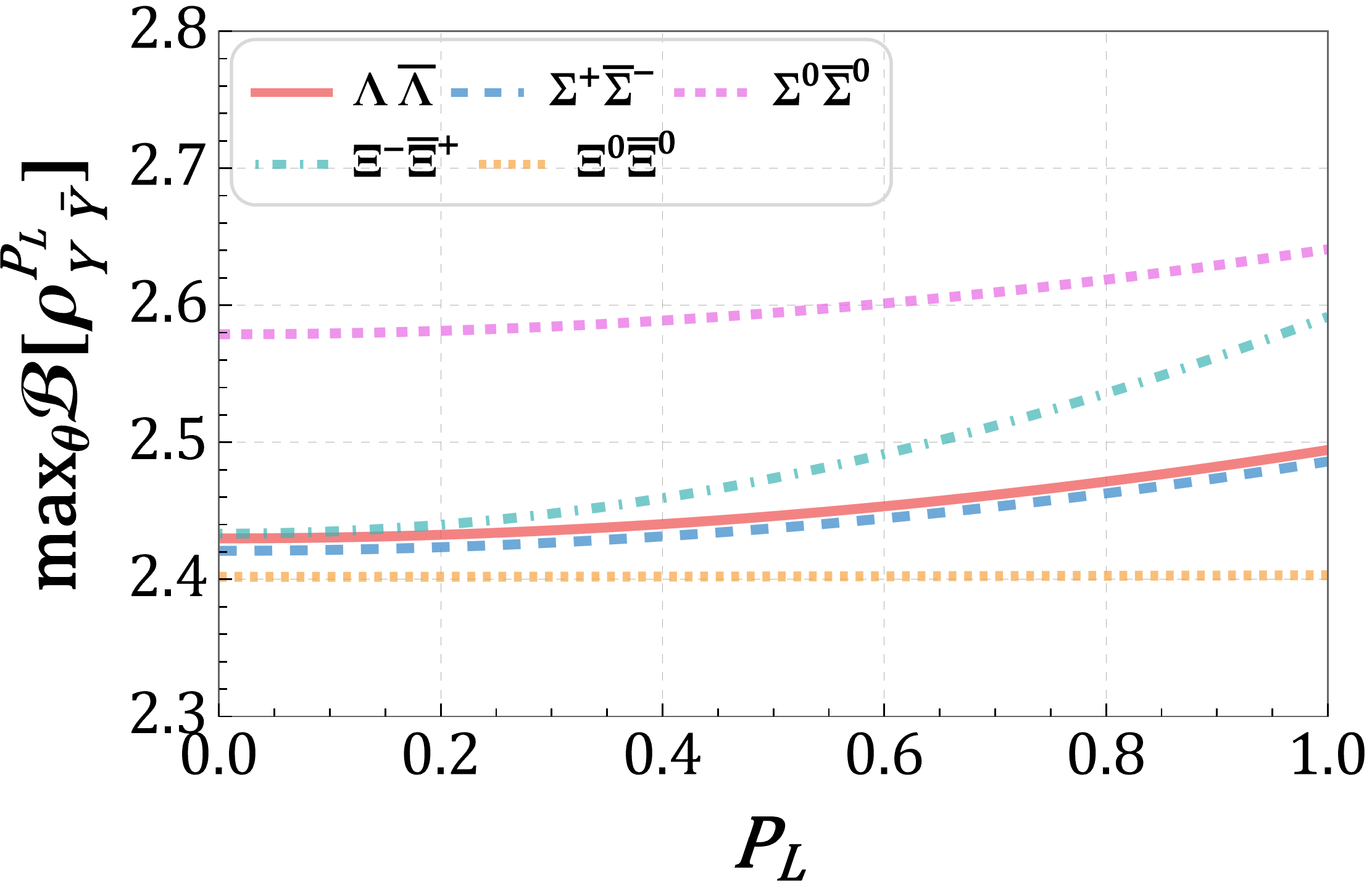}
  		\includegraphics[width = 0.325 \linewidth]{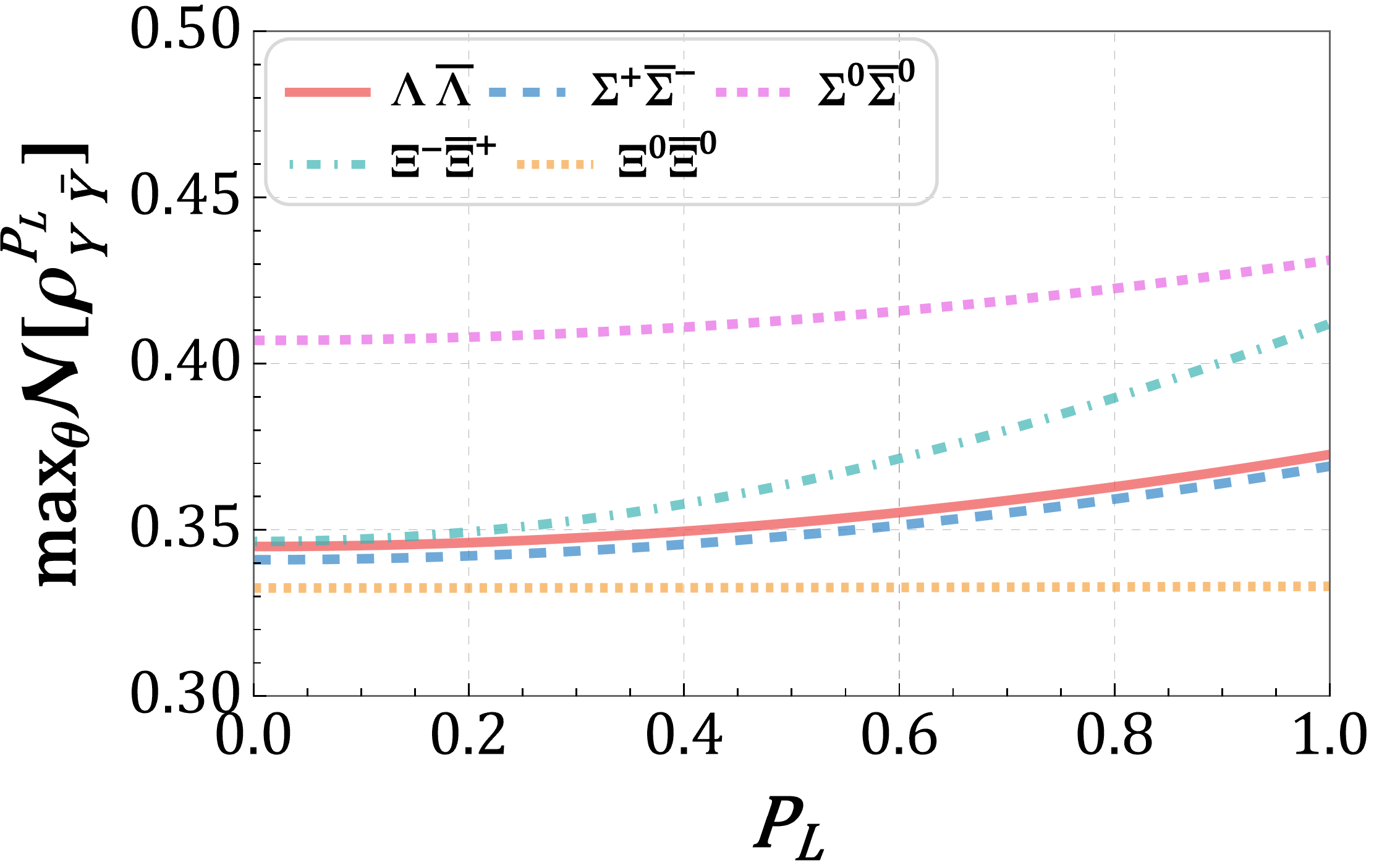}
      \includegraphics[width = 0.325 \linewidth]{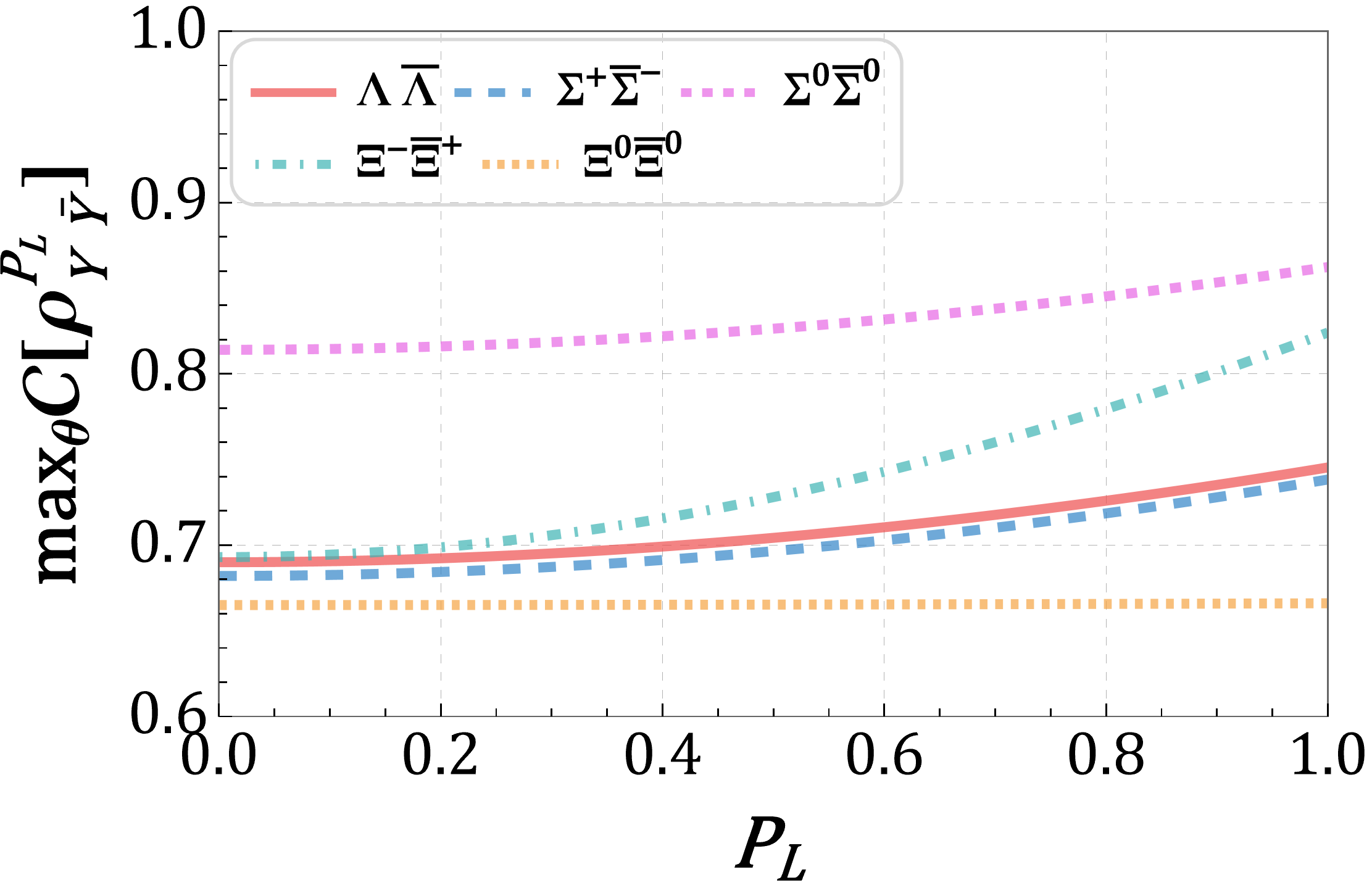}
    \caption{The $\max_{\theta}\mathcal{B}[\rho_{L}]$ (left), negativity $\max_{\theta}\mathcal{N}[\rho_{L}]$ (middle), and concurrence $\max_{\theta}\mathcal{C}[\rho_{L}]$ (right) as a function $P_L$ and $\theta$ in $\psi(3686) \rightarrow Y\bar{Y}$ for $Y = \Lambda$, $\Sigma^{+}$, $\Sigma^{0}$, $\Xi^{-}$ and $\Xi^{0}$. \label{fig:largestpsiPL}
}
\end{figure*}

\begin{figure*}[h]
  		\includegraphics[width = 0.195 \linewidth]{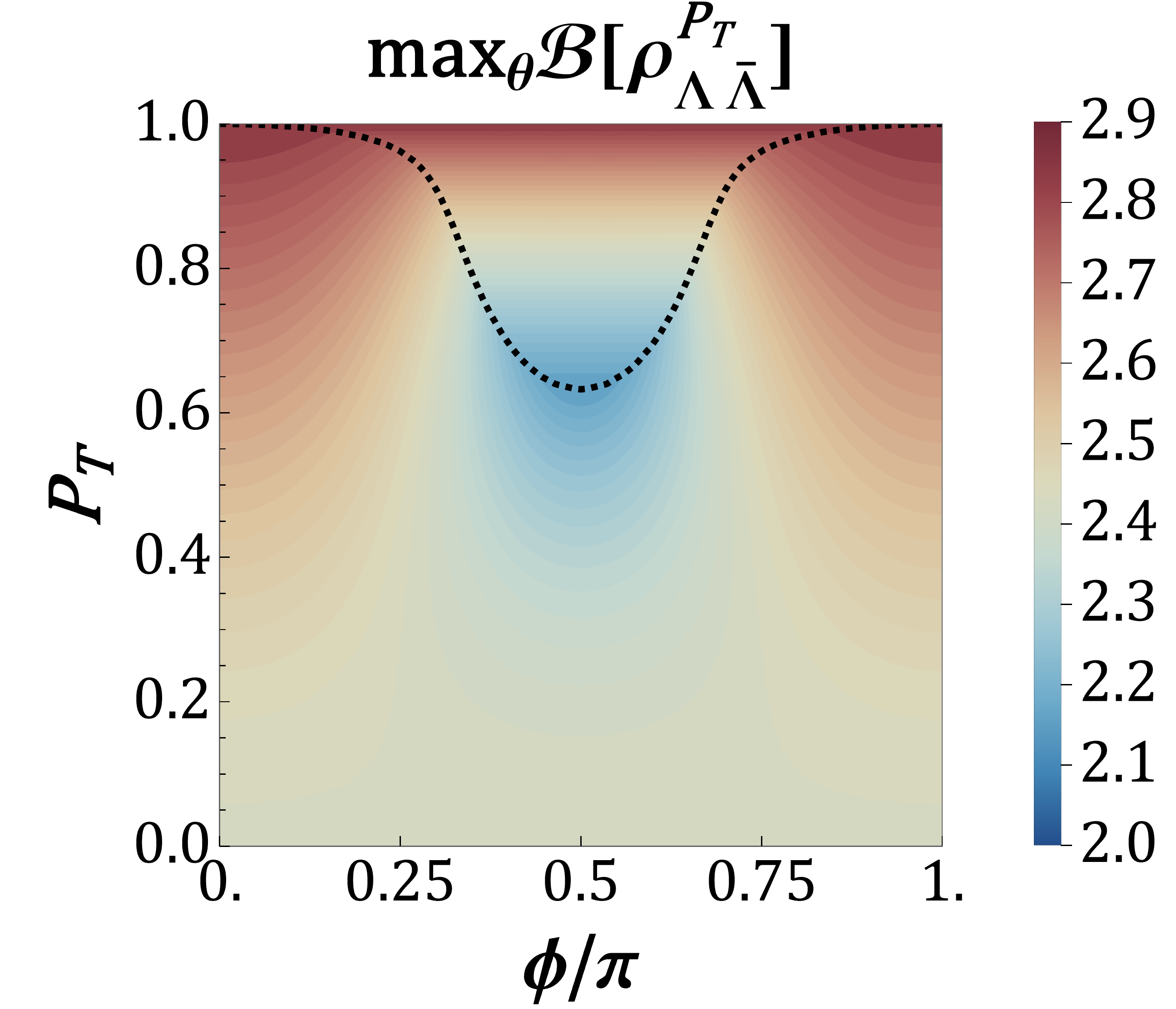}
      \includegraphics[width = 0.195 \linewidth]{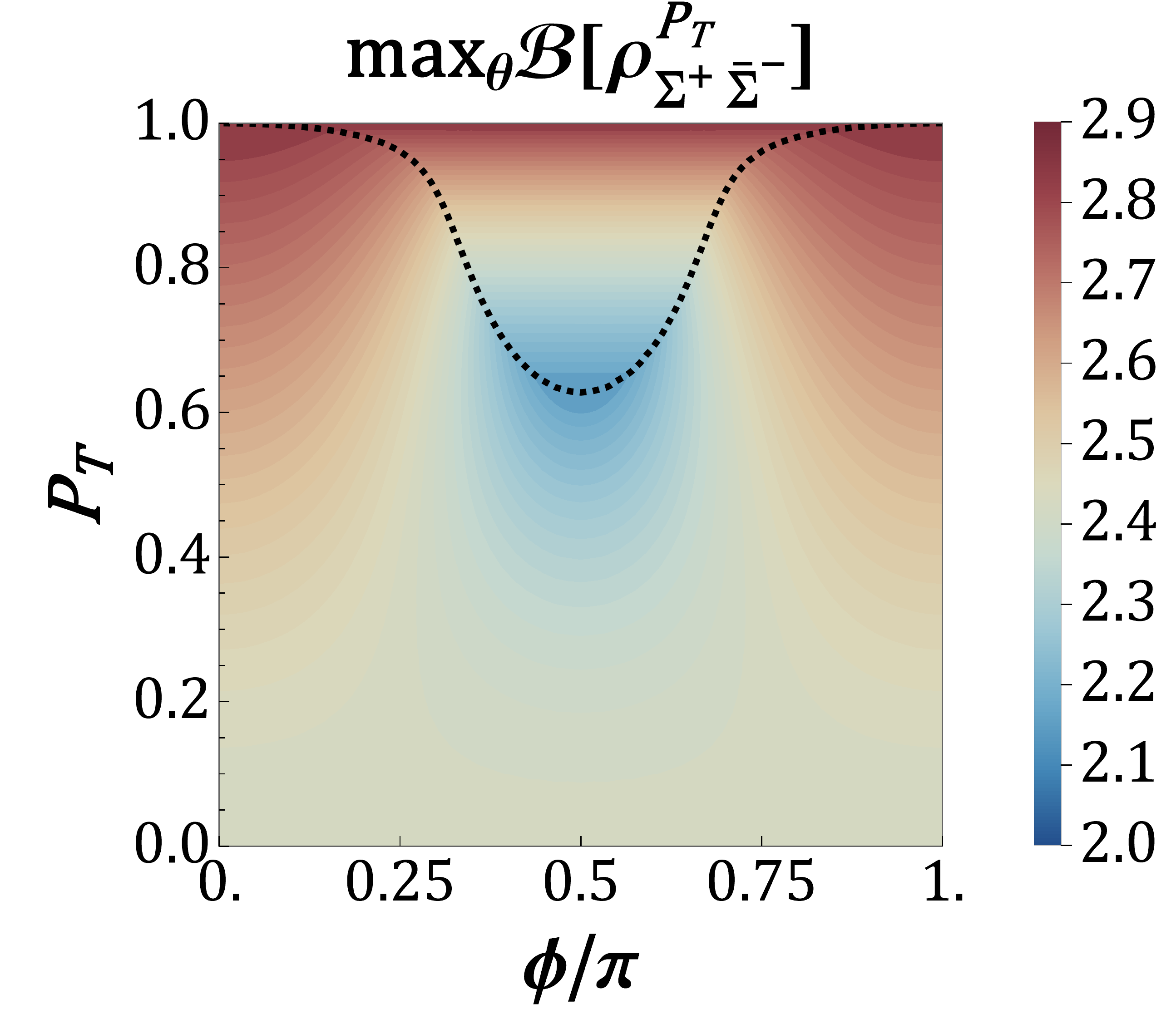}
      \includegraphics[width = 0.195 \linewidth]{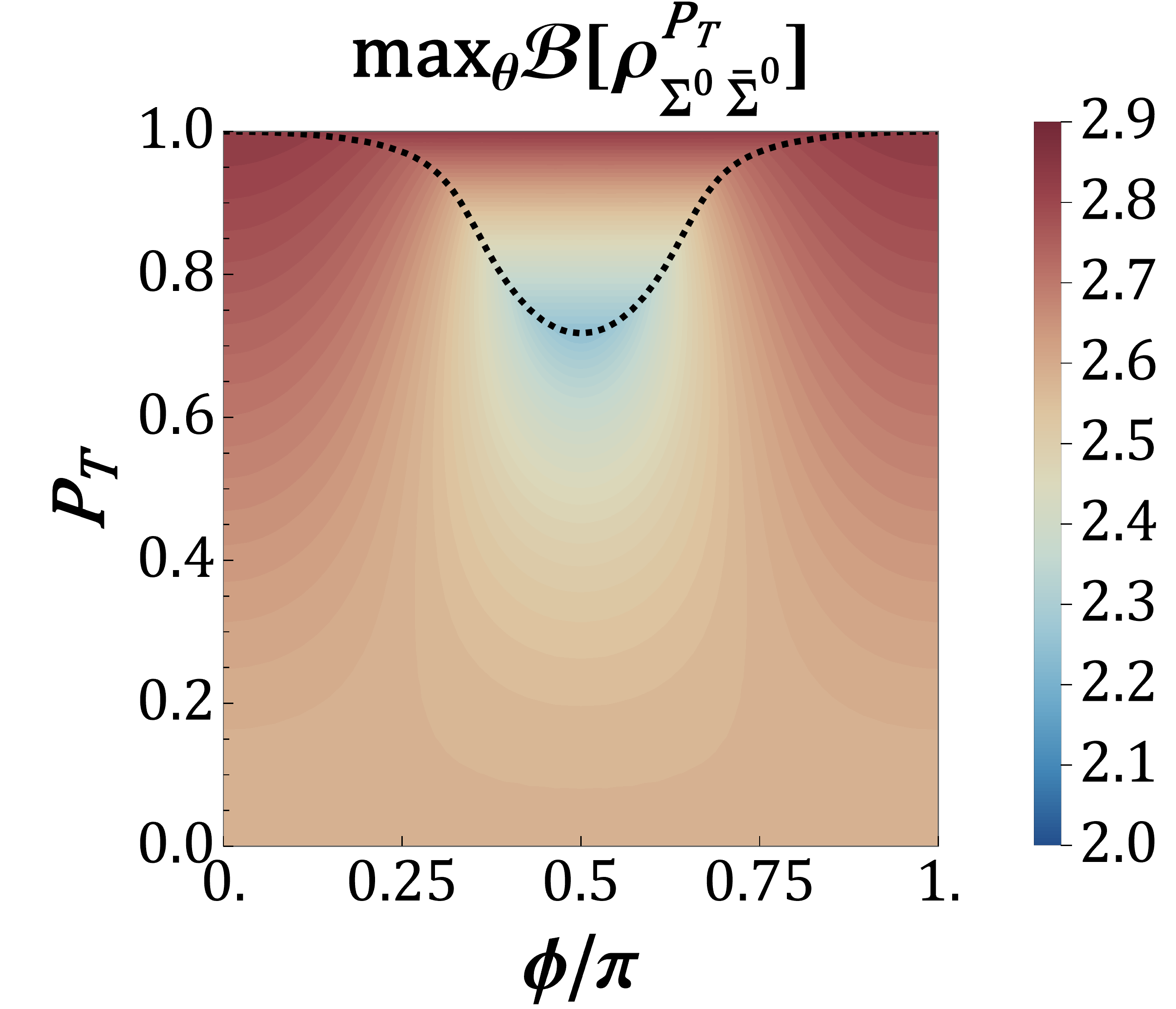}
      \includegraphics[width = 0.195 \linewidth]{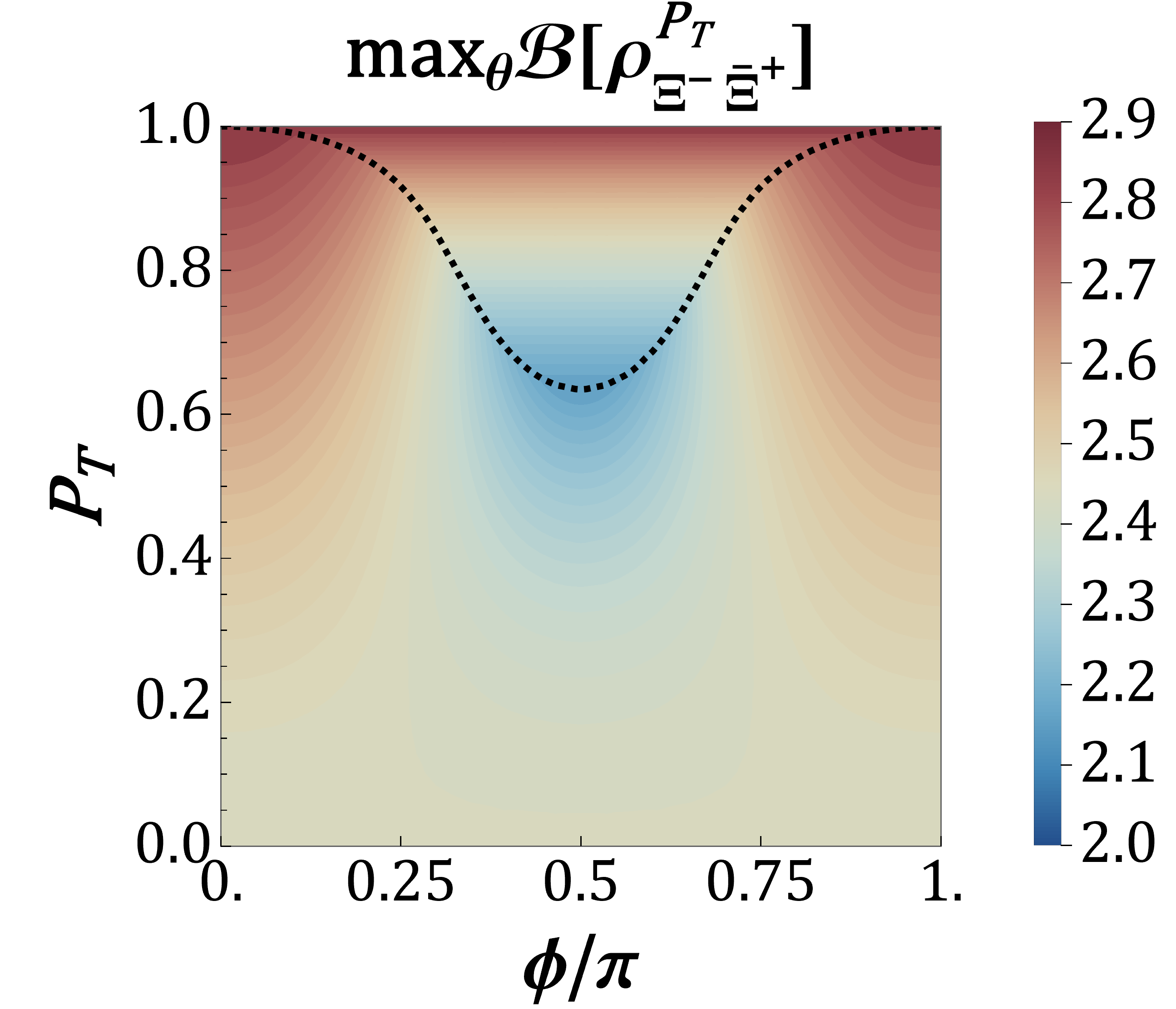}
      \includegraphics[width = 0.195 \linewidth]{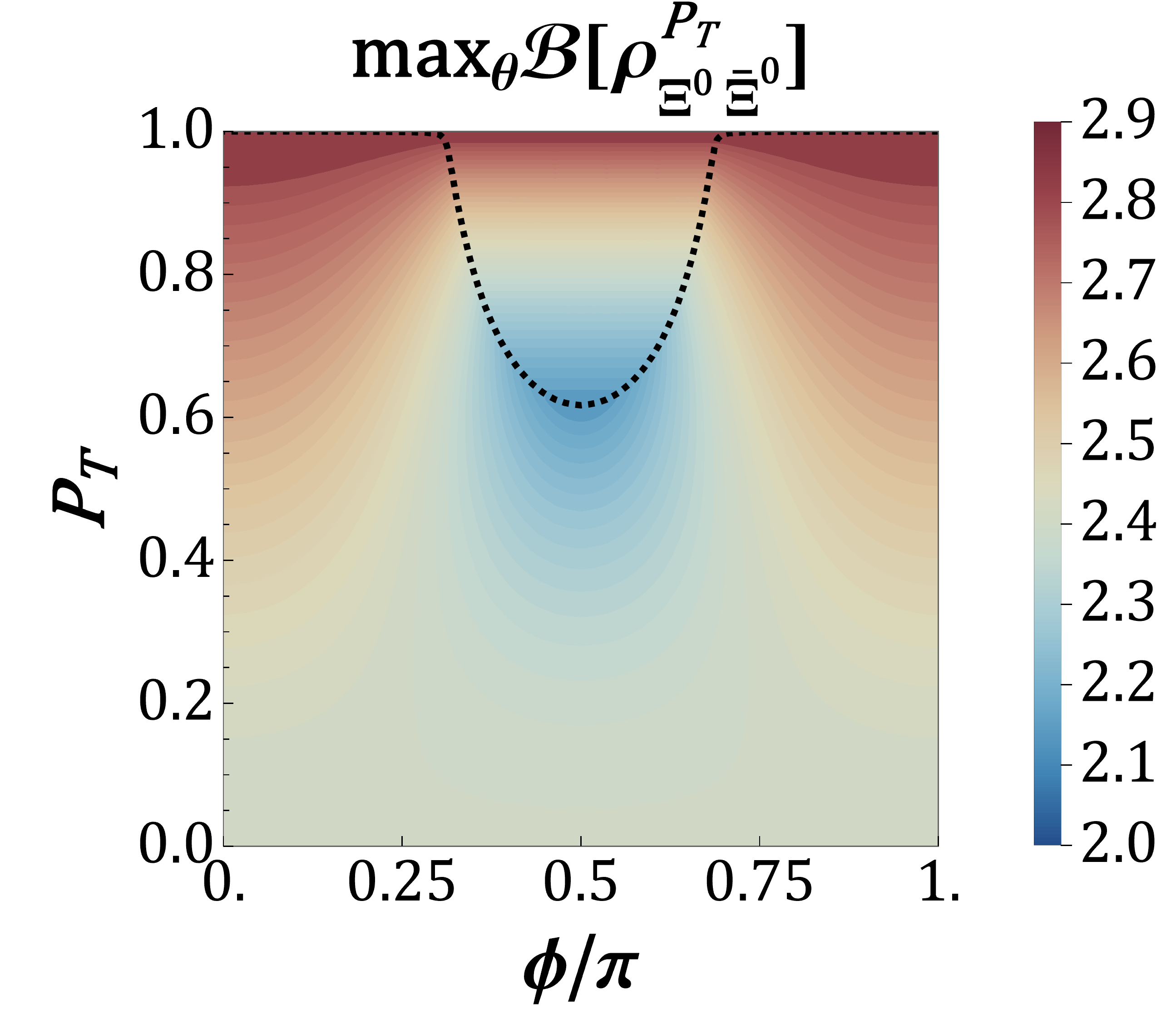}
      \includegraphics[width = 0.195 \linewidth]{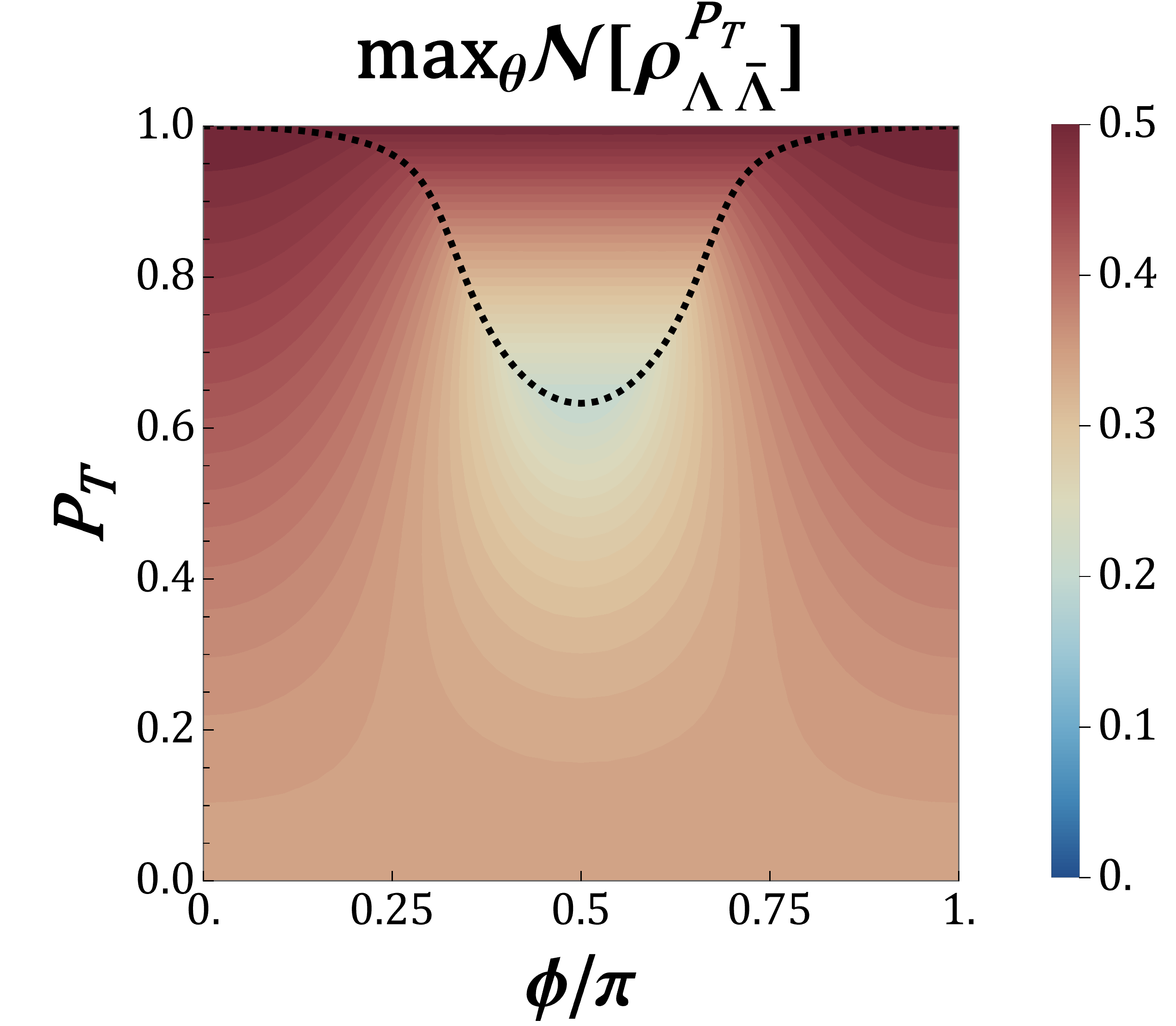}
  		\includegraphics[width = 0.195 \linewidth]{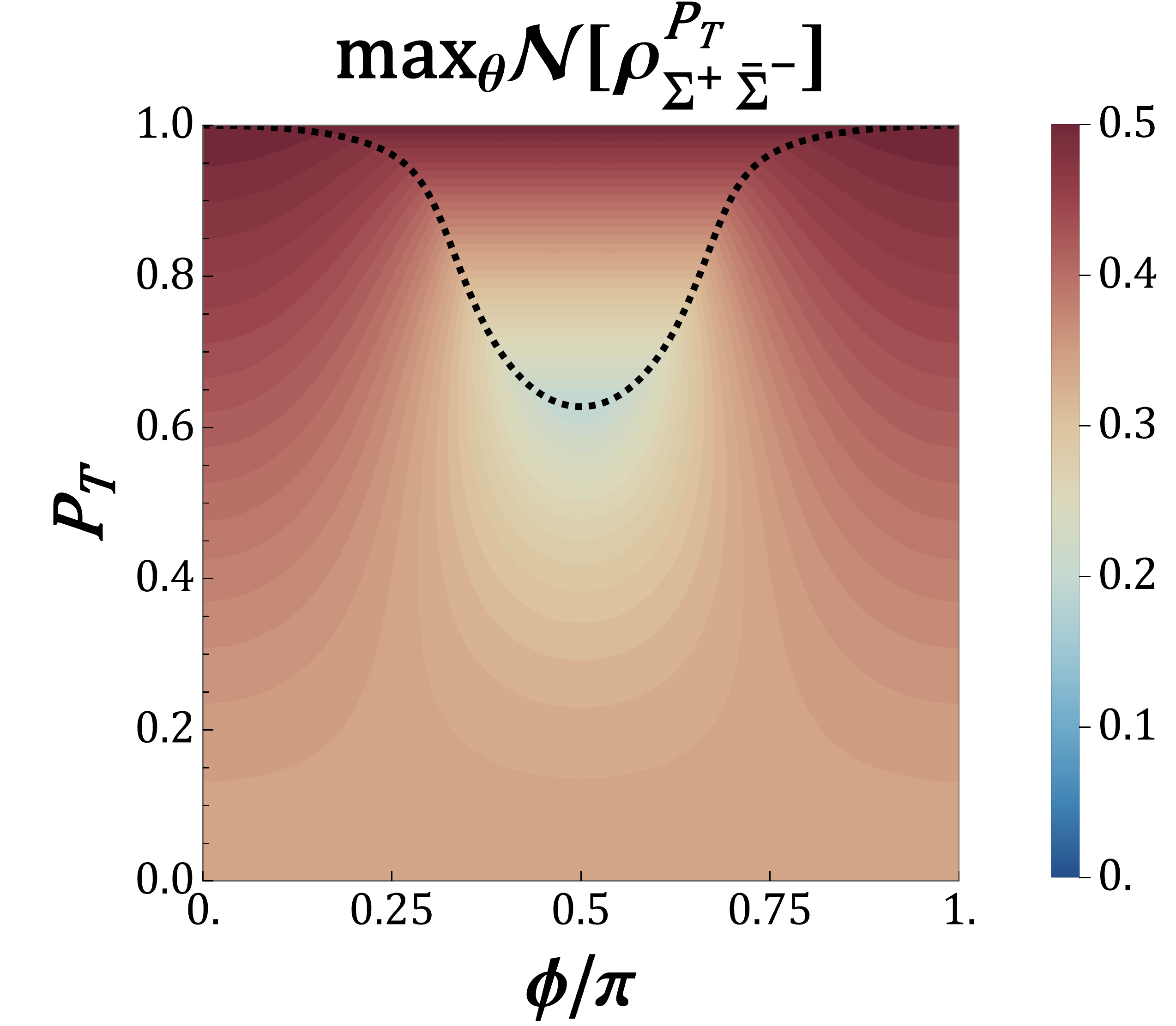}
      \includegraphics[width = 0.195 \linewidth]{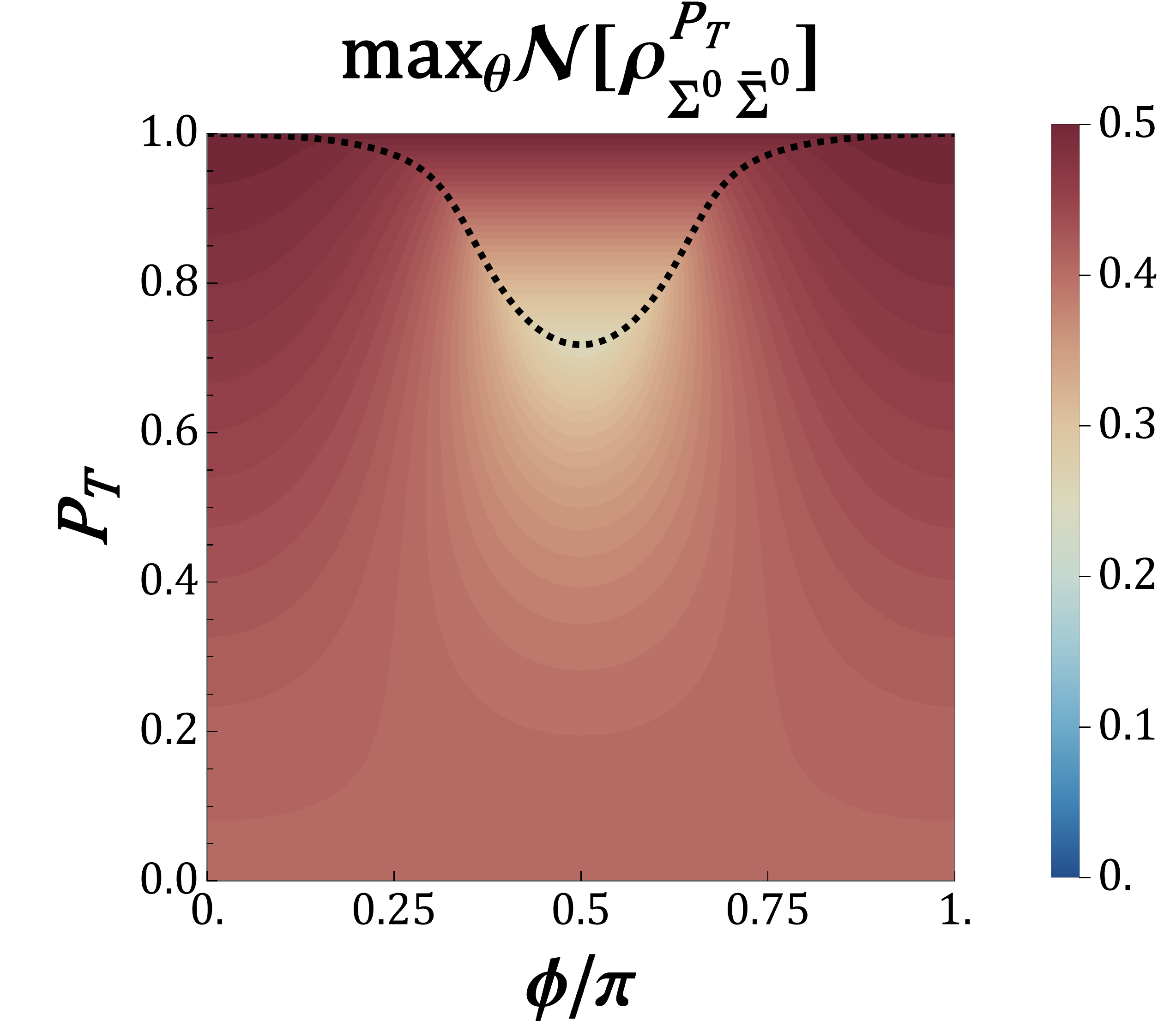}
      \includegraphics[width = 0.195 \linewidth]{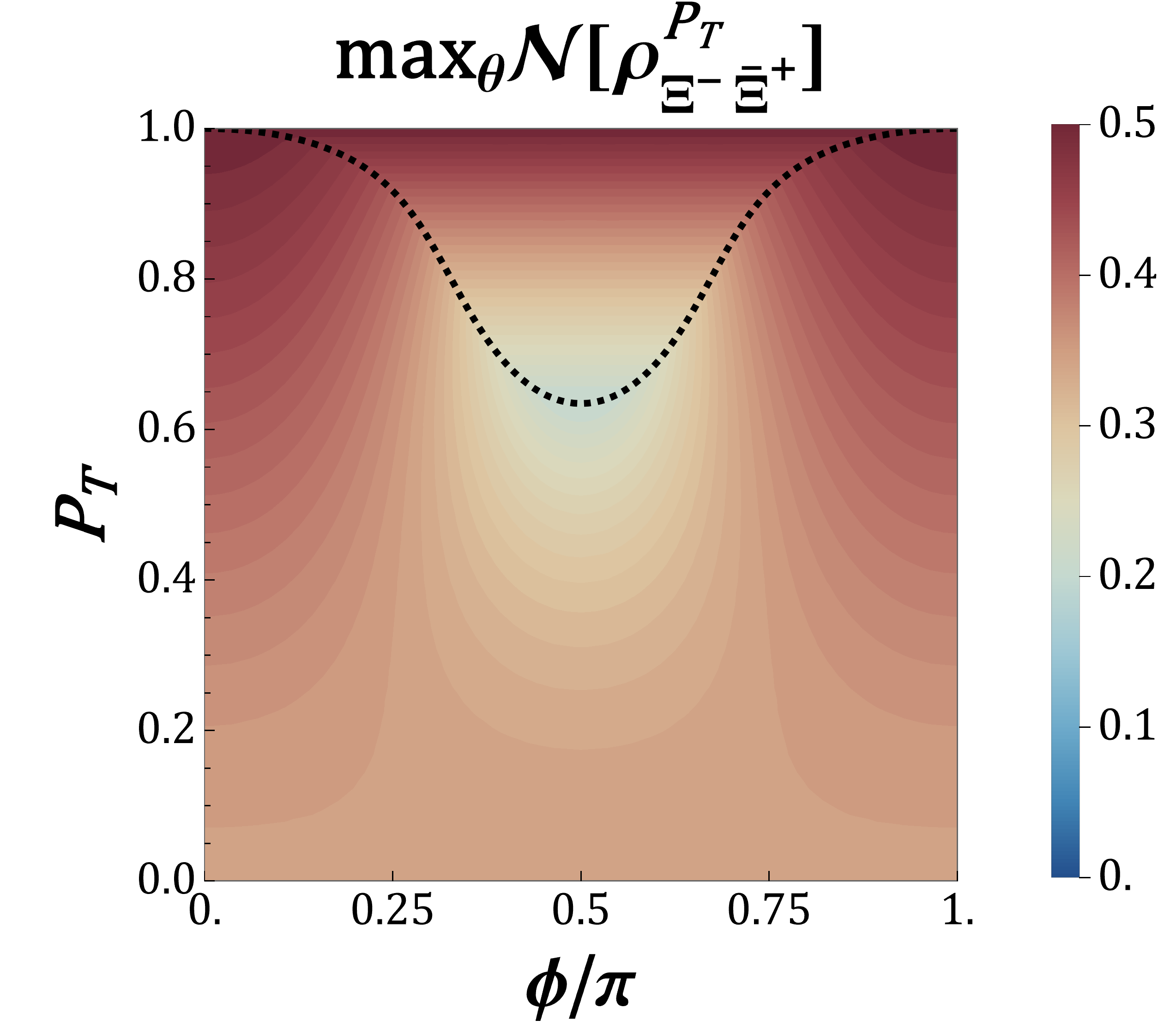}
  		\includegraphics[width = 0.195 \linewidth]{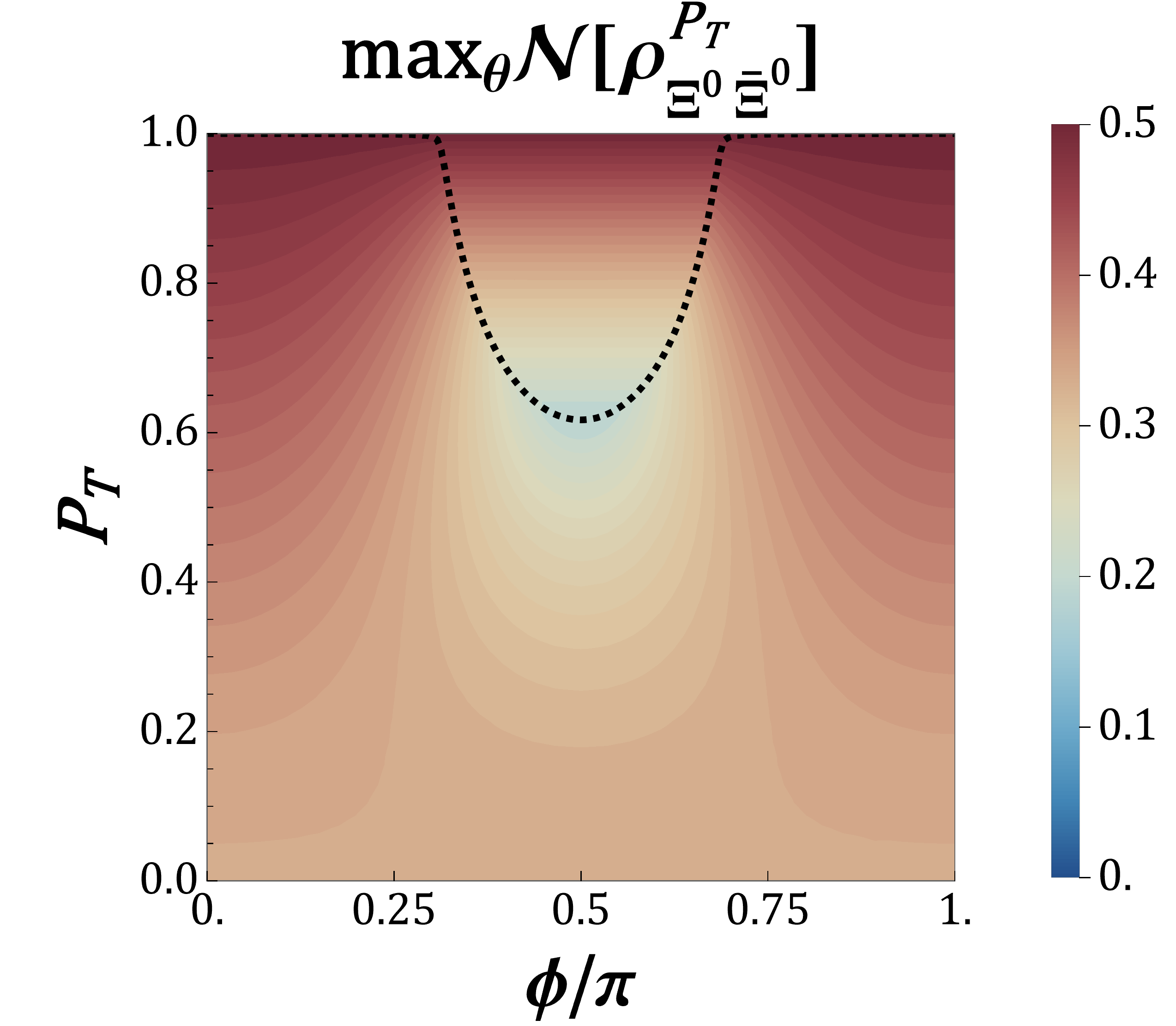}
  		\includegraphics[width = 0.195 \linewidth]{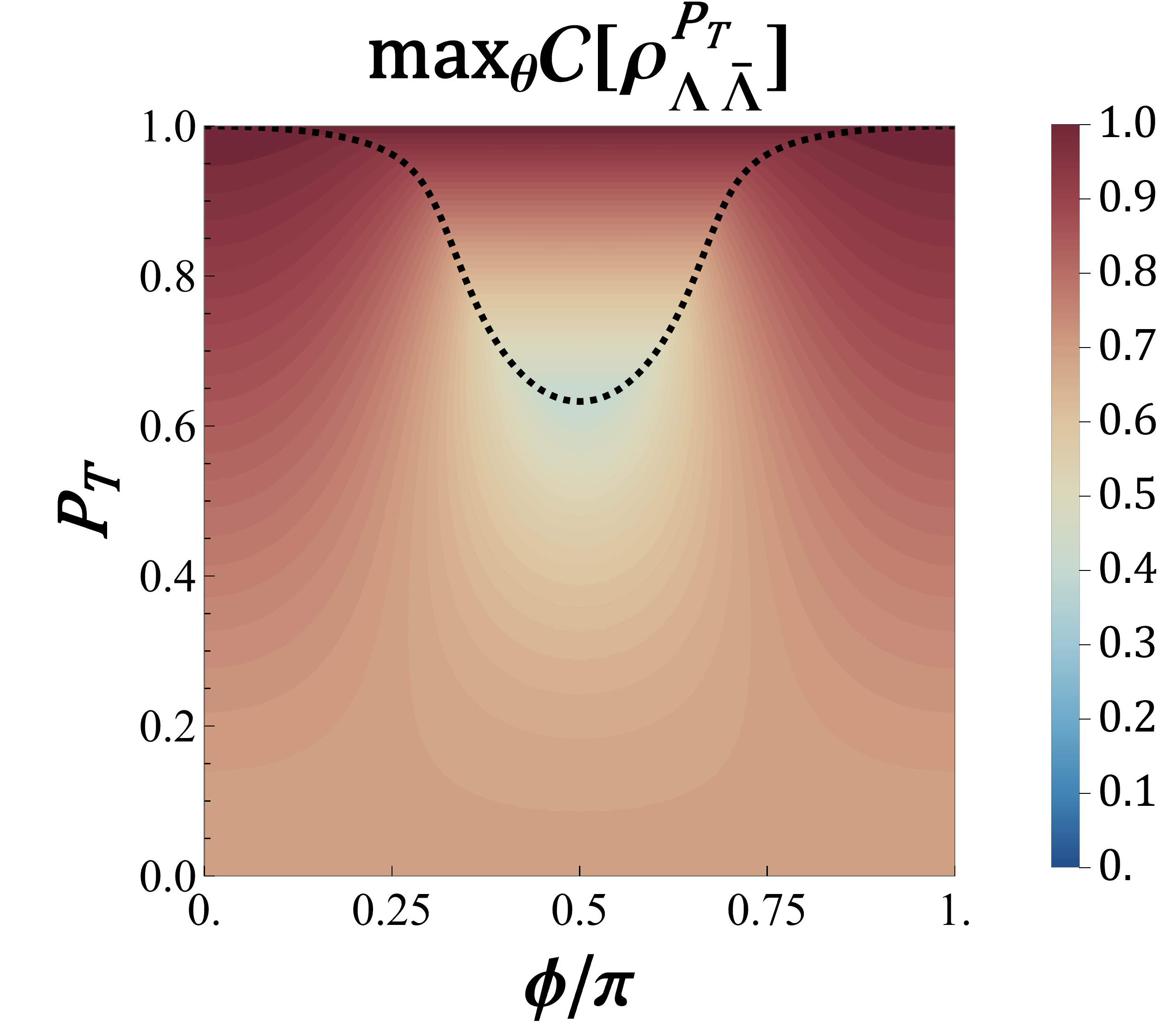}
      \includegraphics[width = 0.195 \linewidth]{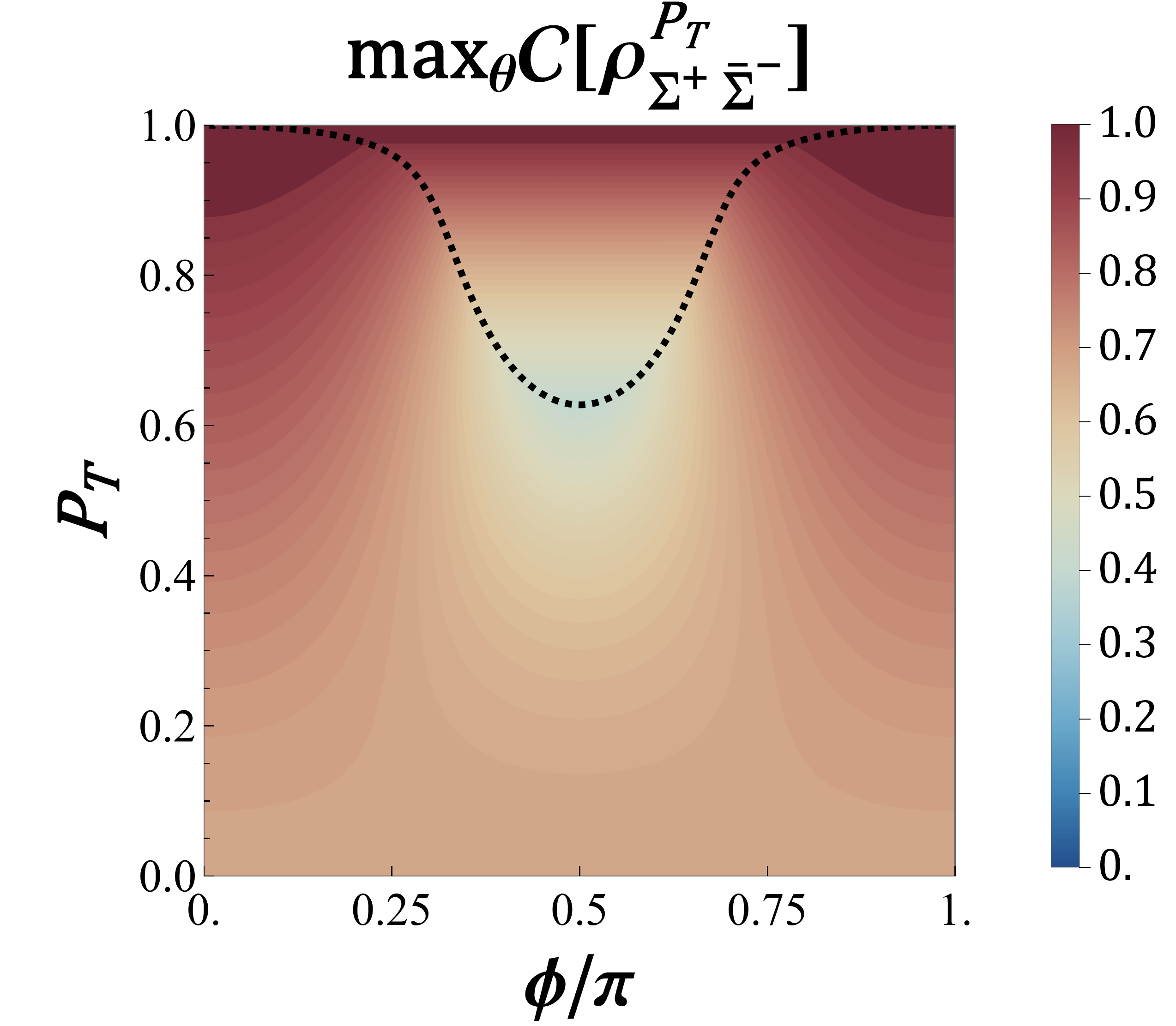}
      \includegraphics[width = 0.195 \linewidth]{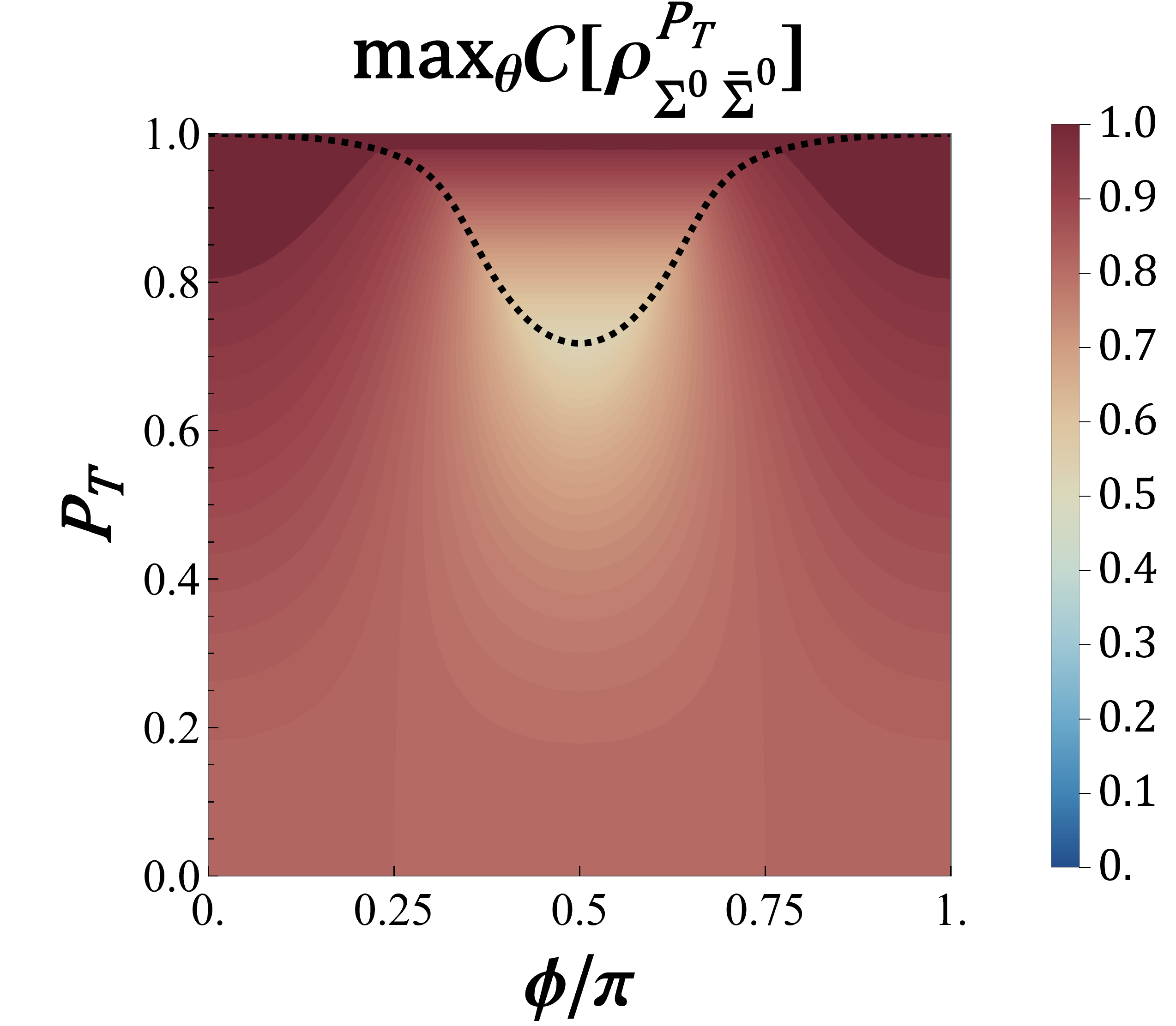}
      \includegraphics[width = 0.195 \linewidth]{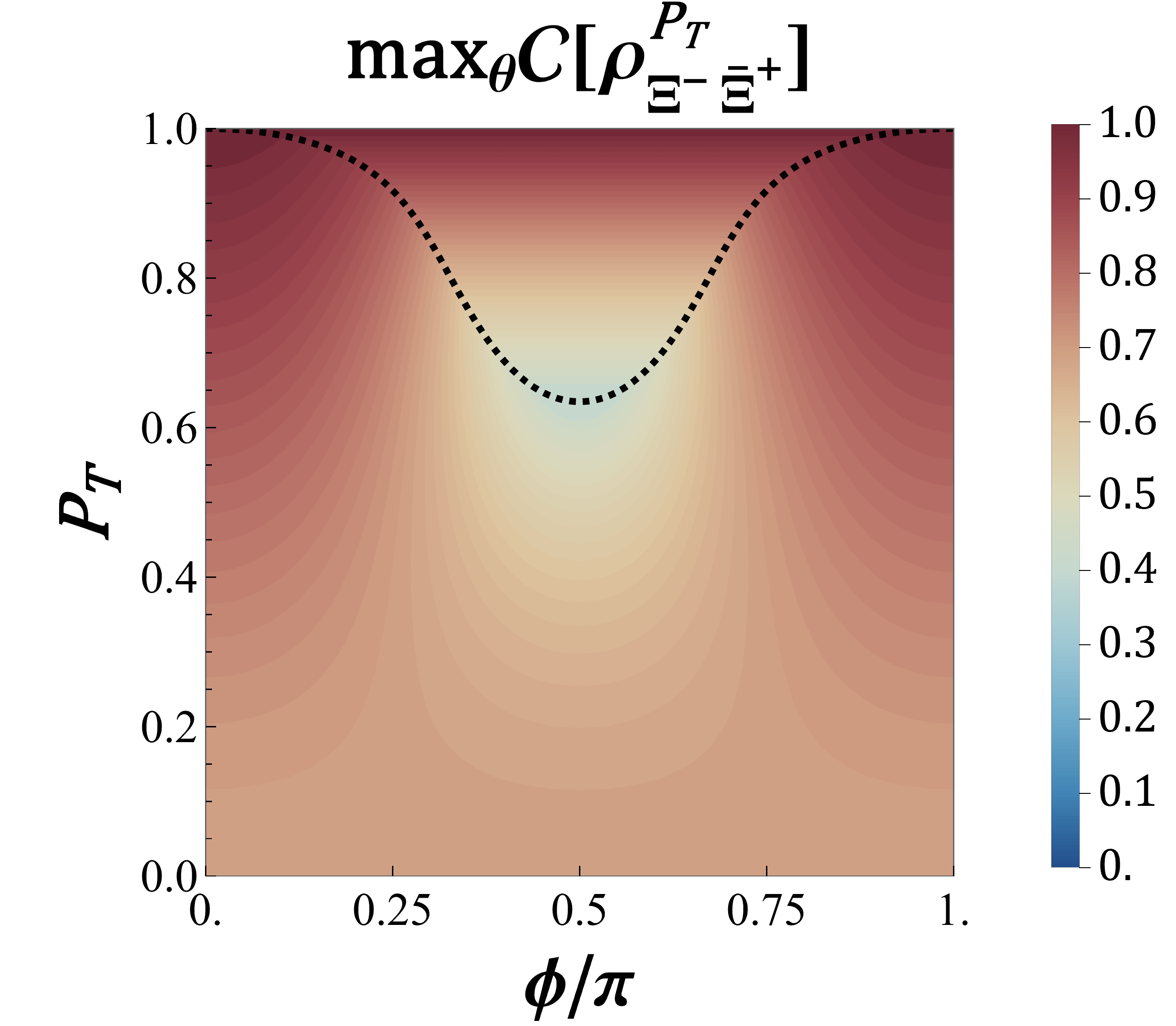}
      \includegraphics[width = 0.195 \linewidth]{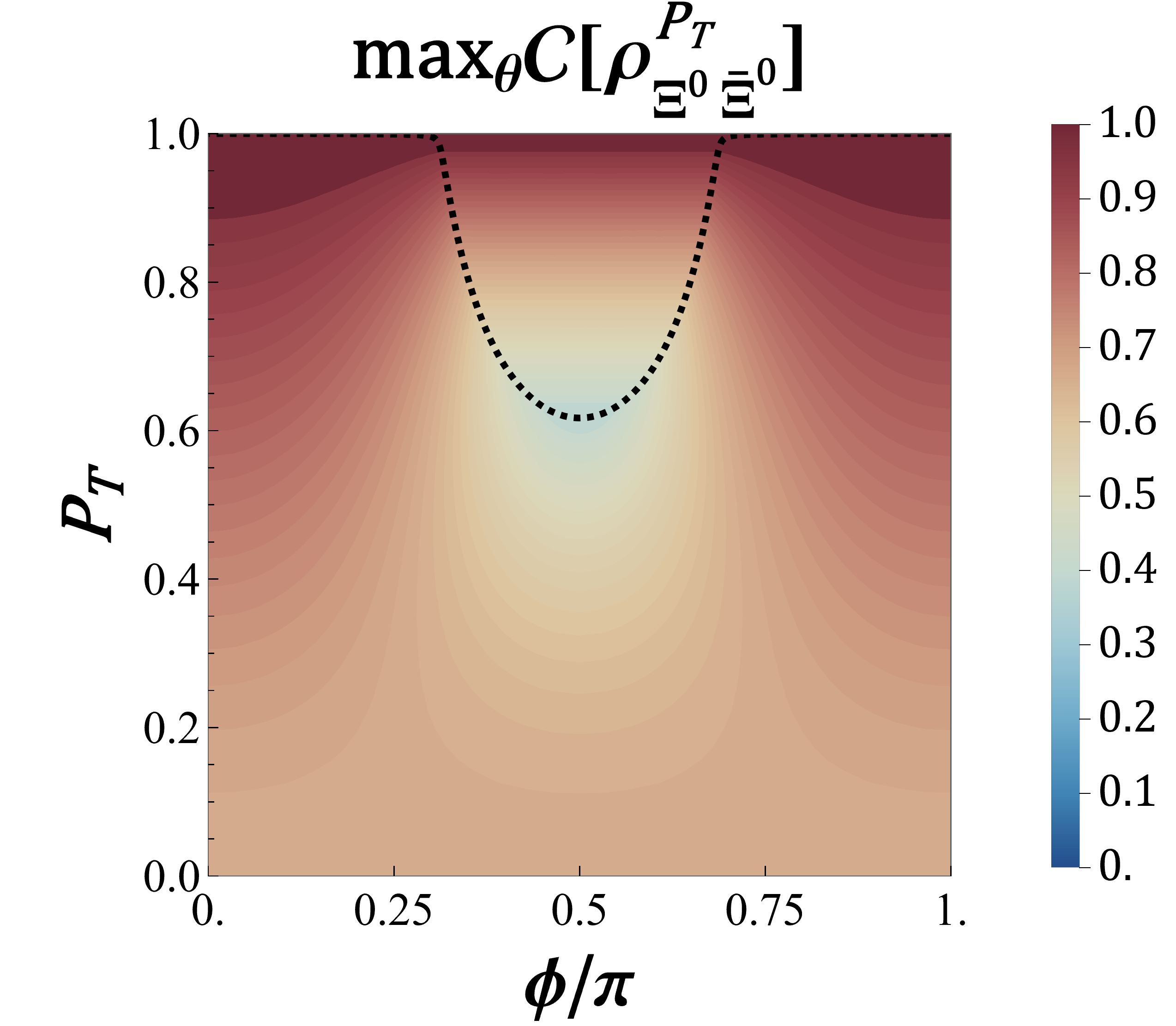}
\caption{
The $\max_{\theta}\mathcal{C}[\rho_{T}]$ (upper panels), $\max_{\theta}\mathcal{N}[\rho^{P_L}_{Y \bar Y}]$ (middle panels), and $\max_{\theta}\mathcal{N}[\rho^{P_T}_{Y \bar Y}]$ (lower panels) as a function of azimuthal angle $\phi$ and transverse polarization degree $P_T$ in $\psi(3686) \rightarrow Y\bar{Y}$ for $Y = \Lambda$, $\Sigma^{+}$, $\Sigma^{0}$, $\Xi^{-}$ and $\Xi^{0}$. The dashed curve is $\mathcal{B,N,C}(\theta=0,\pi)=\mathcal{B,N,C}(\theta=\pi/2)$ in each of upper, middle and lower panels respectively.\label{fig:transpsi}
}
\end{figure*}

\begin{figure*}[h]
      \includegraphics[width = 0.325 \linewidth]{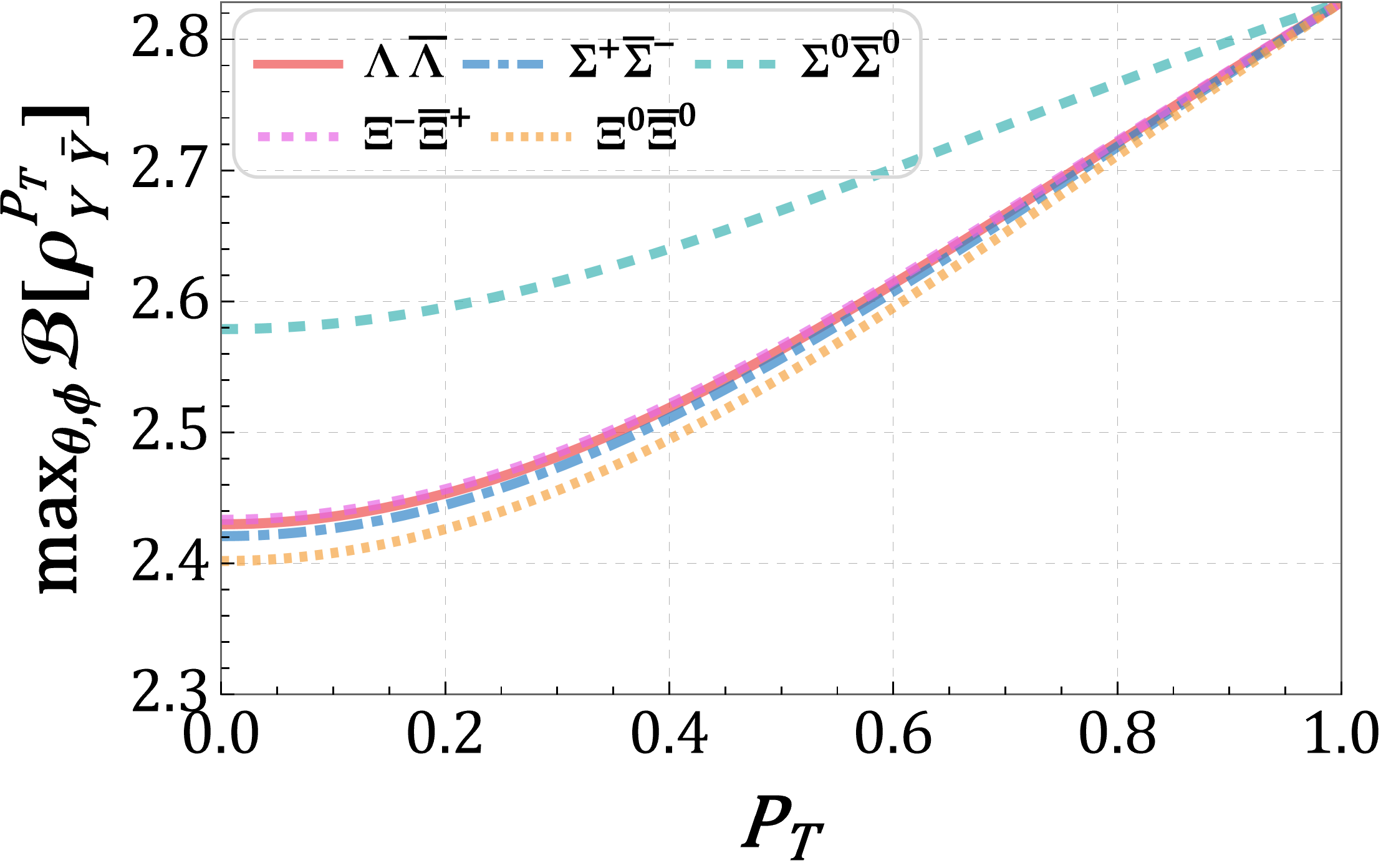}
  		\includegraphics[width = 0.325 \linewidth]{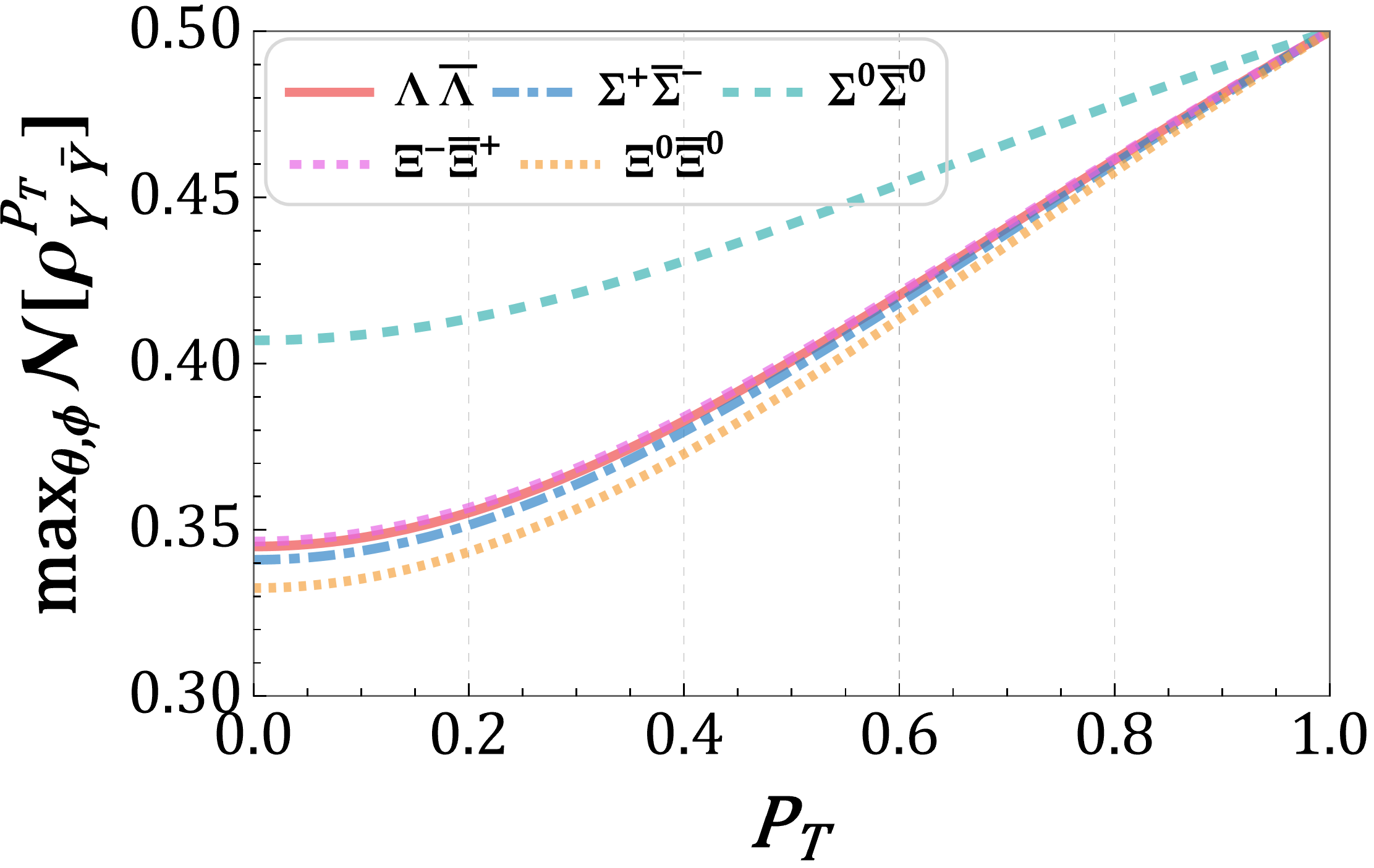}
      \includegraphics[width = 0.325 \linewidth]{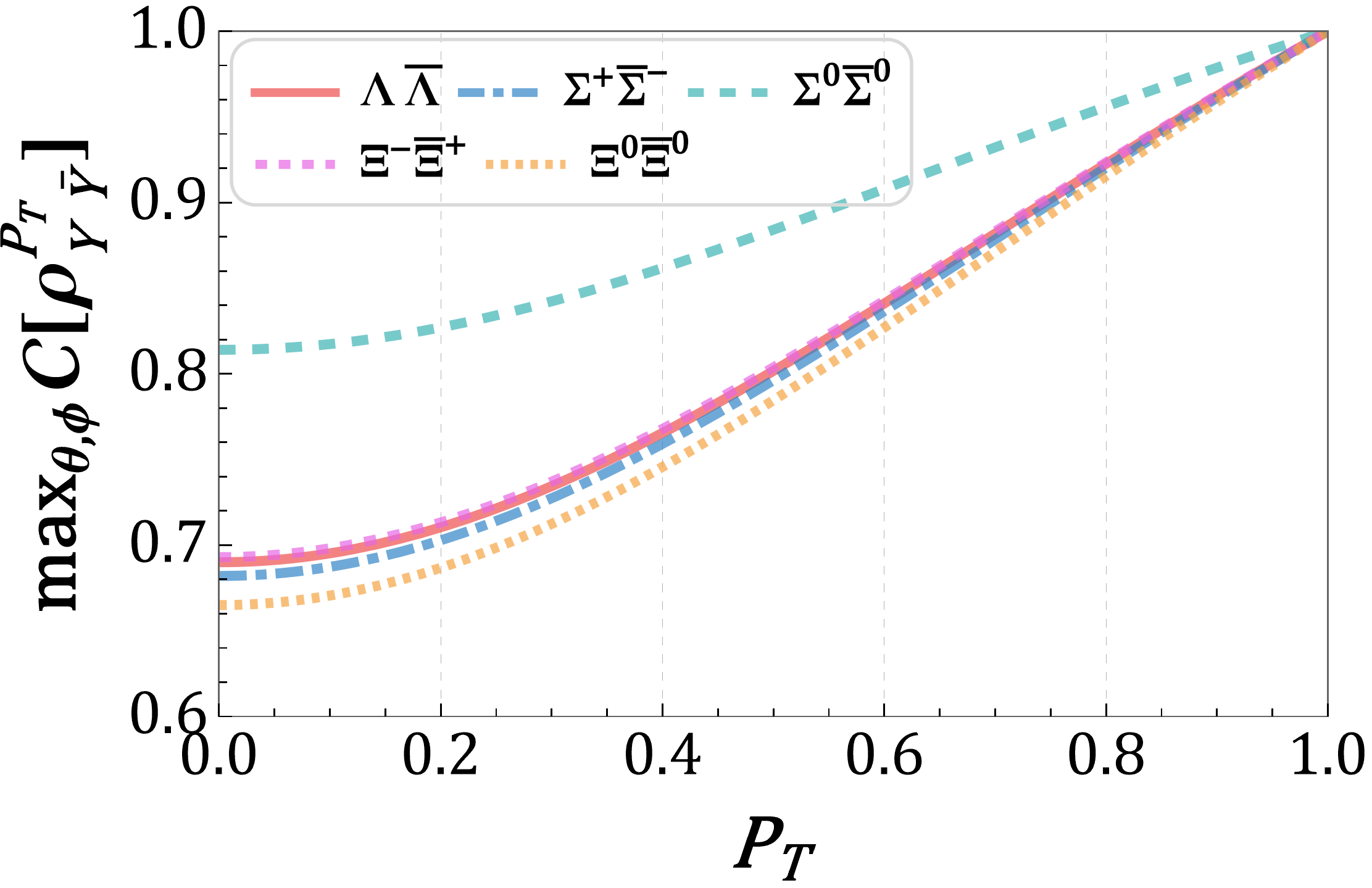}
    \caption{The $\max_{\theta,\phi}\mathcal{B}[\rho_{T}]$ (left), $\max_{\theta,\phi}\mathcal{N}[\rho_{T}]$ (middle), and $\max_{\theta,\phi}\mathcal{C}[\rho_{T}]$ (right) as a function $P_T$ in $\psi(3686) \rightarrow Y\bar{Y}$ for $Y = \Lambda$, $\Sigma^{+}$, $\Sigma^{0}$, $\Xi^{-}$ and $\Xi^{0}$.\label{fig:largestpsiPT}
}
\end{figure*}

In Fig.~\ref{fig:sig0Jpsi} the CHSH parameters and concurrence for $J/\psi \to \Sigma^0{\bar\Sigma}^0$ are shown as a complementary figures in the main text.
In Fig.~\ref{fig:negJpsi} the negativity of $J/\psi\rightarrow Y\bar{Y}$ for $Y = \Lambda$, $\Sigma^{+}$, $\Sigma^{0}$, $\Xi^{-}$ and $\Xi^{0}$ is shown under longitudinal and transverse beam polarization configurations.
For the transverse beam polarization, the dashed curve $\mathcal{N}(\theta = 0,\pi)=\mathcal{N}(\theta =\pi/2)$ in lower panels of Fig.~\ref{fig:negJpsi} divides each panel into two regions.
For the region above the dashed curve, the $\max_{\theta}\mathcal{N}$ locates at $\theta = 0,\pi$.
For the region below the dashed curve, the $\max_{\theta}\mathcal{N}$ locates at $\theta = {\pi/2}$.

The decay parameters of $\psi(3686) \rightarrow Y\bar{Y}$ for $Y = \Lambda$, $\Sigma^{+}$, $\Sigma^{0}$, $\Xi^{-}$ and $\Xi^{0}$ are collected in Table.~\ref{tab:decay_parameters_phi3686}.
The CHSH parameter $\mathcal{B}[\rho^{P_L}_{Y \bar Y}]$, negativity $\mathcal{N}[\rho^{P_L}_{Y \bar Y}]$ and concurrence $\mathcal{C}[\rho^{P_L}_{Y \bar Y}]$ are shown in Figs.~\ref{fig:entanglepsi} and \ref{fig:largestpsiPL} for $\psi(3686) \rightarrow Y\bar{Y}$ under longitudinal beam polarization configuration, and those under transverse beam polarization configuration are shown in Figs.~\ref{fig:transpsi} and \ref{fig:largestpsiPT}.
Under the transverse beam polarization, the dashed curve $\mathcal{B,N,C}(\theta = 0,\pi)=\mathcal{B,N,C}(\theta =\pi/2)$ in Fig.~\ref{fig:transpsi} divides each panel into two regions.
For the region above the dashed curve, the $\max_{\theta}\mathcal{B,N,C}$ locates at $\theta = 0,\pi$.
For the region below the dashed curve, the $\max_{\theta}\mathcal{B,N,C}$ locates at $\theta = {\pi/2}$.

\clearpage

\bibliography{mainv2}

\end{document}